\documentclass[aps,reprint,nofootinbib]{revtex4}
\usepackage{graphicx}  
\usepackage{dcolumn}   
\usepackage{bm}        
\usepackage{amssymb}
\usepackage{hyperref}
\usepackage{multirow}
\usepackage{amsmath}
\usepackage{cases}
\usepackage{placeins}
\usepackage{subfig}
\usepackage{natbib}
\usepackage{color}
\usepackage{comment}
\usepackage[normalem]{ulem} 
\usepackage{float}

\newcommand{ \Op}{O}

\begin{document}

\title{Constraints on lepton flavor universal and non-universal New Physics in $b\, \to\, s\, \ell^+ \ell^-$ decays: a global SMEFT survey}

\author{Md Isha Ali$^{(a)}{}$}
\email{isha.ali080@gmail.com} 
\author{Utpal Chattopadhyay$^{(a)}{}$}
\email{tpuc@iacs.res.in} 
\author{Dilip Kumar Ghosh $^{(a)}{}$}
\email{tpdkg@iacs.res.in}
\author{N Rajeev$^{(b)}{}$}
\email{rajeevneutrino@gmail.com}
\affiliation{\vspace{0.5cm}$^{(a)}$ School of Physical Sciences, Indian Association for the Cultivation of Science, \\ 2A \& 2B, Raja SC Mallick Rd, Jadavpur, Kolkata - 700032, India}
\affiliation{\vspace{0.5cm}$^{(b)}$
Theoretical Physics Division, Physical Research Laboratory, Ahmedabad - 380009, India}

\begin{abstract}
The flavor-changing neutral current semileptonic decays of $B$ mesons provide an excellent platform for indirectly probing New Physics (NP) beyond the Standard Model (SM). Recent measurements of lepton flavor universality (LFU) ratios such as $R_K$, $R_{K^*}$, and $R_\phi$ are consistent with SM predictions, thereby reducing earlier hints of LFU violation. Nevertheless, notable tensions persist in individual branching fractions---for instance $\mathcal{B}(B^+ \to K^+ \mu^+ \mu^-)$ and $\mathcal{B}(B^+ \to K^+ e^+ e^-)$, which deviate at the $4\sigma$--$5\sigma$ level in the low-$q^2$ region ($[1.1,6]$ GeV$^2$). Similarly, the angular observable $P'_5$ in $B \to K^{*} \mu^+ \mu^-$ shows a $3.3\sigma$ deviation, while $\mathcal{B}(B_s \to \phi \mu^+ \mu^-)$ departs from the SM by $3.6\sigma$. Although such discrepancies could in principle originate from underestimated hadronic effects, they may also indicate possible NP contributions. In this work, we perform a global fit to the latest $b \to s \ell^+ \ell^-$ data within the framework of dimension-6 SMEFT operators. 
Our analysis systematically incorporates hadronic uncertainties from form factors, 
non-factorizable corrections, and input parameters, thereby providing a balanced assessment of the NP interpretation. We allow for NP contributions not only in $b \to s \mu^+ \mu^-$ transitions but also in $b \to s e^+ e^-$ channels, thus testing both lepton flavor universal (LFU) and lepton flavor universality violating (LFUV) NP scenarios. The interplay between these possibilities yields distinctive signatures in branching fractions, angular observables, and $\Delta$-observables, offering a comprehensive view of the current flavor anomalies.
\end{abstract}

\maketitle

\section{Introduction}
The hierarchical structure of the Standard Model (SM) Yukawa couplings remain one of the biggest puzzles in particle physics. The resolution of this puzzle calls for the physics beyond the SM (BSM). To this end, various scenarios beyond the SM have been proposed \cite{Crivellin:2023zui}. While constructing these BSM scenarios, one should take utmost care to satisfy various limits on flavor-changing neutral current (FCNC) processes.
In other words, it is very crucial to achieve high precision in both theoretical predictions and experimental measurements of flavor-violating processes in relation to constraining BSM physics.

The long-standing B-anomalies traditionally attract a lot of attention in the flavor physics arena~\cite{Descotes-Genon:2012isb,Isidori:2010kg,Bobeth:2007dw,Altmannshofer:2008dz,Bobeth:2010wg,Matias:2012xw,Beaujean:2012uj,Altmannshofer:2014rta,Descotes-Genon:2015uva,Ciuchini:2015qxb,Beaujean:2013soa,Horgan:2013pva,Hurth:2013ssa,Descotes-Genon:2013wba,Rajeev:2021ntt,Mohapatra:2021ynn,Rajeev:2020aut,Das:2023kch,Dutta:2019wxo}. Apart from the difference in masses, the three charged leptons are not characteristically different in the SM. This is because the
electroweak couplings to gauge bosons are the same for all the lepton flavors leading to lepton flavor universality (LFU), arising out of an accidental symmetry. The long-standing B-anomalies traditionally attract a lot of attention in the flavor physics arena~\cite{London:2021lfn}. The lepton universality is tested in both charged current $b\to c\, \ell \nu$ $(\ell \in \mu,\tau)$ and neutral current $b\to s \ell^+\ell^-$ $(\ell \in e, \mu)$ transitions by comparing the decay rates of the bottom particle into different leptons. The previous results from the BaBar~\cite{BaBar:2008jdv,BaBar:2012mrf,BaBar:2016wgb}, Belle~\cite{Belle:2009zue,BELLE:2019xld} and LHCb~\cite{LHCb:2012bin,LHCb:2012juf,LHCb:2014cxe,LHCb:2014auh,LHCb:2014vgu,LHCb:2016due,LHCb:2021trn} collaborations revealed hints of deviations from lepton universality, however, none of them was statistically significant enough to constitute a clear signature of any new physics (NP) on their own. Among several observables, special attention was given to the ratios in which a large amount of theoretical uncertainties including the hadronic form factors and CKM matrix elements are reduced to a greater extent. 

The LFU tests by comparing the decay rates of B-mesons into second-generation lepton with those into first-generation lepton can conveniently be probed in $b\to s \ell^+\ell^-$ FCNC transitions. Previous measurements of the ratios of branching fractions $R_K$ and $R_{K^*}$ of the decays of $B \to K^{(*)} \ell^+ \ell^-$ ($\ell \in e,\mu$) indicated a hint of LFU break, with a significance of the $3.1\sigma$ standard deviation in $R_K$ and the same coherent pattern of deviations in $R_{K^*}$ but with a lower statistical sensitivity~\cite{LHCb:2021trn}. However, the latest updates on $R_K$ and $R_{K^*}$ from the LHCb experiment using the full Run 1 and Run 2 data samples are found to be consistent with the SM expectation at the level of the $0.2\sigma$ standard deviation~\cite{LHCb:2022vje}. The new $R_K$ and $R_{K^*}$ are reported in two different $q^2$ intervals as mentioned below,
\begin{equation}
 R_K=\frac{\mathcal{B}(B^+ \to K^+ \mu^+ \mu^-)}{\mathcal{B}(B^+ \to K^+ e^+ e^-)} =
 \begin{cases}
      0.994^{+0.090}_{-0.082} \text{(stat)} ^{+0.027}_{-0.029} \text{(syst)} & \text{for $q^2 \in [0.1, 1.1]$ $\rm GeV^2$}\\
      0.949^{+0.042}_{-0.041} \text{(stat)} ^{+0.023}_{-0.023} \text{(syst)} & \text{for $q^2 \in [1.1, 6]$ $\rm GeV^2$}
    \end{cases}
\end{equation}

\begin{equation}
 R_{K^*}=\frac{\mathcal{B}(B^0 \to K^{0*} \mu^+ \mu^-)}{\mathcal{B}(B^0 \to K^{0*} e^+ e^-)} =
 \begin{cases}
      0.927^{+0.093}_{-0.087} \text{(stat)} ^{+0.034}_{-0.033} \text{(syst)} & \text{for $q^2 \in [0.1, 1.1]$ $\rm GeV^2$}\\
      1.027^{+0.072}_{-0.068} \text{(stat)} ^{+0.027}_{-0.027} \text{(syst)} & \text{for $q^2 \in [1.1, 6]$ $\rm GeV^2$}
    \end{cases}
\end{equation}

The interesting fact about this measurement is that although the $R_K$ and $R_{K^*}$ values are close to SM, the individual branching fractions for $\mathcal{B}(B^+ \to K^+ \mu^+ \mu^-)$~\cite{LHCb:2014cxe} and $\mathcal{B}(B^+ \to K^+ e^+ e^-)$~\cite{LHCb:2022vje} and similarly $\mathcal{B}(B^0 \to K^{0*} \mu^+ \mu^-)$~\cite{LHCb:2016ykl} and $\mathcal{B}(B^0 \to K^{0*} e^+ e^-)$~\cite{LHCb:2022vje} deviate from their SM counterparts~\cite{LHCb:2022vje}. The most precise measurements of the differential branching fractions of $\mathcal{B}(B^+ \to K^+ \mu^+ \mu^-)$ and $\mathcal{B}(B^0 \to K^{0*} \mu^+ \mu^-)$ have been made using a data set corresponding to 3 $\rm fb^{-1}$ of integrated luminosity collected by the LHCb detector. However, the corresponding differential branching fractions with electrons in the final state are obtained by combining the latest $R_K$ and $R_{K^*}$ measurements at central $q^2$ with the known differential branching fractions of $B \to K^{(*)} \mu^+ \mu^-$, such that
\begin{eqnarray}
 \mathcal{B}(B^+ \to K^+ \mu^+ \mu^-) &=& (1.186 \pm 0.034 \pm 0.059) \times 10^{-7}, \\ \nonumber
 \mathcal{B}(B^0 \to K^{0*} \mu^+ \mu^-)&=& (2.018 \pm 0.100 \pm 0.053) \times 10^{-7}, \\ \nonumber
 \mathcal{B}(B^+ \to K^+ e^+ e^-)&=& (1.25 \pm 0.064 \pm 0.054) \times 10^{-7}, \\ \nonumber
 \mathcal{B}(B^0 \to K^{0*} e^+ e^-)&=& (1.63 \pm 0.13 \pm 0.11) \times 10^{-7}.
\end{eqnarray}

Apart from $R_K$, $R_{K^*}$ and $\mathcal{B}(B \to K^{(*)}\, \ell^+ \ell^-)$, there are more observables in $b\to s\, \ell^+\ell^-$ transitions that still exhibit consistent patterns of deviations from the SM predictions. The LHCb collaboration also studied the isospin partners of $B \to K^{(*)} \ell^+ \ell^-$ ($\ell \in e,\mu$) decays. In this measurement, the new ratios were labeled to be $R_{KS^0}$ and $R_{K^{*+}}$ which are consistent with SM at $1.4\sigma$ and $1.5\sigma$, respectively, and their combination at $2\sigma$ level~\cite{LHCb:2021lvy}. Interestingly, this result shows that the central values exhibit the same coherent pattern of deviation from lepton universality performed by earlier LHCb measurements. Another interesting test of LFU was performed by constructing angular observable $P'_5$ in $B \to K^* \mu^+ \mu^-$ decays. This observable was proposed based on the symmetries in the angular distribution of $B \to K^* \mu^+ \mu^-$ decays. Like $R_K$ and $R_{K^*}$, the $P'_5$ is considered a clean observable with a reduced level of hadronic uncertainties. Here, the dependence on
soft form factors cancels exactly at the leading order. The LHCb~\cite{LHCb:2020lmf}, ATLAS~\cite{ATLAS:2018gqc}, Belle~\cite{Belle:2016xuo} and CMS~\cite{CMS:2017rzx} have reported the measurement of $P'_5$ in different $q^2$ bins. A discrepancy of more than $3\sigma$ from SM~\cite{Descotes-Genon:2013vna,Descotes-Genon:2012isb,Descotes-Genon:2014uoa} is reported in the bin $q^2\in[4,6]\, \rm GeV^2$ as of LHCb~\cite{LHCb:2020lmf} and ATLAS~\cite{ATLAS:2018gqc} results are concerned. The LHCb collaboration also measured the branching fraction of $\mathcal{B}(B_s \to \phi \mu^+ \mu^-)$ decays which proceed via similar $b\to s \ell^+\ell^-$ quark level transitions. The reported values of the branching fraction deviate from SM predictions~\cite{Bobeth:2013uxa,Beneke:2019slt} at $3\sigma$ concerning the Run I data and $3.6\sigma$ with respect to both the Run I and Run II data from the LHCb collaboration~\cite{LHCb:2021zwz,LHCb:2013tgx,LHCb:2015wdu}. Recently, the LHCb collaboration for the first time reported the measurement of the LFU sensitive ratio $R_\phi$\footnote{In practice, $R_\phi^{-1}$ is measured rather than $R_\phi$ such that the small $B_s \to \phi\, e^+ e^-$ yield appears in the numerator and the statistical behaviour of the observable more closely follows a Gaussian distribution~\cite{LHCb:2024rto}.} in $B_s \to \phi \ell^+ \ell^-$ decays in three different $q^2$ bins i.e $[0.1, 1.1]$, $[1.1, 6]$ and $[15, 19]$ GeV$^2$ leading to:

\begin{equation}
 R_\phi^{-1}=\frac{\mathcal{B}(B_s \to \phi e^+ e^-)}{\mathcal{B}(B_s \to \phi \mu^+ \mu^-)} =
 \begin{cases}
      1.57^{+0.28}_{-0.25} \pm 0.05 & \text{for $q^2 \in [0.1, 1.1]$ $\rm GeV^2$},\\
      0.91^{+0.20}_{-0.19} \pm 0.05 & \text{for $q^2 \in [1.1, 6]$ $\rm GeV^2$},\\
      0.85^{+0.24}_{-0.23} \pm 0.10 & \text{for $q^2 \in [15, 19]$ $\rm GeV^2$}.
    \end{cases}
\end{equation}

The result agrees with the SM expectations~\cite{LHCb:2024rto}. With this measurement, LHCb also reports the first observation of $B_s \to \phi\, e^+ e^-$ decay. All the numerical values of different observables are tabulated in Table~\ref{status}.

\begin{table}[h!]
\centering
\setlength{\tabcolsep}{6pt} 
\renewcommand{\arraystretch}{1.25} 
\resizebox{\columnwidth}{!}{
\begin{tabular}{|c|c|c|c|c|}
\hline
       & $q^2$ bins $\rm (GeV^2)$ & Theoretical predictions & Experimental measurements & Deviation\\
\hline
\hline
\multicolumn{5}{|c|}{$b \to s \ell^+ \ell^-$ observables} \\
\hline
\multirow{2}{*}{$R_K$}  & [0.1, 1.1] & $1 \pm 0.01$~\cite{Bordone:2016gaq,Hiller:2003js} & $0.994^{+0.090+0.027}_{-0.082-0.029} $~\cite{LHCb:2022vje} & - \\
\cline{2-4}
                           & [1.1, 6.0] & $1 \pm 0.01$~\cite{Bordone:2016gaq,Hiller:2003js} & $0.949^{+0.042+0.023}_{-0.041-0.023}$~\cite{LHCb:2022vje} & \\
\hline
$R_{K_S^0}$  & [1.1, 6.0] & $1 \pm 0.01$~\cite{Bordone:2016gaq,Hiller:2003js} & $0.66^{+0.20}_{-0.14}$ (stat) $^{+0.02}_{-0.04}$ (syst)~\cite{LHCb:2021lvy} & $1.4\sigma$ \\
\hline
\multirow{2}{*}{$R_{K^*}$} & [0.045, 1.1] & $1 \pm 0.01$~\cite{Bordone:2016gaq,Hiller:2003js} & $0.927^{+0.093+0.034}_{-0.087-0.033} $~\cite{LHCb:2022vje} & \multirow{2}{*}{-}\\
                                                         \cline{2-4}
                           & [1.1, 6.0]   & $1 \pm 0.01$~\cite{Bordone:2016gaq,Hiller:2003js} & $1.027^{+0.072+0.027}_{-0.068-0.027}$ (stat) $\pm 0.047$ (syst)~\cite{LHCb:2022vje} & \\
                                                       \hline
$R_{K^{*+}}$  & [0.045, 6.0] & $1 \pm 0.01$~\cite{Bordone:2016gaq,Hiller:2003js} & $0.70^{+0.18}_{-0.13}$ (stat) $^{+0.03}_{-0.04}$ (syst)~\cite{LHCb:2021lvy} & $1.5\sigma$ \\
\hline
\multirow{3}{*}{$R_\phi^{-1}$} & [0.1, 1.1] & $1.016$~\cite{Straub:2018kue} & $1.57^{+0.28}_{-0.25} \pm 0.05$~\cite{LHCb:2024rto} & \multirow{3}{*}{-}\\
                                                         \cline{2-4}
                           & [1.1, 6.0] & $1.003$~\cite{Straub:2018kue} & $0.91^{+0.20}_{-0.19} \pm 0.05$~\cite{LHCb:2024rto} & \\
                           \cline{2-4}
                           & [15, 19]   & $1.002$~\cite{Straub:2018kue} & $0.85^{+0.24}_{-0.23} \pm 0.10$~\cite{LHCb:2024rto} & \\
                                                       \hline
\multirow{3}{*}{$P_{5}^{\prime}$} & \multirow{1}{*}{[4.0, 6.0]} & $-0.759 \pm 0.071$~\cite{Descotes-Genon:2013vna} & $-0.439 \pm 0.111 \pm 0.036$~\cite{LHCb:2020lmf} & $3.3\sigma$\\ \cline{2-5}
                                  & \multirow{1}{*}{[4.3, 6.0]} & $-0.795 \pm 0.065$~\cite{Descotes-Genon:2012isb} & $-0.96^{+0.22}_{-0.21}$ (stat) $\pm 0.25$ (syst)~\cite{CMS:2017rzx} & $ 1.0\sigma$\\  \cline{2-5}
                                  & \multirow{1}{*}{[4.0, 8.0]} & $-0.795 \pm 0.054$~\cite{Descotes-Genon:2014uoa} & $-0.267^{+0.275}_{-0.269}$ (stat) $\pm 0.049$ (syst)~\cite{Belle:2016xuo} & $2.1\sigma$\\
\hline
$\mathcal{B}(B^+ \to K^+ \mu^+ \mu^-)$  & {[1.1, 6.0]} & $(1.708 \pm 0.283) \times 10^{-7}$~\cite{Parrott:2022zte} & $(1.186 \pm 0.034 \pm 0.059) \times 10^{-7}$~\cite{LHCb:2022vje} & $4.2\sigma$ \\
\hline
$\mathcal{B}(B^0 \to K^{0*} \mu^+ \mu^-)$ & {[1.1, 6.0]} & $(2.323 \pm 0.381)\times 10^{-7}$~\cite{Bharucha:2015bzk} & $(2.018 \pm 0.100 \pm 0.053) \times 10^{-7}$~\cite{LHCb:2016ykl} & - \\
\hline
$\mathcal{B}(B_s\, \to\, \phi\, \mu^+\, \mu^-)$ & {[1.1, 6.0]} & $(2.647 \pm 0.319)\times 10^{-7}$~\cite{Aebischer:2018iyb,Bharucha:2015bzk} & $(1.41 \pm 0.073 \pm 0.024 \pm 0.068)\times 10^{-7}$~\cite{LHCb:2021zwz,LHCb:2013tgx,LHCb:2015wdu} & $3.6\sigma$\\
\hline
$\mathcal{B}(B_s\, \to \mu^+\, \mu^-)$ & - & $(3.672 \pm 0.154)\times 10^{-9}$~\cite{Bobeth:2013uxa,Beneke:2019slt} & $(3.361 \pm 0.028)\times 10^{-9}$~\cite{Greljo:2022jac} & - \\
\hline
$\mathcal{B}(B^+ \to K^+ e^+ e^-)$  & {[1.1, 6.0]} & $(1.707 \pm 0.290) \times 10^{-7}$~\cite{Parrott:2022zte} & $(1.25 \pm 0.064 \pm 0.054) \times 10^{-7}$~\cite{LHCb:2022vje} & $4.8\sigma$\\
\hline
$\mathcal{B}(B^0 \to K^{0*} e^+ e^-)$ & {[1.1, 6.0]} & $(2.331 \pm 0.377)\times 10^{-7}$~\cite{Bharucha:2015bzk} & $(1.63 \pm 0.13 \pm 0.11) \times 10^{-7}$~\cite{LHCb:2022vje} & - \\
\hline
$\mathcal{B}(B^0 \to \phi\, e^+ e^-)$ & {[1.1, 6.0]} & $(2.622 \pm 0.327) \times 10^{-7}$ ~\cite{Bharucha:2015bzk} & $(1.274 \pm 0.294 \pm 0.049 \pm 0.049 \pm 0.049) \times 10^{-7}$~\cite{LHCb:2024rto} & - \\
\hline
\hline
\multicolumn{5}{|c|}{$b \to s\, \nu\, \bar{\nu}$ observables} \\
\hline
$\mathcal{B}(B^+ \to K^+ \nu \nu)$ & - & $(4.29 \pm 0.13)\times 10^{-6}$~\cite{Becirevic:2023aov,Buras:2022wpw} & $(2.3 \pm 0.7) \times 10^{-5}$~\cite{Belle-II:2023esi} & $2.7\sigma$ \\
                                                  \hline
$\mathcal{B}(B^0 \to K^0 \nu \nu)$ & - & $(4.1 \pm 0.5)\times 10^{-6}$~\cite{Felkl:2021uxi} & $< 2.6 \times 10^{-5}$~\cite{Belle:2017oht} & - \\
\hline
$\mathcal{B}(B^0 \to K^{0*} \nu \nu)$ & - & $(11.6 \pm 1.1)\times 10^{-6}$~\cite{Felkl:2021uxi} & $< 1.8 \times 10^{-5}$~\cite{Belle:2017oht} & - \\
\hline
$\mathcal{B}(B^+ \to K^{+*} \nu \nu)$ & - & $(12.4 \pm 1.2)\times 10^{-6}$~\cite{Felkl:2021uxi} & $< 4.0 \times 10^{-5}$~\cite{Belle:2013tnz} & - \\
\hline
\end{tabular}}
\caption{Current status of the prominent observables in $b\,\to\,s\,\ell^+\,\ell^-$ and $b \, \to \,s\, \nu \, \bar{\nu}$ decays.}
\label{status}
\end{table}

The impact of hadronic uncertainties on the observables of B decay has been a topic of intense debate for a long time~\cite{Ball:1998kk,Beneke:2000wa,Bauer:2000yr,Beneke:2003pa,Gubernari:2018wyi,FlavourLatticeAveragingGroupFLAG:2021npn,Beneke:2001at,Gubernari:2020eft}. The hadronic uncertainties may have local and non-local contributions~\cite{Descotes-Genon:2014uoa,Capdevila:2017ert}. Form factors are associated with local contributions and parameterize the transition of the initial B meson into the final state $K$ or $K^*$ meson. On the other hand, the nonlocal contributions are connected with the charm-quark loops.
Despite the recent progress in lattice calculations, the theoretical predictions of these decay rates suffer from relatively larger uncertainties in $B \to K^{(*)}$ form factor calculations. The determination of form factors based on the light-cone sum rule (LCSR) applies at low values of $q^2$, while the lattice QCD calculations have only been applied to high $q^2$ regions~\cite{Khodjamirian:2010vf,Gubernari:2018wyi,Bharucha:2015bzk}.
The recently improved lattice QCD determination of form factors from HPQCD collaboration has reported the differential branching fraction for $B \to K \ell^+ \ell^-$ ($\ell \in e,\mu$) decays~\cite{Parrott:2022zte}. These form factors are calculated for the full $q^2$ range of decay and have less uncertainties than the previous work, particularly in the low $q^2$ zone~\cite{Parrott:2022rgu}. The branching fraction of SM is reported to exceed LHCb results with tension as high as $4.2\sigma$ for $B \to K \mu^+ \mu^-$ in the low $q^2$ and $2.7\sigma$ in the high $q^2$ region~\cite{Parrott:2022zte}. Similarly, for $B \to K e^+ e^-$ the SM branching fraction exceeds the LHCb~\cite{LHCb:2014cxe} measurement with a deviation of $4.8\sigma$ in the low $q^2$ domain~\cite{Parrott:2022zte}. On the other hand, there is no issue as regards the Belle~\cite{BELLE:2019xld} results, which, however, are in tension with LHCb~\cite{LHCb:2014cxe} for $B \to K \mu^+ \mu^-$ and also have larger uncertainties than those from LHCb and SM. A more detailed discussion can be found in~\cite{Parrott:2022zte}. Similarly, very precise estimations of $B \to K^*$ form factors were obtained in the LCSR~\cite{Khodjamirian:2010vf,Ball:2004rg,Bharucha:2015bzk} and LQCD~\cite{Horgan:2013hoa,Liu:2011raa,Gubernari:2018wyi,Horgan:2013pva}methods. The LCSR form factors are usually considered valid in $q^2 \in [0, 14]$ GeV$^{2}$. It is customary that the lattice computations are extrapolated to low $q^2$ to reduce the uncertainty in the LCSR computations. In that sense, the combined estimation of $B \to K^*$ form factors by combining LCSR with LQCD is also reported in Ref~\cite{Bharucha:2015bzk}. The SM branching fraction for $B^0 \to K^{0*} \mu^+ \mu^-$ exceeds the LHCb~\cite{LHCb:2016ykl} result in the low $q^2$.

Regarding non-local effects, it is challenging to distinguish whether the contributions to $C_9^{\rm eff}$ are due to charm loop effects or NP. Knowing these charm-loop effects is important, particularly near the regions of charmonium resonances, as they may affect branching fractions or angular observables in the presence of NP. These are often treated via model-dependent methods (e.g., dispersion relations)~\cite{Capdevila:2017ert,Khodjamirian:2010vf,Gubernari:2020eft,Blake:2017fyh,Bobeth:2017vxj} or parametrized using nuisance parameters in tools like \texttt{flavio}~\cite{Straub:2018kue}. To mitigate these, theoretical predictions are generally restricted to the low-recoil window $q^2 \in [1.1, 6]$ GeV$^2$, where such effects are relatively suppressed. In our analysis we have taken a conservative
approach by focusing on observables and kinematic windows restricted to the low-recoil window $q^2 \in [1.1, 6]$ GeV$^2$. 

In addition, there might be contributions arising from virtual charm loops emitting soft gluons ($|k|^2 << 4m_c^2$). Previous LCSR-based analyses using B-meson LCDAs estimated these to be significant~\cite{Khodjamirian:2010vf,Gubernari:2020eft}. However, revised calculations (including sub-leading terms) find the effects to be suppressed by up to three orders of magnitude. Recent studies even show vanishing contributions up to twist-4 for $B \to K\, \ell^+ \ell^-$, with minimal impact from kaon mass and higher-twist corrections. Similar results hold for $B \to K^*\, \ell^+ \ell^-$, suggesting that these soft charm-loop effects can be safely neglected~\cite{Mahajan:2024xpo}.

Interestingly, the treatment of hadronic uncertainties has been approached in distinct ways in the literature. Ciuchini et al.~\cite{Ciuchini:2022wbq} perform their analysis adopting a fully data-driven treatment of charming penguin contributions. They show that once the hadronic effects are flexibly absorbed in the fit, the so-called anomalies essentially vanish. 
In this case, the SM predictions remain well within the $68\%$ probability region, leading to the conclusion that there is no compelling need for NP. This represents a conservative, SM-friendly position.

On the other hand, in our analysis we adopt a balanced but NP-leaning stance, 
where we systematically explore uncertainties from form factors, non-factorizable hadronic (charm-loop) corrections, and SM input parameters. With moderate assumptions---that is, uncertainties adding up to $20$--$25\%$---our global fits show a considerable preference for an NP contribution. A similar viewpoint is taken in the recent work of Hurth et al.~\cite{Hurth:2025vfx}, where the analysis is based on QCD factorization but systematically explores uncertainties from form factors, low-$q^2$ bin choices, and non-factorizable power corrections (varied at the $10\%$, $50\%$, and $100\%$ level). With $10$--$20\%$ corrections, the global fits display a $\sim 3$--$6\sigma$ preference for an NP contribution, notably with $\delta C_9 < 0$. They also argue that eliminating this NP preference would require assuming unrealistically large power corrections ($\sim 100\%$), which undermines the plausibility of a purely hadronic explanation.

Based on available theoretical and experimental inputs, global fit analyses were performed by several groups~\cite{Descotes-Genon:2015uva,Alguero:2019ptt,Capdevila:2017bsm,Alguero:2021anc,Altmannshofer:2021qrr,Hurth:2021nsi}. Most of the global analyses are consistent with one another and also they identify the most preferable NP solutions in the vector and axial-vector WCs $C_{9\mu}$ and $C_{10\mu}$ respectively. Many of the above works include both 1D and 2D scenarios in which the most preferred NP explanations were found in $C_{9\mu}$ and $C_{9\mu}=-C_{10\mu}$ 1D hypothesis. and $(C_{9\mu}, C_{10\mu})$, $(C_{9\mu}, C'_{10\mu})$, $(C_{9\mu}=-C_{10\mu}$, $C'_{9\mu}=-C'_{10\mu})$ and $(C_{9\mu}, C'_{9\mu}=-C'_{10\mu}$ 2D hypotheses. Moreover, the authors in Ref.~\cite{Capdevila:2017bsm} perform a fit with 6 couplings at a time (6D)
which shows consistency with both 1D and 2D fits, and particularly confirms more dominant NP effects in $C_{9\mu}$. As discussed in the previous paragraph, additionally these NP effects could also be mimicked by the non-local charm loop effects. Interestingly, the SM-like signature of $\mathcal{B}(B_s \to \mu^+ \mu^-)$ indicates absence of NP in the axial-vector current appearing in $C_{10}$. Moreover, the new SM-like results in $R_K$ and $R_{K^*}$, when interpreted under reasonable hadronic assumptions, are consistent with the SM and remain compatible with scenarios of lepton flavor universal NP (LFU-NP) affecting both $b \to s \mu^+ \mu^-$ and $b \to s e^+ e^-$ transitions. ~\cite{Alguero:2019ptt,Alguero:2021anc,Alguero:2018nvb,Hurth:2021nsi,Kumar:2019qbv,Datta:2019zca}.

To estimate the NP scale behind these anomalies, in particular the process of $b \to s \mu^+ \mu^-$, we will use the Standard Model Effective Field Theory 
(SMEFT) formalism with the relevant higher dimensional operators consisting of SM fields. For example the dimension-6 operator $(1/\Lambda^2) (\bar{b}\gamma^\mu s)(\bar{\mu}\gamma_\mu \mu)$ requires $\Lambda$ to be roughly of the order of $\sim 40$ TeV~\cite{Blanke:2024itg} to satisfy the current anomalies. Although such a large NP scale is beyond the reach of LHC, it can offer promising physics prospects at future multi-TeV muon colliders~\cite{Huang:2021biu,Azatov:2022itm}. The prospects of $\mu^+ \mu^- \to bs$ at 10 TeV muon collider have been studied in Ref~\cite{Altmannshofer:2023uci}. 

In this article, motivated by the new SM-like results in $R_K$ and $R_{K^*}$, 
we revisit the global SMEFT interpretation of $b \to s \ell^+ \ell^-$ processes. 
While these LFU ratios are consistent with the SM, persistent deviations remain in individual branching fractions and angular observables. Under moderate hadronic assumptions, such tensions can still be indicative of possible NP effects, thereby motivating a systematic exploration within the SMEFT framework. To this end, we perform global fits to the dimension-6 SMEFT coefficients, treating the muon and electron flavor indices as distinct. 
We construct several benchmark scenarios, classified into LFU and LFU-violating types, 
and estimate the corresponding bounds on the low-energy Wilson coefficients at the $m_b$ scale through tree-level matching. Our fits include a comprehensive set of observables, such as branching fractions and angular distributions (forward-backward asymmetry, longitudinal polarization fraction, and optimized $P_i^{(\prime)}$ parameters) for both electron and muon final states. We further test lepton flavor universality through $\Delta$-observables.

The paper is organized as follows: In Sec.~\ref{smeftphenomenology} we begin with a brief overview of SMEFT phenomenology. We continue the discussion by writing down the low energy effective Hamiltonian for $b \to s \ell^+ \ell^-$ transitions. Then we provide the matching relations of SMEFT operators to LEFT operators at tree level. In Sec.~\ref{globalfitsmeft} we discuss the global fit results of SMEFT coefficients and bounds on the respective LEFT coefficients. In Sec.~\ref{results1} we discuss the results of various observables in $B \to K^{(*)} \ell^+ \ell^-$ decays such as the branching fractions, angular observables including the longitudinal polarization fraction, the forward-backward asymmetry and some $P_i$ observables. In Sec.~\ref{results2} we study the sensitivity of LFUV in $\Delta$-observables. Finally, we summarize and conclude our discussion in Sec.~\ref{summary}.


\section{SMEFT Phenomenology}
\label{smeftphenomenology}
The SM can be regarded as an effective field theory (EFT) which is valid up to a certain NP scale $\Lambda$. At this scale, one can consider higher-dimensional operators as given below~\cite{Buchmuller:1985jz}, 

\begin{equation}
    \mathcal{L}_{\rm SMEFT} = \mathcal{L}_{\rm SM}^{(4)} + \frac{1}{\Lambda} C_{\nu\nu}^{(5)} O_{\nu\nu}^{(5)} + \frac{1}{\Lambda^2} \sum_k C_{k}^{(6)} O_{k}^{(6)} + \mathcal{O} \bigg(\frac{1}{\Lambda^3} \bigg),
\end{equation}
where, $\mathcal{L}_{\rm SM}^{(4)}$ is the renormalizable SM Lagrangian containing operators with with dimension-2 and dimension-4, $O_{\nu\nu}^{(5)}$ is the dimension-5 Weinberg operator which gives masses to the neutrinos~\cite{Weinberg:1979sa}, and, $O_{k}^{(6)}$ is the set of dimension-6 operators~\cite{Buchmuller:1985jz,Grzadkowski:2010es}. For three generations of fermions, there are 2499 independent dimension-6 operators that do not violate lepton and baryon numbers~\cite{Alonso:2013hga}. The NP effects are parameterized in terms of the Wilson coefficients (WCs) associated with these operators. Therefore, a model-independent search of NP is useful by studying SM extended with higher-dimensional gauge-invariant operators~\cite{Buchmuller:1985jz}. In contrast to testing specific NP models at the LHC for viability, an EFT framework allows us to obtain the constraints on the WCs of higher-dimensional operators. Later, one can map a model to the EFT to examine  
whether the model is compatible with the experiment. 

In EFT, one considers the running and mixing of the Wilson coefficients from $\Lambda$ to the scale of appropriate collider experiments and further down to the scale of low-energy experiments. The leading effects are obtained from the result of the divergent part of the one-loop calculations~\cite{Jenkins:2017jig}. The one-loop anomalous dimension matrix of dimension-6 operators is then computed in the SMEFT framework~\cite{Alonso:2013hga,Alonso:2014zka,Grojean:2013kd,Jenkins:2013zja,Jenkins:2013wua}. The flavor structure of SMEFT can have significant implications, particularly because of the mixing arising from renormalization group evolutions. 
One evolves the SMEFT RGEs that may also involve mixing of WCs from the BSM scale down to the electroweak scale. Below the electroweak scale, the running and mixing should be calculated in low energy effective theory (LEFT)~\cite{Buchmuller:1985jz,Aebischer:2015fzz,Jenkins:2017dyc,Jenkins:2017jig}. Using the complete information of SMEFT and LEFT along with the matching of WCs at the tree level, one can deduce the low energy consequences of NP. We start with SMEFT operators at a high scale and run down to $M_Z$, we match to LEFT and then we again run the LEFT operators down to the desired low energy experimental observable~\cite{Jenkins:2017jig}~\footnote{In our analysis, we do not assume any specific flavor model at the UV scale. For operator matching and RGE evolution, we work in the down-quark mass basis (diagonal $Y_d$). SM parameters are evolved using known RGEs: gauge couplings at 3-loop, and Yukawa couplings and quark masses at 2-loop. Input values such as $\alpha_s$, $y_t$, and quark masses are taken in the $\overline{\text{MS}}$ scheme at reference scales (typically $m_Z$ or $m_t$) and run up to the SMEFT scale. The CKM matrix elements remain unaltered~\cite{Straub:2018kue}.}.

\subsection{Low energy effective Hamiltonian}
The most general effective Hamiltonian describing (${\Delta B = \Delta S =1}$) $b \to s \ell^+ \ell^-$ transition can be written as~\cite{Aebischer:2015fzz}

\begin{equation}
  \mathcal{H}_{\rm eff} =
  -\frac{4G_F}{\sqrt{2}}
  \left(   \sum_{i}   C_i\, \Op_i+C'_i\, \Op'_i
  +\sum_{i} \sum_q  C_i^q \, \Op_i^q +C_i^{'q} \, \Op_i^{'q} \right)\,,
  \label{eqn:Heff}
\end{equation}
where $G_F$ is the Fermi coupling constant, the primed operators represent the right-chiral structure and are obtained by interchanging $P_L \leftrightarrow P_R$. We group the set of operators in Hamiltonian into two parts based on their structure. 
We consider the first set of operators as,

\begin{align}
  \Op_1 &= (\bar{s} \, T^{A} \gamma_\mu P_L c) \; 
  (\bar{c} \, T^{A} \gamma^\mu P_L b)\,,&
  \Op_2 &=(\bar{s} \gamma_\mu P_L  c) \;
  (\bar{c} \gamma^\mu P_L b),\notag\\
  \Op_7 &= \frac{e}{16\pi^2} m_b \,  ( \bar{s}  \, \sigma_{\mu\nu} P_R \,  b ) \; F^{\mu\nu},&
  \Op_8 &= \frac{g_s}{16\pi^2} m_b \,
  ( \bar{s}\, T^{A} \sigma_{\mu\nu} P_R \, \, b ) \; G^{\mu\nu \, A}\,,\notag \\
  \Op_9 &= \frac{e^2}{16 \pi^2}  (\bar{s}\,\gamma_\mu P_L b) \;
    (\bar{\ell} \gamma^\mu \ell)\,, &
  \Op_{10} &=\frac{e^2}{16 \pi^2}
   (\bar{s} \gamma_\mu P_L b) \;    (\bar{\ell} \gamma^\mu \gamma_5 \ell)\,, \label{eqn:O10}\notag \\
  \Op_S &= (\bar{s} P_R b) \; (\bar{\ell} \ell)\,,&
  \Op_P &= (\bar{s} P_R b) \; (\bar{\ell} \gamma_5 \ell)\,,\notag\\
  \Op_T & =(\bar{s} \sigma^{\mu\nu} b) \; (\bar{\ell} \sigma_{\mu\nu} \ell)\,,&
  \Op_{T5} & =(\bar{s} \sigma^{\mu\nu} b) \; (\bar{\ell} \sigma_{\mu\nu} \gamma_5 \ell)\,.
\end{align}

In the second part of the Hamiltonian, we exclusively have four-quark operators containing vectorial $(\Op_3^q - \Op_6^q)$, scalar and tensor $(\Op_{15}^q - \Op_{20}^q)$~\cite{Borzumati:1999qt} Lorentz structures. The summation over $q$ in Eq.\ref{eqn:Heff} represents all the light quarks $q=u,d,c,s,b$. The explicit forms of these operators are given by,
\begin{align}
  \Op_3^q &= (\bar{s} \gamma_\mu P_Lb)\;   ( \bar{q} \gamma^\mu q ) &
  \Op_4^q &=(\bar{s}\, T^{A} \gamma_\mu P_L b)\;
  (\bar{q}\, T^{A} \gamma^\mu q )\,, \notag\\
  \Op_5^q&=(\bar{s}\gamma_\mu \gamma_\nu \gamma_\rho P_L b)\;
  ( \bar{q} \gamma^\mu \gamma^\nu \gamma^\rho q)\,, &
  \Op_6^q &=(\bar{s}\, T^{A}\gamma_\mu \gamma_\nu \gamma_\rho P_L b)\;
  (\bar{q}\, T^{A} \gamma^\mu \gamma^\nu \gamma^\rho q)\,,\notag\\
  \Op_{15}^q &=  (\bar{s}P_R b) (\bar{q} P_R q)\,, &
  \Op_{16}^q &=  (\bar{s}_\alpha P_R b_\beta) (\bar{q}_\beta P_R q_\alpha)\,, \notag \\
  \Op_{17}^q &=  (\bar{s}P_R b) (\bar{q} P_L q)\,, &
  \Op_{18}^q &=  (\bar{s}_\alpha P_R b_\beta) (\bar{q}_\beta P_L q_\alpha)\,, \notag\\
  \Op_{19}^q &=  (\bar{s} \sigma^{\mu\nu} P_R b) (\bar{q} \sigma_{\mu\nu} P_R q)\,, &
  \Op_{20}^q &=  (\bar{s}_\alpha \sigma^{\mu\nu} P_R b_\beta) (\bar{q}_\beta \sigma_{\mu\nu} P_R q_\alpha)\,.
\end{align}

\subsection{Matching of SMEFT to LEFT at tree level}
In relation to studying low energy observables in an SMEFT framework there are three different energy scales like i) the high scale $\Lambda$ where the SMEFT WCs corresponding to higher dimensional gauge invariant operators composed of SM fields are given as inputs, ii) the electroweak scale like that of W-boson mass $m_W$, and iii) a low energy scale corresponding to the observables under study like the mass of the bottom quark $m_b$ or any further low scale like that of lepton masses. In the "match and run" procedure for the RGEs, at $\Lambda$ one starts with unit values of the WCs under study while all other coefficients are put to zero. The RGEs at one-loop level evolved up to the electroweak scale, which may also include mixing effects, leading to coefficients appropriate to the same scale physics such as that involving Higgs or Z-bosons~\cite{Alonso:2013hga,Alonso:2014zka,Grojean:2013kd,Jenkins:2013zja,Jenkins:2013wua}. Below this scale, the heavy particles of SM are integrated out, and the fields of light particles form the higher dimensional LEFT operators that do not obey the electroweak symmetry~\cite{Jenkins:2017jig}. For example, the effective Hamiltonian written in Eq.~\ref{eqn:Heff} governing $b \to s$ transitions is invariant under only $SU(3)_{\rm C} \times U(1)_{\rm EM}$~\cite{Buras:1998raa,Buchalla:1995vs}. 
The SMEFT and LEFT operator coefficients are appropriately matched at the electroweak scale in this match-and-run procedure at the tree level. Multiple SMEFT operator coefficients may contribute to a given LEFT coefficient at the matching scale. The LEFT operators are then evolved at one-loop level to a low energy scale like $m_b$ and the coefficients are used to compute relevant low energy processes~\cite{Crivellin:2013hpa,Crivellin:2014cta,Pruna:2014asa,Ali:2023kua,Bhattacharya:2014wla, Alonso:2015sja,Calibbi:2015kma,Alonso:2014csa,Buras:2014fpa}.

In Table~\ref{tab:operators1} we list all the gauge invariant operators contributing to $b \to s$ transitions at tree level~\cite{Aebischer:2015fzz}. All operators in Table~\ref{tab:operators1} involve the quark fields and are on the interaction basis as these SMEFT operators are defined at a scale higher than that of electroweak symmetry breaking. The operators are classified into various categories: for example, 
 $(\bar{L}L)(\bar{L}L)$ and $(\bar{R}R)(\bar{R}R)$ represent the product of two left-handed and right-handed vector currents respectively, whereas,
$(\bar{L}L)(\bar{R}R)$ represents the product of left-handed and a right-handed vector current operators and, $(\bar{L}R)(\bar{R}L)$ or $(\bar{L}R)(\bar{L}R)$ represents the product of either a scalar or tensor current. The operators in all these classes are divided into purely leptonic, semileptonic, and nonleptonic operators. The remaining classifications including $\psi^2 X \phi$, $\psi^2 \phi^3 $ and $\psi^2 \phi^2 D $ represent the product of left or right-handed quark currents and boson fields~\cite{Grzadkowski:2010es}. 
\begin{table}[ht]
  \centering
  \renewcommand{\arraystretch}{1.2}
  \small
  \begin{tabular}{|c|c|c|c|c|c|}
    \hline
    \multicolumn{2}{|c|}{$(\bar{L}R)(\bar{R}L)$ or $(\bar{L}R)(\bar{L}R)$}&
    \multicolumn{2}{|c|}{$(\bar{L}L)(\bar{L}L)$} &
    \multicolumn{2}{|c|}{$\psi^2 X \phi$} \\
    \hline
    $Q_{\ell edq}$ &
    $(\bar{\ell }_i^a e_j) (\bar{d}_k q_l^a)$ &
    $Q_{qq}^{(1)}$  &
    $(\bar q_i \gamma_\mu q_j)(\bar q_k \gamma^\mu q_l)$ &
    $Q_{dW}$ &
    $( \bar{q}_i \sigma^{\mu\nu} d_j ) \tau^{\small I} \phi W^{\small I}_{\mu\nu}$ \\
    $Q_{quqd}^{(1)}$ &
    $(\bar q^a_i u_j) \epsilon_{ab} (\bar q^b_k d_l)$ &
    $Q_{\ell q}^{(1)}$                &
    $(\bar \ell_i \gamma_\mu \ell_j)(\bar q_k \gamma^\mu q_l)$ &
    $Q_{dB}$ &
    $( \bar{q}_i \sigma^{\mu\nu} d_j ) \phi B_{\mu\nu}$ \\
    $Q_{quqd}^{(8)}$ &
    $(\bar q^a_i T^{\small A}u_j) \epsilon_{ab} (\bar q^b_k T^{\small A}d_l)$   &
    $Q_{qq}^{(3)}$  &
    $(\bar q_i \gamma_\mu \tau^I q_j)(\bar q_k \gamma^\mu \tau^I q_l)$ &
    $Q_{dG}$ &
    $(\bar{q}_i \sigma^{\mu\nu} T^{\small A}d_j) \phi G^{\small A}_{\mu\nu}$ \\\cline{1-2}\cline{5-6}
    \multicolumn{2}{||c|}{$(\bar{L}L)(\bar{R}R)$} &
    $Q_{\ell q}^{(3)}$                &
    $(\bar \ell_i \gamma_\mu \tau^I \ell_j)(\bar q_k \gamma^\mu \tau^I q_l)$  &
   \multicolumn{2}{|c||}{$\psi^2 \phi^3 $} \\\cline{1-2}\cline{5-6}
    $Q_{\ell d}$ &
    $(\bar \ell_i \gamma_\mu \ell_j)(\bar d_k \gamma^\mu d_l)$ &
    & &
    $Q_{d\phi}$  &
    $(\phi^{\dag}\phi)(\bar q_i\,d_j\,\phi)$  \\ \cline{3-6}
    $Q_{qe}$    &
    $(\bar q_i \gamma_\mu q_j)(\bar e_k \gamma^\mu e_l)$ &
    \multicolumn{2}{|c|}{$(\bar{R}R)(\bar{R}R)$} &
    \multicolumn{2}{|c||}{$\psi^2 \phi^2 D $}
    \\ \cline{3-6}
    $Q_{qu}^{(1)}$         &
    $(\bar q_i \gamma_\mu q_j)(\bar u_k \gamma^\mu u_l)$ &
    $Q_{dd}$        &
    $(\bar d_i \gamma_\mu d_j)(\bar d_k \gamma^\mu d_l)$ &
    $Q_{\phi q}^{(1)}$ &
    $( \phi^\dagger i D_\mu \phi )
    ( \bar{q}_i \gamma^\mu q_j)$ \\
    $Q_{qd}^{(1)}$         &
    $(\bar q_i \gamma_\mu q_j)(\bar d_k \gamma^\mu d_l)$ &
    $Q_{ed}$  &
    $(\bar e_i \gamma_\mu e_j)(\bar d_k \gamma^\mu d_l)$ &
    $Q_{\phi q}^{(3)}$ &
    $( \phi^\dagger i D_\mu \! ^{\small I} \phi )
    ( \bar{q}_i \tau^{\small I} \gamma^\mu q_j )$ \\
    $Q_{qu}^{(8)}$         &
    $(\bar q_i \gamma_\mu T^{\small A} q_j)(\bar u_k \gamma^\mu T^{\small A} u_l)$ &
    $Q_{ud}^{(1)}$                &
    $(\bar u_i \gamma_\mu u_j)(\bar d_k \gamma^\mu d_l)$ &
    $Q_{\phi d}$ &
    $( \phi^\dagger i D_\mu \phi )
    ( \bar{d}_i \gamma^\mu d_j )$    \\
    $Q_{qd}^{(8)}$         &
    $(\bar q_i \gamma_\mu T^{\small A} q_j)(\bar d_k \gamma^\mu T^{\small A} d_l)$ &
    $Q_{ud}^{(8)}$                &
    $(\bar u_i \gamma_\mu T^{\small A}u_j)(\bar d_k \gamma^\mu T^{\small A} d_l)$ &
    $Q_{\phi ud}$ &
    $ i( \tilde{\phi}^\dagger D_\mu \phi)(\bar{u}_i \gamma^\mu d_j) $\\
  \hline 
  \end{tabular}
  \caption{List of the dimension-six SMEFT operators that contribute to
  $b \to s$ transitions at tree level.}
  \label{tab:operators1}
\end{table}

In our analysis, we focus on the subset of dimension-6 SMEFT operators from Table~\ref{tab:operators1} that contribute at tree level to $b \to s\, \ell^+ \ell^-$ transitions, specifically generating vector and axial-vector semileptonic operators after matching onto LEFT. These include $C_{\ell q}^{(1),(3)}$, $C_{qe}$, $C_{\ell d}$, $C_{ed}$, $C_{\phi q}^{(1),(3)}$, and $C_{\phi d}$. Our choice is motivated by global fits~\cite{Altmannshofer:2014rta,Capdevila:2017bsm,Alguero:2021anc,Geng:2021nhg} which consistently point towards vector/axial-vector structures being favored by the data. Scalar and pseudoscalar operators are tightly constrained by $B_s \to \mu^+\mu^-$ measurements~\cite{Alonso:2014csa}, while tensor operators are not generated at dimension-6 in SMEFT and would imply light new particles excluded by current collider bounds~\cite{Buchmuller:1985jz,Grzadkowski:2010es,Alonso:2014csa}. Loop-induced SMEFT contributions are also neglected, assuming they are subleading compared to the dominant tree-level effects. The matching relations are written as follows~\cite{Aebischer:2015fzz}:
\begin{align}
  [C_9]_{ii} &=
  \frac{\pi}{\alpha}  \frac{v^2}{\Lambda^2}     \left[ [\tilde{C}^{(1)}_{\ell  q}]_{ii23}
  + [\tilde{C}^{(3)}_{\ell  q}]_{ii23}     + [\tilde{C}_{qe}]_{23ii}  \right]\,, \\
  [C_{10}]_{ii} &=  \frac{\pi}{\alpha}  \frac{v^2}{\Lambda^2}
  \left[[\tilde{C}_{qe}]_{23ii}     -[\tilde{C}^{(1)}_{\ell  q}]_{ii23}
  - [\tilde{C}^{(3)}_{\ell  q}]_{ii23}    \right]\,,
  \label{eqn:Wilson9-10}
\end{align}
where the $i = 1,2$, correspond to $e$ and $\mu$ respectively\footnote{"tilde" symbol in the following equations represent the definition of operators in the mass basis}. Similar contributions appear as in the following for the coefficients of operators $\Op'_9,\Op'_{10}$ from vector-currents involving right-handed quarks:
\begin{align}
  [C'_9]_{ii} &=
 \frac{\pi}{\alpha} \frac{v^2}{\Lambda^2}
 \left[  [\tilde{C}_{\ell  d}]_{ii23} + [\tilde{C}_{ed}]_{ii23} \right]\,,\\
  [C'_{10}]_{ii} &=  \frac{\pi}{\alpha}  \frac{v^2}{\Lambda^2}
  \left[     [\tilde{C}_{ed}]_{ii23}-    [\tilde{C}_{\ell  d}]_{ii23}   \right]\,.
  \label{eqn:Wilson9-10prime}
\end{align}

The operators $\Op_9$ and $\Op_{10}$, and similarly $\Op^{\prime}_9$ and $\Op^{\prime}_{10}$, receive the following lepton flavor conserving tree-level contribution through the effective $\bar{s}$-$b$-$Z$ coupling appearing in the operators $Q_{\phi d}, Q_{\phi q}^{(1)}$ and $Q_{\phi q}^{(3)}$~\cite{Aebischer:2015fzz}. Thus, the associated coefficients are given below.
\begin{align}
  [C_9]_{ii} &=  \frac{\pi}{\alpha}
   \frac{v^2}{\Lambda^2}  \left(   [\tilde{C}_{\phi q}^{(1)}]_{23}   +[\tilde{C}_{\phi q}^{(3) }]_{23}   \right)
  \left( -1+4\, \rm sin^2 \theta_W \right)\,,\\
  [C_{10}]_{ii} &=\frac{\pi}{\alpha}
  \frac{v^2}{\Lambda^2}    \left(   [\tilde{C}_{\phi q}^{(1)}]_{23}   + [\tilde{C}_{\phi q}^{(3)}]_{23} \right)\,, \\
  [C'_9]_{ii} &= \frac{\pi}{\alpha}
  \frac{v^2}{\Lambda^2}  \, [\tilde{C}_{\phi d}]_{23}
  \left( -1+4\, \rm sin^2 \theta_W \right)  \,,\\
  [C'_{10}]_{ii} &= \frac{\pi}{\alpha}  \frac{v^2}{\Lambda^2}\,   [\tilde{C}_{\phi d}]_{23}\,.
\end{align}

The hadronic matrix elements for $B \to K$ and $B \to K^*$, and the differential decay distributions for $B \to K^{(*)} \ell^+ \ell^-$ decays followed by the definitions of various observables can be seen in Appendix~\ref{ap1}.

\section{Global fit of SMEFT coefficients and bounds on LEFT WCs}
\label{globalfitsmeft}
To identify the pattern of deviations in most of the $b\, \to\, s\, \ell^+ \ell^-$ observables, we perform a global fit on the subset of dimension-6 SMEFT operators that contribute to the $b \to s\, \ell^+ \ell^-$ transitions at tree level, specifically to vector and axial-vector operators. These include the operators $[C_{\ell q}^{(1,3)}]$, $[C_{qe}]$, $[C_{\ell d}]$, $[C_{ed}]$, $[C_{\phi q}^{(1,3)}]$, and $[C_{\phi d}]$, which are chosen based on their direct relevance to neutral current B-anomalies. We assume these to be the only non-zero Wilson coefficients at the new physics scale $\Lambda = 10$ TeV, and neglect other SMEFT operators that would contribute only at the loop level. For our global fit, we consider measurements from various $b\, \to\, s\, \ell^+ \ell^-$ observables\footnote{The definitions of these observables may be seen in Appendix~\ref{ap1}} as enumerated below. 
\begin{itemize}
    \item The branching fractions of $\mathcal{B}(B^{+,0} \to K \mu^+ \mu^-)$, $\mathcal{B}(B^{0,+} \to K^* \mu^+ \mu^-)$, $\mathcal{B}(B_s \to \phi \mu^+ \mu^-)$, $\mathcal{B}(\Lambda_b \to \Lambda \mu^+ \mu^-)$ and $\mathcal{B}(B_s \to \mu^+ \mu^-)$ evaluated in different $q^2$ bins.
    \item The branching fractions of $\mathcal{B}(B^+ \to K e^+ e^-)$ and $\mathcal{B}(B^0 \to K^* e^+ e^-)$ for $q^2 \in [1.1, 6]$ GeV$^2$.
    \item The angular coefficients $P_1$, $P_2$, $P_3$, $P'_4$, $P'_5$, $P'_6$, $P'_8$, $A_{FB}$ and $F_L$ at various $q^2$ bins of $B^{0,+} \to K^* \mu^+ \mu^-$ decay.
    \item The longitudinal and horizontal components of forward-backward asymmetry $A_{FB}$ in $\Lambda_b \to \Lambda \mu^+ \mu^-$ decays reported at $q^2 \in [15, 20]$ GeV$^2$.
    \item Few measurements of $P_1$, $P_2$ and $F_L$ in $\mathcal{B}(B^0 \to K^* e^+ e^-)$ decays.
    \item The isospin partner ratios $R_{KS0}$ and $R_{K^{*+}}$.
    \item The latest measurements of LFU sensitive observables $R_K$ and $R_{K^*}$ and also $\Delta P'_4$ and $\Delta P'_5$.
\end{itemize}

Assuming the lepton universality that gives equal status to $e$ and $\mu$, the
set of SMEFT coefficients involved in the global fit has a specific flavor index: $[C_{qe}]_{23ii}$, $[C_{\ell q}^{(1),(3)}]_{ii23}$, $[C_{ed}]_{ii23}$ and $[C_{\ell d}]_{ii23}$, where i = 1, 2 represents the electron and muon, respectively. These four WCs refer to four-fermion operators with a $2q2\ell$ operator-structure. Furthermore, the Higgs quark interactions quantified by
$[C_{\phi q}^{(1),(3)}]_{23}$ and $[C_{\phi d}]_{23}$ are universal for interactions with both electron and muon. 
The contributions of all these WCs to the low-energy WCs namely, $C_{9,10}^{(\prime)}$ have already been discussed at the end of Section~\ref{smeftphenomenology}. Among these six SMEFT WC types, we construct several scenarios involving one, two, three, or four operators at a time, and these appear as 1D, 2D, 3D, and 4D scenarios in
Table~\ref{tab:scenario_listing} where we have labeled the scenarios from S1 to S68. 
Notably, we classify the above scenarios into three sets based on the NP contribution type, namely, LFU-NP, LFUV-NP, and LFU+LFUV NP. 

 \begin{table}[h]
 \centering
\begin{tabular}{|c|c|c|c|}
 \hline
 Scenario & SMEFT coupling &   Scenario & SMEFT coupling \\
\hline \hline
S1 & $[C_{qe}]_{2311}=[C_{qe}]_{2322}$ & S2 &$[C_{\ell q}^{(1)}]_{1123}=[C_{\ell q}^{(1)}]_{2223}$  \\
\hline
S3 & $[C_{\ell q}^{(3)}]_{1123}=[C_{\ell q}^{(3)}]_{2223}$ &S4 & $[C_{\ell d}]_{1123}=[C_{\ell d}]_{2223}$\\
 \hline
 S5 & $[C_{ed}]_{1123}=[C_{ed}]_{2223}$ & S6 & $[C_{\phi q}^{(1)}]_{23}$  \\
 \hline
S7 & $[C_{\phi q}^{(3)}]_{23}$ & S8 & $[C_{\phi d}]_{23}$   \\
 \hline
\hline
S9 & $[C_{qe}]_{2311}=[C_{qe}]_{2322}$, $[C_{\phi q}^{(3)}]_{23}$  & S10 & $[C_{\ell q}^{(3)}]_{1123}=[C_{\ell q}^{(3)}]_{2223}$, $[C_{\phi q}^{(3)}]_{23}$ \\
 \hline
S11 & $[C_{\ell d}]_{1123}=[C_{\ell d}]_{2223}$, $[C_{\phi q}^{(3)}]_{23}$ & S12 & $[C_{ed}]_{1123}=[C_{ed}]_{2223}$, $[C_{\phi q}^{(3)}]_{23}$  \\
 \hline
 S13 & $[C_{qe}]_{2311}=[C_{qe}]_{2322}$, $[C_{\phi d}]_{23}$ & S14 & $[C_{\ell q}^{(3)}]_{1123}=[C_{\ell q}^{(3)}]_{2223}$, $[C_{\phi d}]_{23}$ \\
 \hline
S15 & $[C_{\ell d}]_{1123}=[C_{\ell d}]_{2223}$, $[C_{\phi d}]_{23}$ & S16 & $[C_{ed}]_{1123}=[C_{ed}]_{2223}$, $[C_{\phi d}]_{23}$ \\
\hline
S17 & $[C_{qe}]_{2311}$, $[C_{qe}]_{2322}$  & S18 & $[C_{\ell q}^{(3)}]_{1123}$, $[C_{\ell q}^{(3)}]_{2223}$ \\
  \hline  
S19 & $[C_{ed}]_{1123}$, $[C_{ed}]_{2223}$ & S20 & $[C_{\ell d}]_{1123}$, $[C_{\ell d}]_{2223}$  \\
 \hline
S21 & $[C_{qe}]_{2311}$, $[C_{\ell q}^{(3)}]_{2223}$ & S22 & $[C_{qe}]_{2311}$, $[C_{ed}]_{2223}$ \\
  \hline
S23 & $[C_{qe}]_{2311}$, $[C_{\ell d}]_{2223}$ & S24 & $[C_{qe}]_{2322}$, $[C_{\ell q}^{(3)}]_{1123}$ \\
  \hline
  S25 & $[C_{qe}]_{2322}$, $[C_{ed}]_{1123}$ & S26 & $[C_{qe}]_{2322}$, $[C_{\ell d}]_{1123}$  \\
  \hline
S27 & $[C_{\ell q}^{(3)}]_{1123}$, $[C_{ed}]_{2223}$ & S28 & $[C_{\ell q}^{(3)}]_{1123}$, $[C_{\ell d}]_{2223}$ \\
  \hline
S29 & $[C_{\ell q}^{(3)}]_{2223}$, $[C_{ed}]_{1123}$ &  S30 & $[C_{\ell q}^{(3)}]_{2223}$, $[C_{\ell d}]_{1123}$ \\
  \hline
S31 & $[C_{ed}]_{1123}$, $[C_{\ell d}]_{2223}$ & S32 & $[C_{ed}]_{2223}$, $[C_{\ell d}]_{1123}$  \\
\hline
S33 & $[C_{qe}]_{2311}$, $[C_{\phi q}^{(3)}]_{23}$ & S34 & $[C_{qe}]_{2322}$, $[C_{\phi q}^{(3)}]_{23}$  \\
  \hline
S35 & $[C_{\ell q}^{(3)}]_{1123}$, $[C_{\phi q}^{(3)}]_{23}$ & S36 & $[C_{\ell q}^{(3)}]_{2223}$, $[C_{\phi q}^{(3)}]_{23}$\\
  \hline
S37 & $[C_{ed}]_{1123}$, $[C_{\phi q}^{(3)}]_{23}$ &  S38 & $[C_{ed}]_{2223}$, $[C_{\phi q}^{(3)}]_{23}$ \\
  \hline
  S39 & $[C_{\ell d}]_{1123}$, $[C_{\phi q}^{(3)}]_{23}$ &  S40 & $[C_{\ell d}]_{2223}$, $[C_{\phi q}^{(3)}]_{23}$ \\
  \hline
S41 & $[C_{qe}]_{2311}$, $[C_{\phi d}]_{23}$ &  S42 & $[C_{qe}]_{2322}$, $[C_{\phi d}]_{23}$\\
  \hline
S43 & $[C_{\ell q}^{(3)}]_{1123}$, $[C_{\phi d}]_{23}$ &  S44 & $[C_{\ell q}^{(3)}]_{2223}$, $[C_{\phi d}]_{23}$\\
  \hline
  S45 & $[C_{ed}]_{1123}$, $[C_{\phi d}]_{23}$ & S46 & $[C_{ed}]_{2223}$, $[C_{\phi d}]_{23}$  \\
  \hline
S47 & $[C_{\ell d}]_{1123}$, $[C_{\phi d}]_{23}$ &  S48 & $[C_{\ell d}]_{2223}$, $[C_{\phi d}]_{23}$\\
  \hline
 \hline
S49 & $[C_{qe}]_{2311}$, $[C_{qe}]_{2322}$, $[C_{\phi q}^{(3)}]_{23}$ &  S50 & $[C_{\ell q}^{(3)}]_{1123}$, $[C_{\ell q}^{(3)}]_{2223}$, $[C_{\phi q}^{(3)}]_{23}$ \\
  \hline
S51 & $[C_{ed}]_{1123}$, $[C_{ed}]_{2223}$, $[C_{\phi q}^{(3)}]_{23}$ & S52 & $[C_{\ell d}]_{1123}$, $[C_{\ell d}]_{2223}$, $[C_{\phi q}^{(3)}]_{23}$  \\
  \hline
S53 & $[C_{qe}]_{2311}$, $[C_{qe}]_{2322}$, $[C_{\phi d}]_{23}$ &  S54 & $[C_{\ell q}^{(3)}]_{1123}$, $[C_{\ell q}^{(3)}]_{2223}$, $[C_{\phi d}]_{23}$ \\
  \hline
S55 & $[C_{ed}]_{1123}$, $[C_{ed}]_{2223}$, $[C_{\phi d}]_{23}$ &  S56 & $[C_{\ell d}]_{1123}$, $[C_{\ell d}]_{2223}$, $[C_{\phi d}]_{23}$   \\
  \hline
S57 & $[C_{qe}]_{2311}$, $[C_{\ell q}^{(3)}]_{2223}$, $[C_{\phi q}^{(3)}]_{23}$ & S58 & $[C_{qe}]_{2322}$, $[C_{\ell q}^{(3)}]_{1123}$, $[C_{\phi q}^{(3)}]_{23}$  \\
  \hline
S59 & $[C_{ed}]_{1123}$, $[C_{\ell d}]_{2223}$, $[C_{\phi d}]_{23}$ & S60 & $[C_{\ell d}]_{1123}$, $[C_{ed}]_{2223}$, $[C_{\phi d}]_{23}$  \\
  \hline
   \hline
S61 & $[C_{qe}]_{2311}$, $[C_{ed}]_{2223}$, $[C_{\phi q}]_{23}$, $[C_{\phi d}]_{23}$ & S62 & $[C_{ed}]_{1123}$, $[C_{qe}]_{2322}$, $[C_{\phi q}^{(3)}]_{23}$, $[C_{\phi d}]_{23}$  \\
  \hline
S63 & $[C_{qe}]_{2311}$, $[C_{\ell d}]_{2223}$, $[C_{\phi q}^{(3)}]_{23}$, $[C_{\phi d}]_{23}$ &  
  S64 & $[C_{\ell d}]_{1123}$, $[C_{qe}]_{2322}$, $[C_{\phi q}^{(3)}]_{23}$, $[C_{\phi d}]_{23}$ \\
  \hline
  S65 & $[C_{\ell q}^{(3)}]_{1123}$, $[C_{ed}]_{2223}$, $[C_{\phi q}^{(3)}]_{23}$, $[C_{\phi d}]_{23}$ & S66 & $[C_{ed}]_{1123}$, $[C_{\ell q}^{(3)}]_{2223}$, $[C_{\phi q}^{(3)}]_{23}$, $[C_{\phi d}]_{23}$  \\
  \hline
S67 & $[C_{\ell q}^{(3)}]_{1123}$, $[C_{\ell d}]_{2223}$, $[C_{\phi q}^{(3)}]_{23}$, $[C_{\phi d}]_{23}$ & S68 & $[C_{\ell d}]_{1123}$, $[C_{\ell q}^{(3)}]_{2223}$, $[C_{\phi q}^{(3)}]_{23}$, $[C_{\phi d}]_{23}$  \\
  \hline
\end{tabular}
 \caption{List of scenarios and related Wilson Coefficients that may potentially be used in 1D, 2D, 3D, and 4D analyses.}
 \label{tab:scenario_listing}
 \end{table}
 
The scenarios S1-S8 are used for 1D analyses {\it i.e.} only a given WC is non-vanishing at the scale $\Lambda$. The lepton flavor-specific couplings as appearing in scenarios S1-S5 with the $2q2\ell$ operator structure for both muon and electron are universal, or in other words, S1 to S5 cases refer to LFU-NP operators. The WCs contributing to Higgs-quark couplings as in the scenarios S6-S8 do not distinguish an electron from a muon, indicating that the scenarios S6-S8 are also of LFU-NP type.
 
 The scenarios S9-S16 corresponding to 2D analyses combine one non-zero coupling from S1-S5 and one from S6-S8. Consequently, all NP contributions from S1-S16 are of LFU-NP category. The cases S17-S48 in Table~\ref{tab:scenario_listing},
represent other 2D scenarios. Among them, S17-S32 have two independent couplings, both with $2q2\ell$ operator structure along with lepton flavor-specific couplings indicating their non-universal (LFUV-NP) nature for NP. On the other hand, the scenarios S33-S48 have two independent non-vanishing couplings: one with a $2q2\ell$ operator structure and the other with a Higgs-quark structure, contributing to LFU+LFUV NP. Going to the 3D case the scenarios S49-S60 with three non-vanishing couplings contain combinations of both operator structures, thus contributing to LFU+LFUV NP. The same is true for the 4D scenarios S61-S68.

We perform a global fit to all the above mentioned SMEFT scenarios using the fast-likelihood approach implemented in the code \texttt{Flavio}~\cite{Straub:2018kue}. The likelihood function is approximated as Gaussian in both experimental and theoretical uncertainties, resulting in a $\chi^2$ function that depends only on the NP Wilson coefficients. The theoretical uncertainties—including form factors, CKM elements, quark masses, etc.—are treated as nuisance parameters sampled from normal distributions, according to their known uncertainties and correlations. The resulting theory covariance matrix $C_{\rm th}$ is precomputed at the SM point and held fixed throughout the fit. This approximation is justified since current data do not indicate large NP contributions that would strongly modify these uncertainties. The experimental covariance matrix $C_{\rm exp}$ incorporates the published experimental correlations wherever available (notably from LHCb). On the other hand, for branching ratio measurements without official correlations, we assume uncorrelated statistical errors and fully correlated systematics within the same experiment. This approach allows us to consistently include bin-to-bin and observable-to-observable correlations while keeping the fit computationally efficient. More details about the fast-likelihood is discussed in Appendix~\ref{fastfit}. For all the omitted details one can refer to Ref.~\cite{Straub:2018kue}. 

Whether a given NP scenario describes the experimental data better than the SM can be quantified by the difference between the NP likelihood and the SM likelihood. We obtain the best-fit points corresponding to each NP scenario by minimizing $\chi^2$~\cite{Altmannshofer:2014rta} in
Tables~\ref{tab_fit1} and \ref{tab_fit2}. Expressed in units of standard deviation $\sigma$, the $\rm pull$ values reported for each scenario 
quantify the degree to which the SM is disfavoured. 
We now extract the cases for which $\rm pull \ge 4$ and strictly satisfies $R_{K^{(*)}}$ experimental constraint and construct Table~\ref{tab_fit3} corresponding to 1D, 2D, 3D and 4D scenarios. 
Although some 3D and 4D scenarios have $\rm pull \ge 4$, however, compared to 2D scenarios, their contributions are not unique. They do not show any significant signatures different from the 2D scenarios. This is because we notice that $[C_{qe}]_{23ii}$, $[C_{\ell q}^{(1),(3)}]_{ii23}$ and $[C_{\phi q}^{(1),(3)}]_{23}$ to be the dominant WCs and hence are mainly preferred by the data. Moreover, the presence of the same operators significantly changes the $\rm pull$. Hence we mostly focus on 1D and 2D effects on the observables. The above operators involve the left-handed currents and contribute to the vector/axial-vector coefficients $C_{9,10}$ at low energies. 
Considering both electron and muon we mention $\Delta C_{9,10}^{e/\mu}$ in Table~\ref{tab_fit4} which are the NP contributions to $C_{9,10}$ at the scale of $\mu_b$.
The scenarios S3, S9 and S10 have universal contributions to electrons and muons since they are of LFU-NP type, while S18 shows LFUV-NP and S42 shows LFU+LFUV-NP type contributions. Interestingly, S3 can be related to the very popular scenario $C_9=-C_{10}$ which also appears in $Z'$ and a few LQ models~\cite{Alok:2017jgr}. In the remaining scenarios, we notice $|\Delta C_{9}| > |\Delta C_{10}|$. Except for S42 which affects $C_9^e$ only slightly, all the scenarios mentioned in Table~\ref{tab_fit4} have prominent contributions to $\Delta C_{9,10}^e$.

\begin{table}[h]
 \centering
 \setlength{\tabcolsep}{6pt} 
 \begin{tabular}{|c|c|c|c|c|}
 \hline
 Scenario & SMEFT coupling & Best fit & $\Delta \chi^2$ & $\rm pull$\\
\hline \hline
S1 & $[C_{qe}]_{2311}=[C_{qe}]_{2322}$ & 0.013 & 0.68 & 0.83 \\
 \hline
S2 & $[C_{\ell q}^{(1)}]_{1123}=[C_{\ell q}^{(1)}]_{2223}$ & 0.089 & 22.77 & 4.77\\
 \hline
S3 & $[C_{\ell q}^{(3)}]_{1123}=[C_{\ell q}^{(3)}]_{2223}$ & 0.083 & 18.95 & 4.35\\
 \hline
S4 & $[C_{\ell d}]_{1123}=[C_{\ell d}]_{2223}$ & -0.015 & 2.19 & 1.48\\
 \hline
S5 & $[C_{ed}]_{1123}=[C_{ed}]_{2223}$ & 0.019 & 1.72 & 1.31\\
 \hline
S6 & $[C_{\phi q}^{(1)}]_{23}$ & -0.041 & 4.29 & 2.07\\
 \hline
S7 & $[C_{\phi q}^{(3)}]_{23}$ & -0.035 & 4.05 & 2.01\\
 \hline
S8 & $[C_{\phi d}]_{23}$ & 0.004 & 0.097 & 0.31\\
  \hline
  \hline
S9 & $[C_{qe}]_{2311}=[C_{qe}]_{2322}$, $[C_{\phi q}^{(3)}]_{23}$ & (0.190, -0.238) & 42.76 & 6.21\\
 \hline
S10 & $[C_{\ell q}^{(3)}]_{1123}=[C_{\ell q}^{(3)}]_{2223}$, $[C_{\phi q}^{(3)}]_{23}$ & (0.231, 0.162) & 47.12 & 6.55\\
 \hline
S11 & $[C_{\ell d}]_{1123}=[C_{\ell d}]_{2223}$, $[C_{\phi q}^{(3)}]_{23}$ & (0.030, -0.064) & 8.56 & 2.46\\
 \hline
S12 & $[C_{ed}]_{1123}=[C_{ed}]_{2223}$, $[C_{\phi q}^{(3)}]_{23}$ & (-0.010, -0.044) & 4.32 & 1.57\\
 \hline
S13 & $[C_{qe}]_{2311}=[C_{qe}]_{2322}$, $[C_{\phi d}]_{23}$ & (0.049, 0.041) & 3.28 & 1.30\\
 \hline
S14 & $[C_{\ell q}^{(3)}]_{1123}=[C_{\ell q}^{(3)}]_{2223}$, $[C_{\phi d}]_{23}$ & (0.072, 0.017) & 18.59 & 3.91\\
 \hline
S15 & $[C_{\ell d}]_{1123}=[C_{\ell d}]_{2223}$, $[C_{\phi d}]_{23}$ & (0.111, 0.156) & 10.96 & 2.86\\
 \hline
S16 & $[C_{ed}]_{1123}=[C_{ed}]_{2223}$, $[C_{\phi d}]_{23}$ & (0.059, -0.047) & 3.48 & 1.35\\
  \hline
S17 & $[C_{qe}]_{2311}$, $[C_{qe}]_{2322}$ & (-0.183, 0.011) & 4.25 & 1.56\\
  \hline
S18 & $[C_{\ell q}^{(3)}]_{1123}$, $[C_{\ell q}^{(3)}]_{2223}$ & (0.072, 0.084) & 21.72 & 4.27\\
  \hline
S19 & $[C_{ed}]_{1123}$, $[C_{ed}]_{2223}$ & (0.120, 0.017) & 3.33 & 1.31\\
 \hline
S20 & $[C_{\ell d}]_{1123}$, $[C_{\ell d}]_{2223}$ & (-0.013, 0.003) & 2.41 & 1.03\\
 \hline
S21 & $[C_{qe}]_{2311}$, $[C_{\ell q}^{(3)}]_{2223}$ & (-0.089, 0.030) & 12.55 & 3.12\\
  \hline
S22 & $[C_{qe}]_{2311}$, $[C_{ed}]_{2223}$ &  (-0.183, 0.018) & 5.30 & 1.81\\
  \hline
S23 & $[C_{qe}]_{2311}$, $[C_{\ell d}]_{2223}$ &  (-0.162, 0.005) & 4.13 & 1.53\\
  \hline
S24 & $[C_{qe}]_{2322}$, $[C_{\ell q}^{(3)}]_{1123}$ & (0.014, -0.011) & 1.69 & 0.79\\
  \hline
S25 & $[C_{qe}]_{2322}$, $[C_{ed}]_{1123}$ &  (0.015, 0.123) & 2.94 & 1.20\\
  \hline
S26 & $[C_{qe}]_{2322}$, $[C_{\ell d}]_{1123}$ &  (0.014, -0.015) & 2.99 & 1.21\\
  \hline
S27 & $[C_{\ell q}^{(3)}]_{1123}$, $[C_{ed}]_{2223}$ &  (-0.010, 0.017) & 2.27 & 0.99\\
  \hline
S28 & $[C_{\ell q}^{(3)}]_{1123}$, $[C_{\ell d}]_{2223}$ &  (-0.006, 0.008) & 2.26 & 0.99\\
  \hline
S29 & $[C_{\ell q}^{(3)}]_{2223}$, $[C_{ed}]_{1123}$ &  (0.029, 0.018) & 10.35 & 2.77\\
  \hline
S30 & $[C_{\ell q}^{(3)}]_{2223}$, $[C_{\ell d}]_{1123}$ &  (0.030, -0.001) & 10.81 & 2.84\\
  \hline
S31 & $[C_{ed}]_{1123}$, $[C_{\ell d}]_{2223}$ & (0.090, 0.008) & 2.77 & 1.15\\
  \hline
S32 & $[C_{ed}]_{2223}$, $[C_{\ell d}]_{1123}$ & (0.021, -0.016) & 4.18 & 1.54\\
  \hline
  \hline
 \end{tabular}
 \caption{Part 1: Global fit of SMEFT coefficients in 1D, 2D, 3D and 4D scenarios. We report the best-fit points, the $\Delta \chi^2$, and pull for each SMEFT scenario. The best fit points correspond to the minimum $\chi^2$. The NP scale is assumed at $\Lambda=10$ TeV.}
 \label{tab_fit1}
 \end{table}

 \begin{table}[h]
 \centering
 \setlength{\tabcolsep}{6pt} 
 \begin{tabular}{|c|c|c|c|c|}
 \hline
 Scenario & SMEFT coupling & Best fit & $\Delta \chi^2$ & $\rm pull$ \\
\hline \hline
S33 & $[C_{qe}]_{2311}$, $[C_{\phi q}^{(3)}]_{23}$ & (-0.157, -0.033) & 7.05 & 2.18\\
  \hline
S34 & $[C_{qe}]_{2322}$, $[C_{\phi d}]_{23}$ &  (0.044, 0.041) & 4.20 & 1.54\\
  \hline
S35 & $[C_{\ell q}^{(3)}]_{1123}$, $[C_{\phi q}^{(3)}]_{23}$ & (-0.011, -0.039) & 5.97 & 1.96\\
  \hline
S36 & $[C_{\ell q}^{(3)}]_{2223}$, $[C_{\phi q}^{(3)}]_{23}$ & (0.029, -0.009) & 11.09 & 2.89\\
  \hline
S37 & $[C_{ed}]_{1123}$, $[C_{\phi q}^{(3)}]_{23}$ &  (0.123, -0.037) & 6.61 & 2.09\\
  \hline
S38 & $[C_{ed}]_{2223}$, $[C_{\phi q}^{(3)}]_{23}$ &  (-0.010, -0.047) & 4.94 & 1.72\\
  \hline
S39 & $[C_{\ell d}]_{1123}$, $[C_{\phi q}^{(3)}]_{23}$ &  (-0.015, -0.033) & 5.92 & 1.94\\
  \hline
S40 & $[C_{\ell d}]_{2223}$, $[C_{\phi q}^{(3)}]_{23}$ &  (0.020, -0.053) & 10.07 & 2.72\\
  \hline
S41 & $[C_{qe}]_{2311}$, $[C_{\phi d}]_{23}$ &  (-0.185, 0.007) & 3.97 & 1.49\\
  \hline
S42 & $[C_{qe}]_{2322}$, $[C_{\phi q}^{(3)}]_{23}$ & (0.180, -0.239) & 41.17 & 6.09\\
  \hline
S43 & $[C_{\ell q}^{(3)}]_{1123}$, $[C_{\phi d}]_{23}$ &  (-0.010, -0.002) & 0.93 & 0.48\\
  \hline
S44 & $[C_{\ell q}^{(3)}]_{2223}$, $[C_{\phi d}]_{23}$ &  (0.031, -0.012) & 9.84 & 2.68\\
  \hline
S45 & $[C_{ed}]_{1123}$, $[C_{\phi d}]_{23}$ & (0.115, 0.010) & 2.57 & 1.09\\
  \hline
S46 & $[C_{ed}]_{2223}$, $[C_{\phi d}]_{23}$ & (0.051, -0.038) & 3.15 & 1.26\\
  \hline
S47 & $[C_{\ell d}]_{1123}$, $[C_{\phi d}]_{23}$ & (-0.016, 0.007) & 2.52 & 1.07\\
  \hline
S48 & $[C_{\ell d}]_{2223}$, $[C_{\phi d}]_{23}$ & (0.022, 0.036) & 4.99 & 1.73\\
  \hline
  \hline
S49 & $[C_{qe}]_{2311}$, $[C_{qe}]_{2322}$, $[C_{\phi q}]_{23}$ & (0.187, 0.192, -0.247) & 41.57 & 5.84\\
  \hline
S50 & $[C_{\ell q}^{(3)}]_{1123}$, $[C_{\ell q}^{(3)}]_{2223}$, $[C_{\phi q}^{(3)}]_{23}$ & (0.220, 0.232, 0.161) & 53.50 & 6.75\\
  \hline
S51 & $[C_{ed}]_{1123}$, $[C_{ed}]_{2223}$, $[C_{\phi q}^{(3)}]_{23}$ & (0.122, -0.006, -0.042) & 6.68 & 1.73\\
  \hline
S52 & $[C_{\ell d}]_{1123}$, $[C_{\ell d}]_{2223}$, $[C_{\phi q}^{(3)}]_{23}$ & (0.019, 0.033, -0.065) & 12.25 & 2.72\\
  \hline
S53 & $[C_{qe}]_{2311}$, $[C_{qe}]_{2322}$, $[C_{\phi d}]_{23}$ & (0.018, 0.052, 0.045) & 4.72 & 1.30\\
  \hline
S54 & $[C_{\ell q}^{(3)}]_{1123}$, $[C_{\ell q}^{(3)}]_{2223}$, $[C_{\phi d}]_{23}$ & (0.095, 0.107, -0.035) & 25.55 & 4.38\\
  \hline
S55 & $[C_{ed}]_{1123}$, $[C_{ed}]_{2223}$, $[C_{\phi d}]_{23}$ & (0.143, 0.053, -0.042) & 4.88 & 1.34\\
  \hline
S56 & $[C_{\ell d}]_{1123}$, $[C_{\ell d}]_{2223}$, $[C_{\phi d}]_{23}$ & (0.087, 0.102, 0.145) & 11.48 & 2.60\\
%
\hline
S57 & $[C_{qe}]_{2311}$, $[C_{\ell q}^{(3)}]_{2223}$, $[C_{\phi q}^{(3)}]_{23}$ &  (-0.088, 0.029, -0.004) & 12.72 & 2.79\\
  \hline
S58 & $[C_{qe}]_{2322}$, $[C_{\ell q}^{(3)}]_{1123}$, $[C_{\phi q}^{(3)}]_{23}$ &  (-0.025, 0.187, -0.238) & 48.23 & 6.37\\
  \hline
S59 & $[C_{ed}]_{1123}$, $[C_{\ell d}]_{2223}$, $[C_{\phi d}]_{23}$ & (-0.055, 0.027, 0.040) & 6.43 & 1.68\\
  \hline
S60 & $[C_{\ell d}]_{1123}$, $[C_{ed}]_{2223}$, $[C_{\phi d}]_{23}$ & (-0.019, 0.063, -0.052) & 6.21 & 1.64\\
  \hline
  \hline
S61 & $[C_{qe}]_{2311}$, $[C_{ed}]_{2223}$, $[C_{\phi q}^{(3)}]_{23}$, $[C_{\phi d}]_{23}$ &  (-0.129, 0.046, -0.047, -0.064) & 11.74 & 2.34\\
  \hline
S62 & $[C_{ed}]_{1123}$, $[C_{qe}]_{2322}$, $[C_{\phi q}^{(3)}]_{23}$, $[C_{\phi d}]_{23}$ &  (0.177, 0.180, -0.219, 0.009) & 45.84 & 5.94\\
  \hline
S63 & $[C_{qe}]_{2311}$, $[C_{\ell d}]_{2223}$, $[C_{\phi q}^{(3)}]_{23}$, $[C_{\phi d}]_{23}$ &  (-0.022, 0.020, -0.051, 0.003) & 10.57 & 2.14\\
  \hline
S64 & $[C_{\ell d}]_{1123}$, $[C_{qe}]_{2322}$, $[C_{\phi q}^{(3)}]_{23}$, $[C_{\phi d}]_{23}$ &  (-0.005, 0.173, -0.224, 0.003) & 39.74 & 5.45\\
  \hline
S65 & $[C_{\ell q}^{(3)}]_{1123}$, $[C_{ed}]_{2223}$, $[C_{\phi q}^{(3)}]_{23}$, $[C_{\phi d}]_{23}$ &  (-0.010, 0.046, -0.060, -0.075) & 11.05 & 2.23\\
  \hline
S66 & $[C_{ed}]_{1123}$, $[C_{\ell q}^{(3)}]_{2223}$, $[C_{\phi q}^{(3)}]_{23}$, $[C_{\phi d}]_{23}$ &  (0.031, 0.025, -0.025, -0.018) & 11.09 & 2.23\\
  \hline
S67 & $[C_{\ell q}^{(3)}]_{1123}$, $[C_{\ell d}]_{2223}$, $[C_{\phi q}^{(3)}]_{23}$, $[C_{\phi d}]_{23}$ &  (-0.002, 0.020, -0.061, -0.004) & 11.45 & 2.29\\
  \hline
S68 & $[C_{\ell d}]_{1123}$, $[C_{\ell q}^{(3)}]_{2223}$, $[C_{\phi q}^{(3)}]_{23}$, $[C_{\phi d}]_{23}$ &  (-0.005, 0.023, -0.035, -0.025) & 12.70 & 2.48\\
  \hline
 \end{tabular}
 \caption{Part 2: Global fit of SMEFT coefficients in 1D, 2D, 3D and 4D scenarios. We report the best-fit points, the $\Delta \chi^2$ and pull for each SMEFT scenario. The best-fit points correspond to the minimum $\chi^2$. The NP scale is assumed at $\Lambda=10$ TeV.}
 \label{tab_fit2}
 \end{table}

  \begin{table}[h]
 \centering
 \setlength{\tabcolsep}{6pt} 
 \begin{tabular}{|c|c|c|c|c|}
 \hline
 Scenario & SMEFT coupling & Best fit & $\Delta \chi^2$ & pull \\
\hline \hline
\multicolumn{5}{|c|}{1D scenario} \\
\hline
S3 & $[C_{\ell q}^{(3)}]_{1123}=[C_{\ell q}^{(3)}]_{2223}$ & 0.083 & 18.95 & 4.35\\
 \hline
 \hline
\multicolumn{5}{|c|}{2D scenario} \\
\hline
S9 & $[C_{qe}]_{2311}=[C_{qe}]_{2322}$, $[C_{\phi q}^{(3)}]_{23}$ & (0.190, -0.238) & 42.76 & 6.21\\
 \hline
S10 & $[C_{\ell q}^{(3)}]_{1123}=[C_{\ell q}^{(3)}]_{2223}$, $[C_{\phi q}^{(3)}]_{23}$ & (0.231, 0.162) & 47.12 & 6.55\\
 \hline
S18 & $[C_{\ell q}^{(3)}]_{1123}$, $[C_{\ell q}^{(3)}]_{2223}$ & (0.072, 0.084) & 21.72 & 4.27\\
  \hline
S42 & $[C_{qe}]_{2322}$, $[C_{\phi q}^{(3)}]_{23}$ & (0.180, -0.239) & 41.17 & 6.09\\
  \hline
  \hline
\multicolumn{5}{|c|}{3D scenario} \\
\hline
S49 & $[C_{qe}]_{2311}$, $[C_{qe}]_{2322}$, $[C_{\phi q}^{(3)}]_{23}$ & (0.187, 0.192, -0.247) & 41.57 & 5.84\\
  \hline
S50 & $[C_{\ell q}^{(3)}]_{1123}$, $[C_{\ell q}^{(3)}]_{2223}$, $[C_{\phi q}^{(3)}]_{23}$ & (0.220, 0.232, 0.161) & 53.50 & 6.75\\
  \hline
S54 & $[C_{\ell q}^{(3)}]_{1123}$, $[C_{\ell q}^{(3)}]_{2223}$, $[C_{\phi d}]_{23}$ & (0.095, 0.107, -0.035) & 25.55 & 4.38\\
   \hline
S58 & $[C_{qe}]_{2322}$, $[C_{\ell q}^{(3)}]_{1123}$, $[C_{\phi q}^{(3)}]_{23}$ &  (-0.025, 0.187, -0.238) & 48.23 & 6.37\\
  \hline
  \hline
\multicolumn{5}{|c|}{4D scenario} \\
\hline
S62 & $[C_{ed}]_{1123}$, $[C_{qe}]_{2322}$, $[C_{\phi q}^{(3)}]_{23}$, $[C_{\phi d}]_{23}$ &  (0.177, 0.180, -0.219, 0.009) & 45.84 & 5.94\\
  \hline
S64 & $[C_{\ell d}]_{1123}$, $[C_{qe}]_{2322}$, $[C_{\phi q}^{(3)}]_{23}$, $[C_{\phi d}]_{23}$ &  (-0.005, 0.173, -0.224, 0.003) & 39.74 & 5.45\\
  \hline
 \end{tabular}
 \caption{The global "best" NP scenarios with larger $\Delta \chi^2$ or $\rm pull$ being $\ge 4$ and consistent with the $R_{K^{(*)}}$ measurement picked from the Tables~\ref{tab_fit1} and \ref{tab_fit2}. We report the best-fit points, the $\Delta \chi^2$, and pull for each SMEFT scenario. The best-fit points correspond to the minimum $\chi^2$. The NP scale is assumed at $\Lambda=10$ TeV.}
 \label{tab_fit3}
 \end{table}

 \begin{table}[h]
 \centering
 \setlength{\tabcolsep}{6pt} 
 \begin{tabular}{|c|c|c|c|c|c|}
 \hline
 NP contribution & S3 & S9 & S10 & S18 & S42 \\
 \hline
 \hline
 $\Delta C_9^e$ & \multirow{2}{*}{-0.45} & \multirow{2}{*}{-1.16} & \multirow{2}{*}{-1.21} & -0.39 &  -0.07 \\
 $\Delta C_9^\mu$ &  &  & & -0.46 & -1.12 \\
 \hline
 $\Delta C_{10}^e$ & \multirow{2}{*}{0.46} & \multirow{2}{*}{0.31} & \multirow{2}{*}{0.33} & 0.40 & 1.42 \\
 $\Delta C_{10}^\mu$ &  &  & & 0.47 & 0.37 \\
 \hline
 \end{tabular}
 \caption{The corresponding bounds on low energy WCs $C_{9,10}^{e/\mu}$ at the scale of $\mu_b$.}
 \label{tab_fit4}
 \end{table}

\begin{figure}[htbp]
\centering
\includegraphics[width=0.49\textwidth]{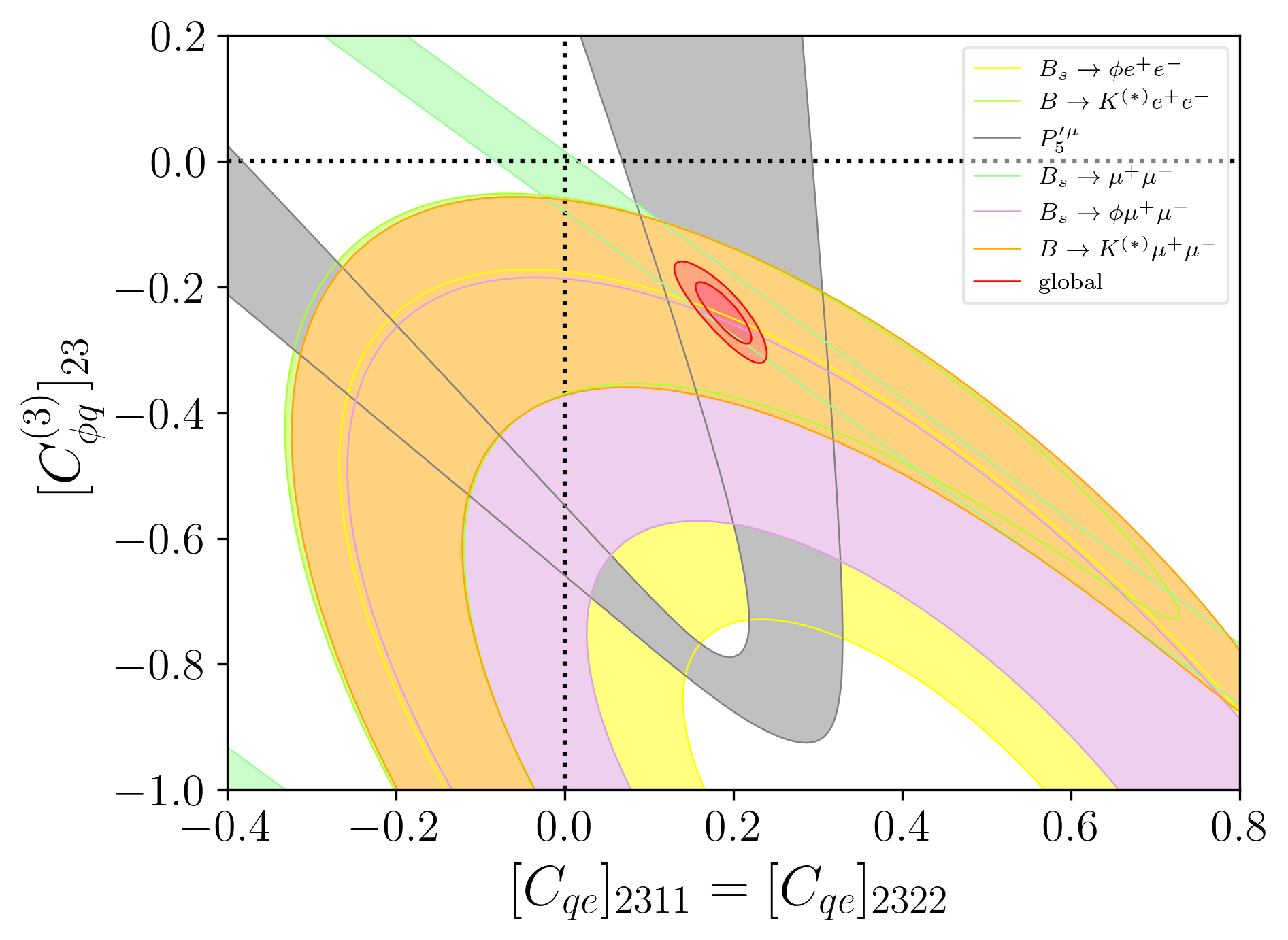}
\includegraphics[width=0.49\textwidth]{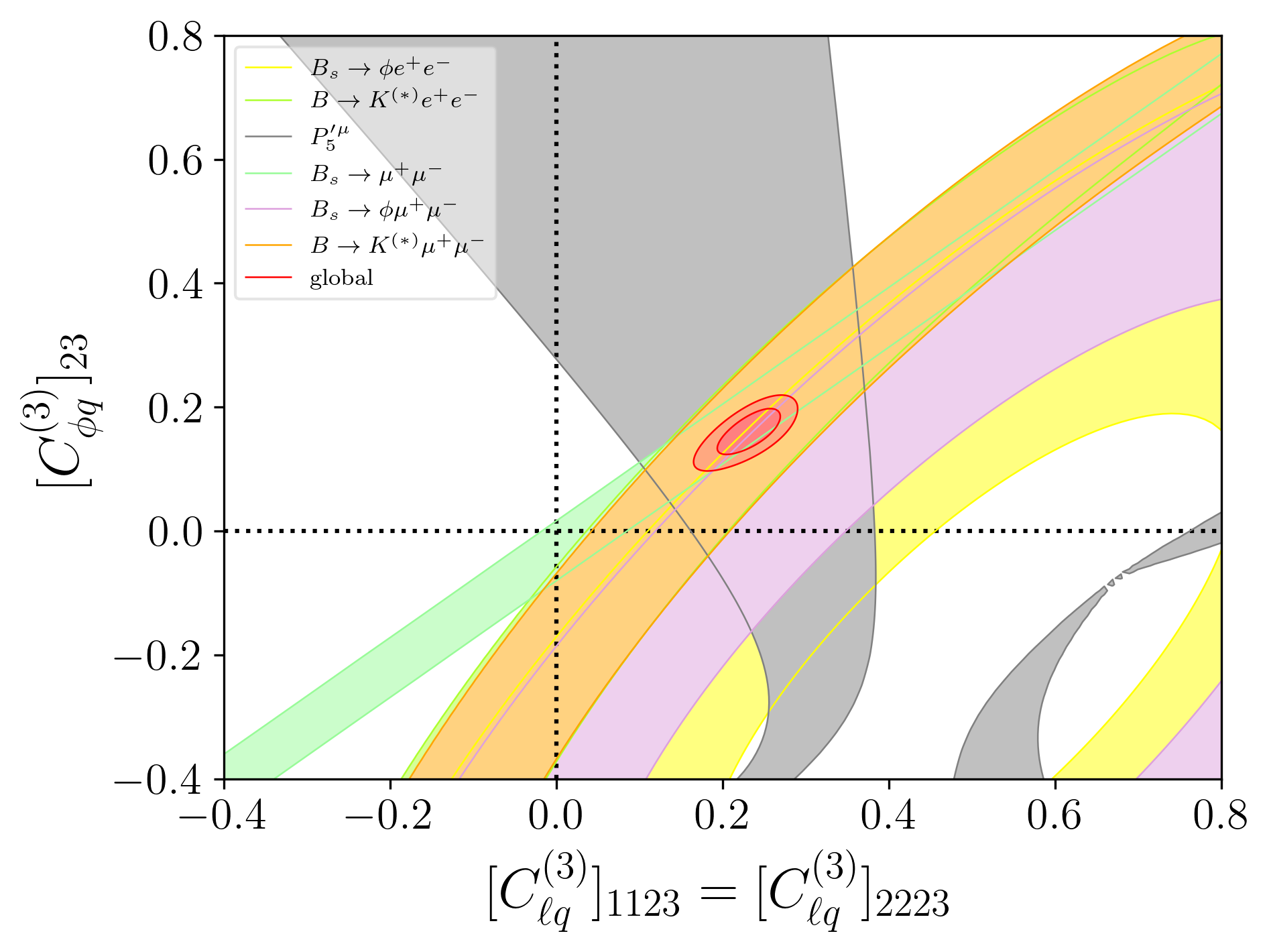}
\includegraphics[width=0.49\textwidth]{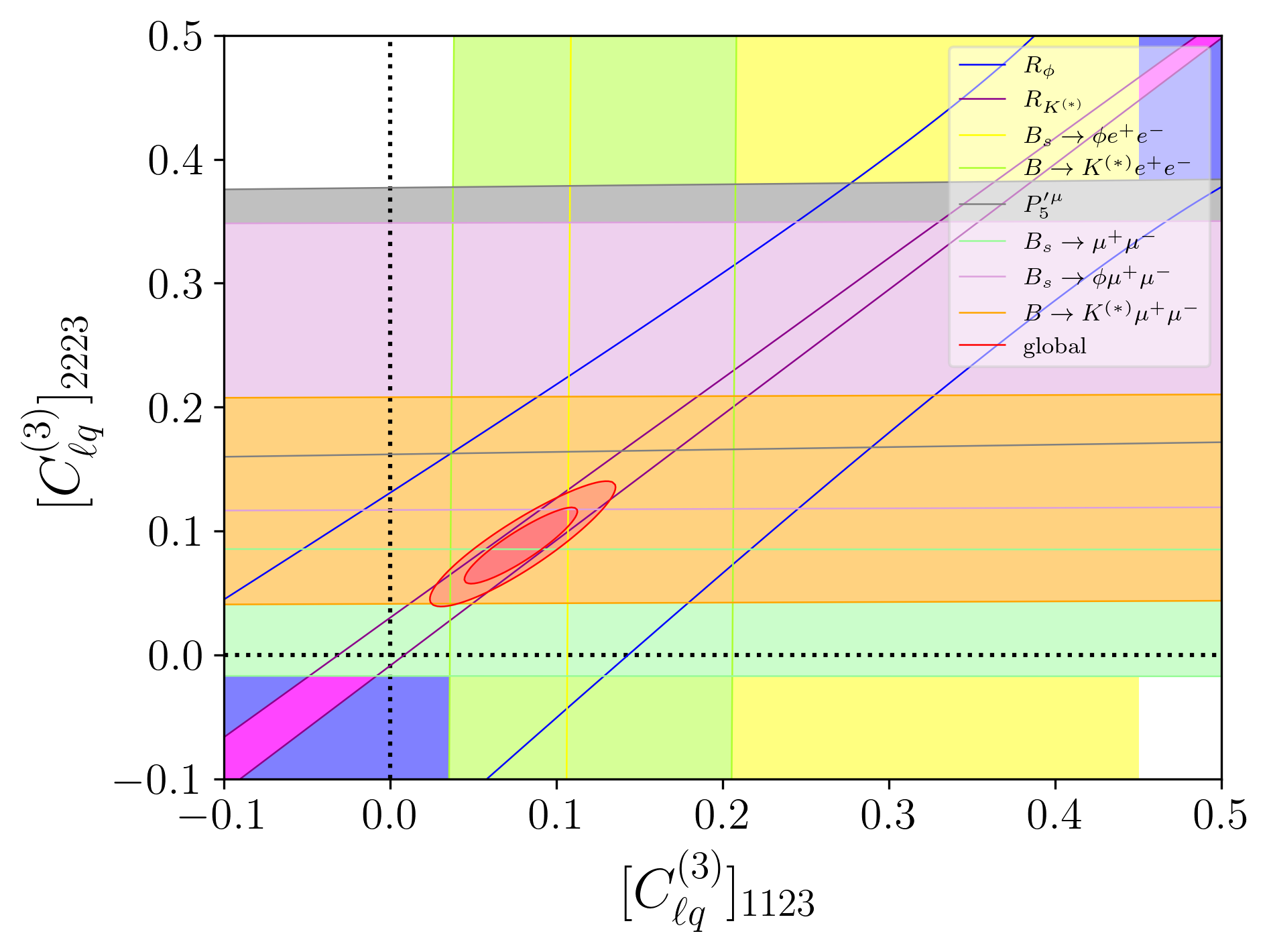}
\includegraphics[width=0.49\textwidth]{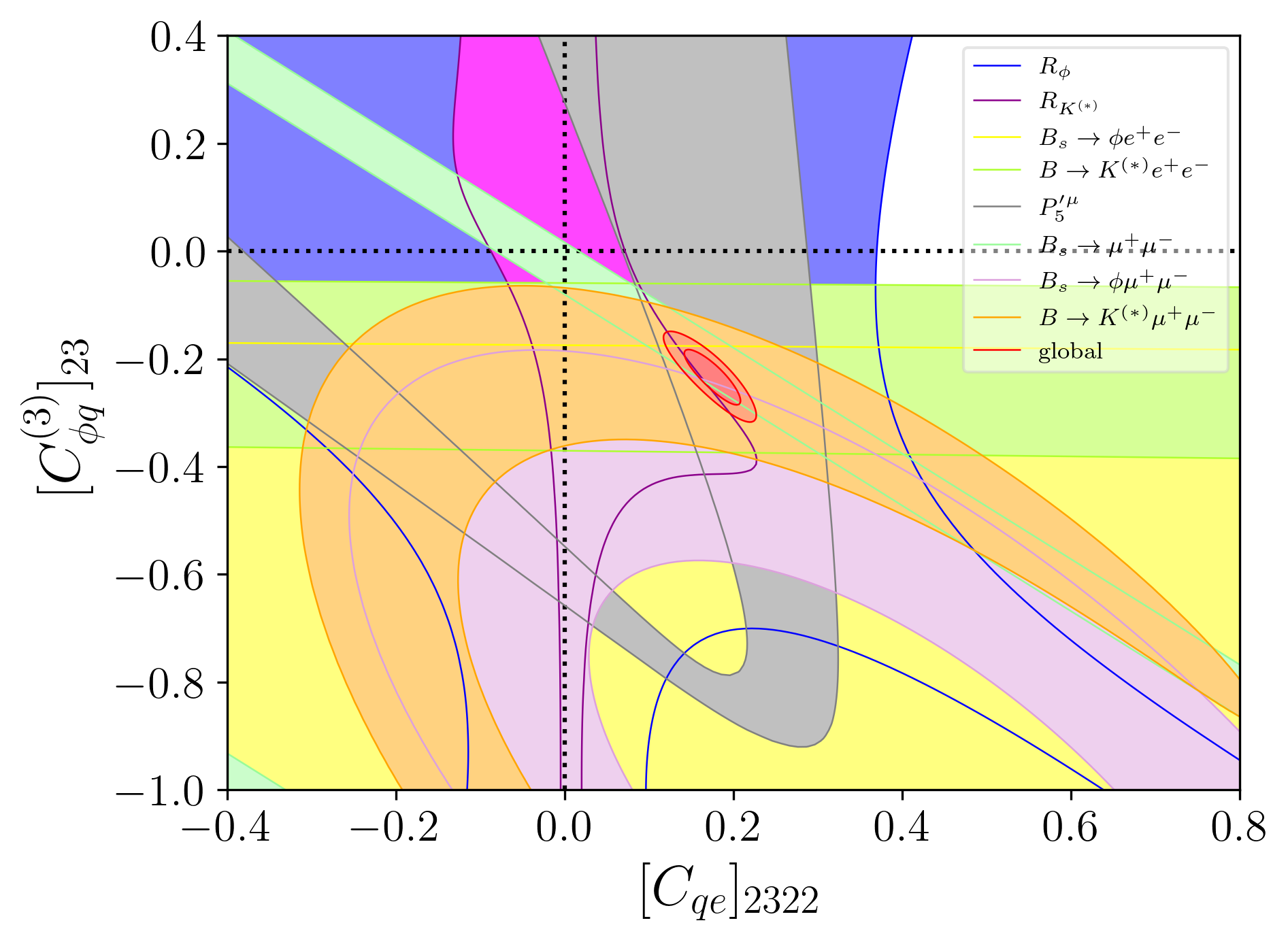}
\caption{The likelihood contours of the global fit and several fits to subsets of observables in the plane of SMEFT coefficients (i) S9: $[C_{qe}]_{2311}=[C_{qe}]_{2322}$, $[C_{\phi q}^{(3)}]_{23}$, (ii) S10: $[C_{\ell q}^{(3)}]_{1123}=[C_{\ell q}^{(3)}]_{2223}$, $[C_{\phi q}^{(3)}]_{23}$, (iii) S18: $[C_{\ell q}^{(3)}]_{1123}$, $[C_{\ell q}^{(3)}]_{2223}$ and (iv) S42: $[C_{qe}]_{2322}$, $[C_{\phi q}^{(3)}]_{23}$.}
\label{fig_contour}
\end{figure}

In Fig.~\ref{fig_contour} we show the likelihood contours of the global fit and several fits to subsets of observables in the plane of SMEFT coefficients (i) S9: $[C_{qe}]_{2311}=[C_{qe}]_{2322}$, $[C_{\phi q}^{(3)}]_{23}$, (ii) S10: $[C_{\ell q}^{(3)}]_{1123}=[C_{\ell q}^{(3)}]_{2223}$, $[C_{\phi q}^{(3)}]_{23}$, (iii) S18: $[C_{\ell q}^{(3)}]_{1123}$, $[C_{\ell q}^{(3)}]_{2223}$ and (iv) S42: $[C_{qe}]_{2322}$, $[C_{\phi q}^{(3)}]_{23}$. The subset of observables includes the $1\sigma$ bounds of the latest results of the average of $R_K$ and $R_{K^*}$ (light blue region), $R_\phi$ (navy blue region), $P'_5 (B^{0} \to K^* \mu^+ \mu^-)$ (gray region), branching fractions of $\mathcal{B}(B \to K^{(*)} \mu^+ \mu^-)$ (orange region), $\mathcal{B}(B_s \to \phi \mu^+ \mu^-)$ (plum region), $\mathcal{B}(B_s \to \mu^+ \mu^-)$ (pale green region), $\mathcal{B}(B \to K^{(*)} e^+ e^-)$ (green-yellow region) $\mathcal{B}(B_s \to \phi e^+ e^-)$ (dark green region). The global fit including the full data set is represented in red corresponding to $1\sigma$ and $2\sigma$ bounds of the best fit values.
\begin{itemize}
    \item The first figure on the left top represents the scenario S9: $[C_{qe}]_{2311}=[C_{qe}]_{2322}$, $[C_{\phi q}^{(3)}]_{23}$ and probes LFU-NP. The SM-like behavior of the latest measurements of the average of $R_K$ and $R_{K^*}$ and the branching fraction of $\mathcal{B}(B_s \to \mu^+ \mu^-)$ are consistent with the global fit. The entire parameter space is consistent with $R_\phi$ and the average of $R_K$ \& $R_{K^*}$. Since S9 is of LFU-NP type, any point in this region of the parameter space can satisfy $R_K$ and $R_{K^*}$. The same argument is true for $R_\phi$. Interestingly, the global fit can simultaneously accommodate the measurements of $P'_5$, $\mathcal{B}(B \to K^{(*)} \mu^+ \mu^-)$, and $\mathcal{B}(B \to K^{(*)} e^+ e^-)$ at $1\sigma$ of their respective errors. However, the $1\sigma$ contours corresponding to $\mathcal{B}(B_s \to \phi \mu^+ \mu^-)$ and $\mathcal{B}(B_s \to \phi e^+ e^-)$ partially overlap with the global contour, so there is hardly any scope to accommodate it in the global fit within $1\sigma$ of its experimental error. 
    
    \item The second figure on the right top represents the scenario S10: $[C_{\ell q}^{(3)}]_{1123}=[C_{\ell q}^{(3)}]_{2223}$, $[C_{\phi q}^{(3)}]_{23}$ which also contributes as LFU-NP. Therefore, any point in this region of parameter space can satisfy the average of $R_K$ \& $R_{K^*}$ and $R_\phi$. The global fit can fully accommodate the average of $R_K$ \& $R_{K^*}$, $R_\phi$, $\mathcal{B}(B_s \to \mu^+ \mu^-)$, $P'_5$, $\mathcal{B}(B \to K^{(*)} \mu^+ \mu^-)$, $\mathcal{B}(B \to K^{(*)} e^+ e^-)$, and partially $\mathcal{B}(B_s \to \phi \mu^+ \mu^-)$ and $\mathcal{B}(B_s \to \phi e^+ e^-)$ considering the $1\sigma$ error of the respective measurements. 
    
    \item The third figure on the left-bottom representing the scenario S18: $[C_{\ell q}^{(3)}]_{1123}$, $[C_{\ell q}^{(3)}]_{2223}$ offers an LFUV-NP. Unlike the previous two 2D scenarios, here both the SMEFT coefficients are completely independent of the lepton flavors. The global fit can simultaneously accommodate the average of $R_K$ \& $R_{K^*}$, $R_\phi$, $\mathcal{B}(B_s \to \mu^+ \mu^-)$, $\mathcal{B}(B \to K^{(*)} \mu^+  \mu^-)$ and $\mathcal{B}(B \to K^{(*)} e^+ e^-)$ within the $1\sigma$ error of the respective measurements. However, it is difficult to accommodate $P'_5$ and $\mathcal{B}(B_s \to \phi \mu^+ \mu^-)$ and $\mathcal{B}(B_s \to \phi e^+ e^-)$ within $1\sigma$.
    
    \item The fourth figure on the right-bottom represents the scenario S42: $[C_{qe}]_{2322}$, $[C_{\phi q}^{(3)}]_{23}$ with a similar set of SMEFT coefficients of the first case, but it offers LFU+LFUV-NP with one of its couplings contributing only to muon. Interestingly, the global fit can fully accommodate $\mathcal{B}(B_s \to \mu^+ \mu^-)$, $P'_5$, $\mathcal{B}(B \to K^{(*)} \mu^+ \mu^-)$ and $\mathcal{B}(B \to K^{(*)} e^+ e^-)$, $R_\phi$ and $\mathcal{B}(B_s \to \phi e^+ e^-)$ at $1\sigma$ of their experimental error. There is a partial overlap of the 1$\sigma$ contour of $\mathcal{B}(B_s \to \phi \mu^+ \mu^-)$ with the global $1\sigma$ contour. There is no complete overlap region of the average of $R_K$ and $R_{K^*}$ $1\sigma$ contour with the global $1\sigma $ except at the boundaries. However, this is not the case when individual $R_K$ and $R_{K^*}$ are considered.
\end{itemize}

\section{Analyzing $b \to s \mu^+ \mu^-$ and $b \to s e^+ e^-$ decay observables}
\label{results1}
In this section, we probe the behavior of several observables such as the branching fractions, the ratios of appropriate branching fractions, several angular observables including the forward-backward asymmetry, the longitudinal polarization fraction, $P_i$-observables for both electron and muon final states of $B \to K^{(*)} \ell^+ \ell^-$ decays. We explore the above in the presence non-vanishing SMEFT coefficients for 1D and 2D NP scenarios with large $\Delta \chi^2$ as given in Table~\ref{tab_fit3}. We depict the $q^2$ distributions of the above observables for muon and electron final states. The observables are evaluated in different $q^2$ bins mostly in the low $q^2$ domain including {\{1.1,6.0\} ${\rm GeV}^2$ along with bins at high $q^2$ regions including \{15.0,19.0\} ${\rm GeV}^2$. All the numerical entries are given in Table~\ref{tab_ap1} -\ref{tab_ap15} of Appendix~\ref{ap3}. The bin sizes are fixed from the LHCb measurements. In addition to $B \to K^{(*)} \ell^+ \ell^-$ decays, we also compute $B_s \to \phi \ell^+ \ell^-$ and $\Lambda_b \to \Lambda \ell^+ \ell^-$ decays in the Appendix~\ref{ap4}. The error associated with the observables (both in SM and NP) are reported for the bin-wise predictions in Fig.~\ref{fig_q2_br}-Fig~\ref{fig_q2_p5p} and Table~\ref{tab_ap1} -\ref{tab_ap15}. However, we do not show errors in $q^2$ distribution plots from [1.1, 6] GeV$^2$. The total error budget on any observables is computed by combining all sources of uncertainties including form factors, charm-loop uncertainties and soft-gluon exchange, CKM elements, decay constants, bag parameters, Wilson coefficients and correlated parameter sets (e.g.form factor fit correlations). Our observations for $B \to K^{(*)} \ell^+ \ell^-$ decays are as follows:

\begin{itemize}
    \item Branching ratio (BR): The integrated values of the branching ratios of $B^+ \to K^+ \ell^+ \ell^-$ and $B^0 \to K^{0*} \ell^+ \ell^-$ ($\ell \in e, \mu$) in different $q^2$ bins are reported in Table~\ref{tab_ap1} and \ref{tab_ap2}. 
    The SM is lepton flavor universal, the BRs in each bin are similar for both electron and muon final states (of $\mathcal{O}(10^{-8})$). The contribution from all the NP scenarios is less than the SM and is distinguishable at the $1\sigma$-$2\sigma$ level at different $q^2$ bins. Since NP scenarios S3, S9, and S10 contribute as LFU-NP, the NP in electron and muon final states cannot be distinguished. Even the contributions from LFUV-NP scenarios such as S18 and S42 for muon and electron final states are hardy distinguishable. Interestingly, all these NP scenarios contribute to BRs in such a way that they keep the ratio of muon to electron branching ratios to unity. 
    
    The left top and left bottom figures in Fig.~\ref{fig_q2_br} show the BR distribution results for the muon final state while considering theoretical error estimates of the same for SM (pink boxes) and the NP  scenario S9 (green boxes). In these two figures, both the SM and NP predictions are compared with the experimental data from LHCb marked with black plus signs. Typically, the width of these boxes and the plus signs denote the bin sizes and their corresponding heights denote the results that include the associated errors from theory and experiment. As is evident from the plots, the theoretical uncertainties of the BR distributions are appreciably large. Certainly, our NP prediction from the global fit is in good agreement with the data and is clearly distinguishable from the same associated with SM beyond the uncertainties in almost all the bins.
    
   Similarly, as shown in the middle and right columns of Fig.~\ref{fig_q2_br} in a smaller $q^2$ domain of $\{1.1, 6\}$ ${\rm GeV}^2$ we show the distributions of the BRs of $B^+ \to K^+ \ell^+ \ell^-$ and $B^0 \to K^{0*} \ell^+ \ell^-$ for $\ell=\mu$ and $e$ respectively. We notice that all the NP distributions are distinguishable from the SM and lie below the SM curve in each of the figures. The scenarios S3 and S18 completely overlap in $B^+ \to K^{+} \mu^+ \mu^-$ whereas a slight distinction is found in $B^+ \to K^{+} e^+ e^-$. Similarly, S9, S10, and S42 are indistinguishable from each other in both decay modes. Coming to the $q^2$ distribution of $B^0 \to K^{0*} \ell^+ \ell^-$ ($\ell \in e, \mu$), a similar outcome applies for the scenarios S3 and S18. Similarly, scenarios S9, S10, and S42 stay close to each other in $B^0 \to K^{0*} \mu^+ \mu^-$. However, scenario S42 exhibits quite different behaviour of the distribution for $B^0 \to K^{0*} e^+ e^-$ exhibiting a clear distinction from S9 and S10. 

    \begin{figure}[h]
 \centering
 \includegraphics[width=5.9cm,height=4.3cm]{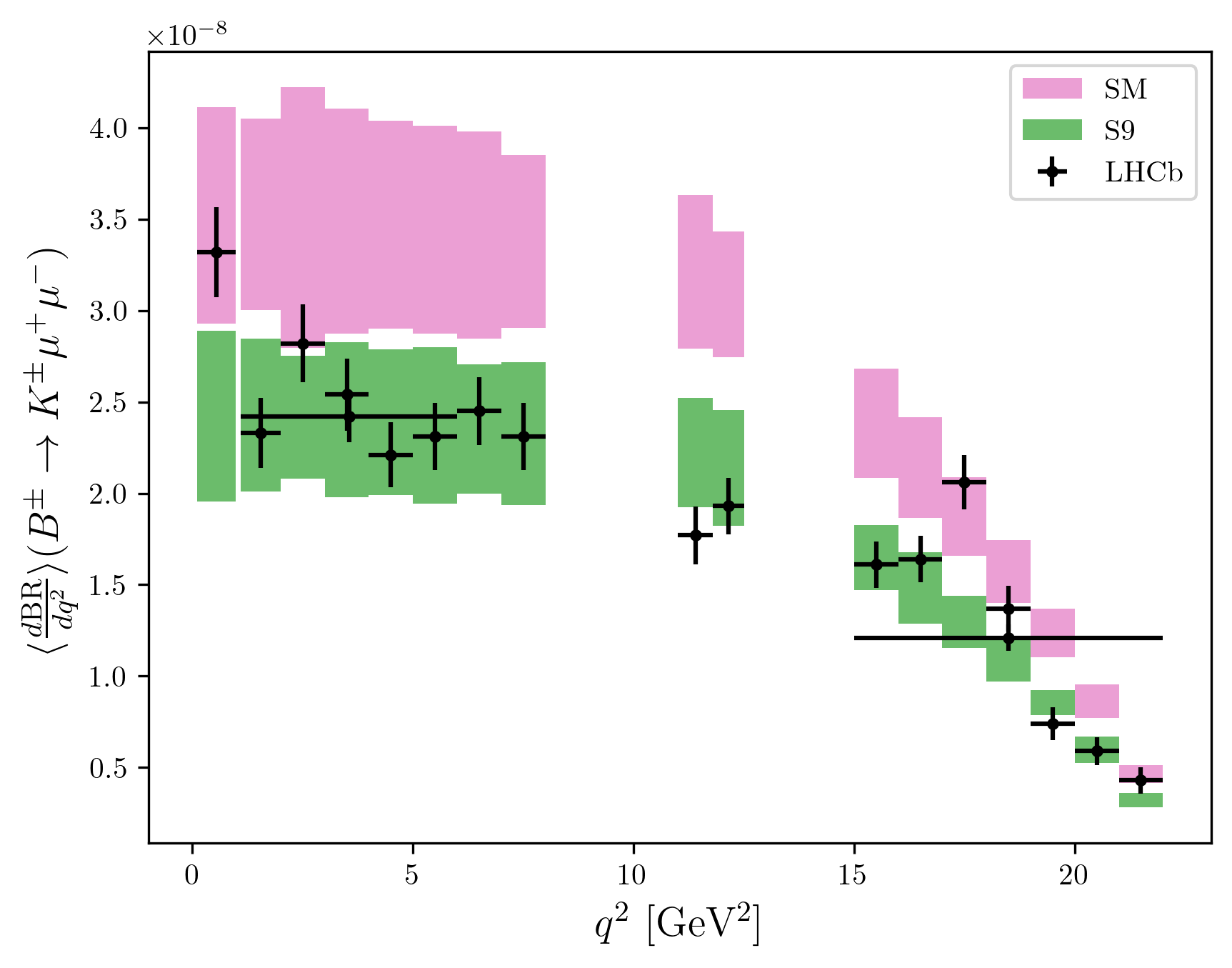}
\includegraphics[width=5.9cm,height=4.3cm]{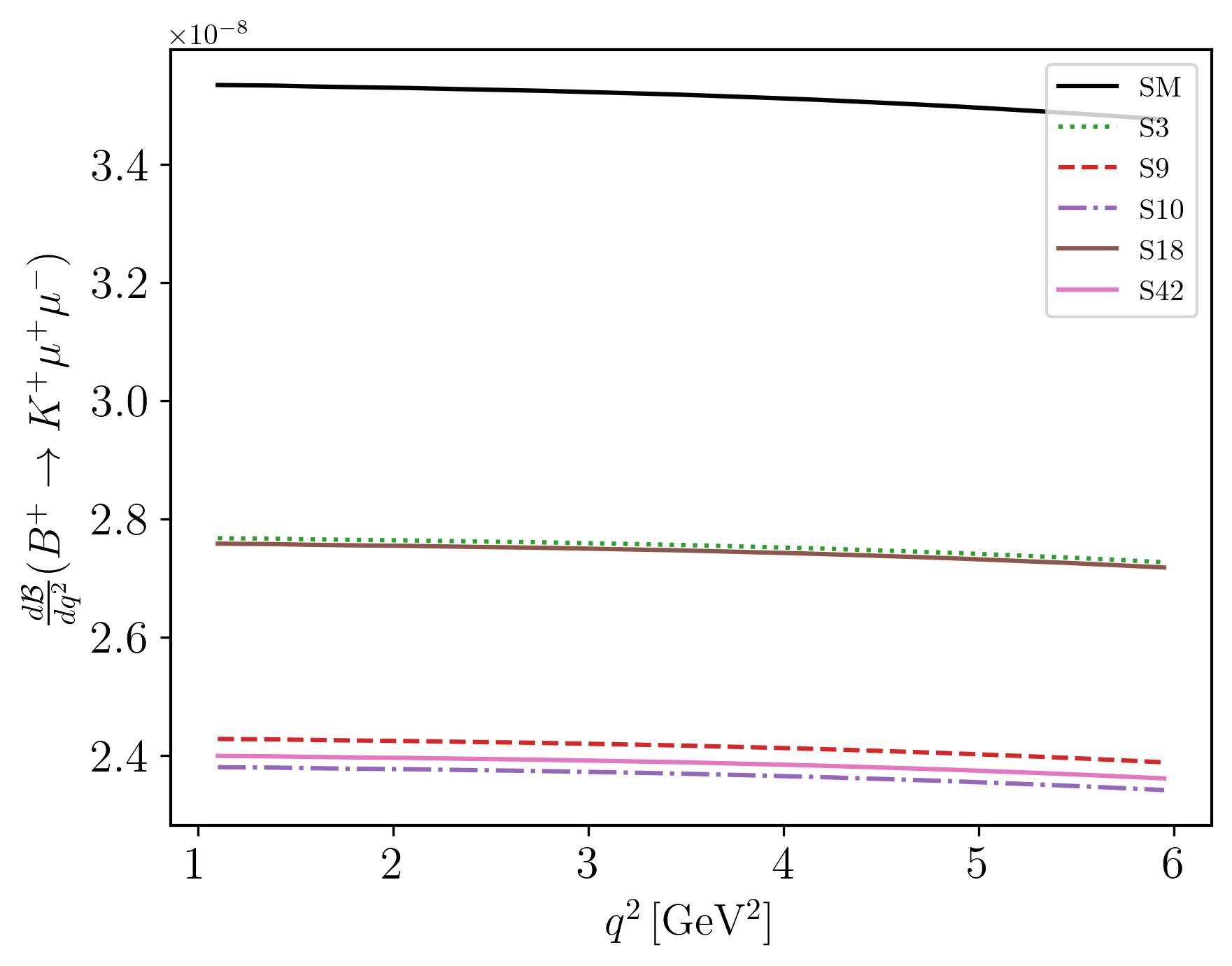}
\includegraphics[width=5.9cm,height=4.3cm]{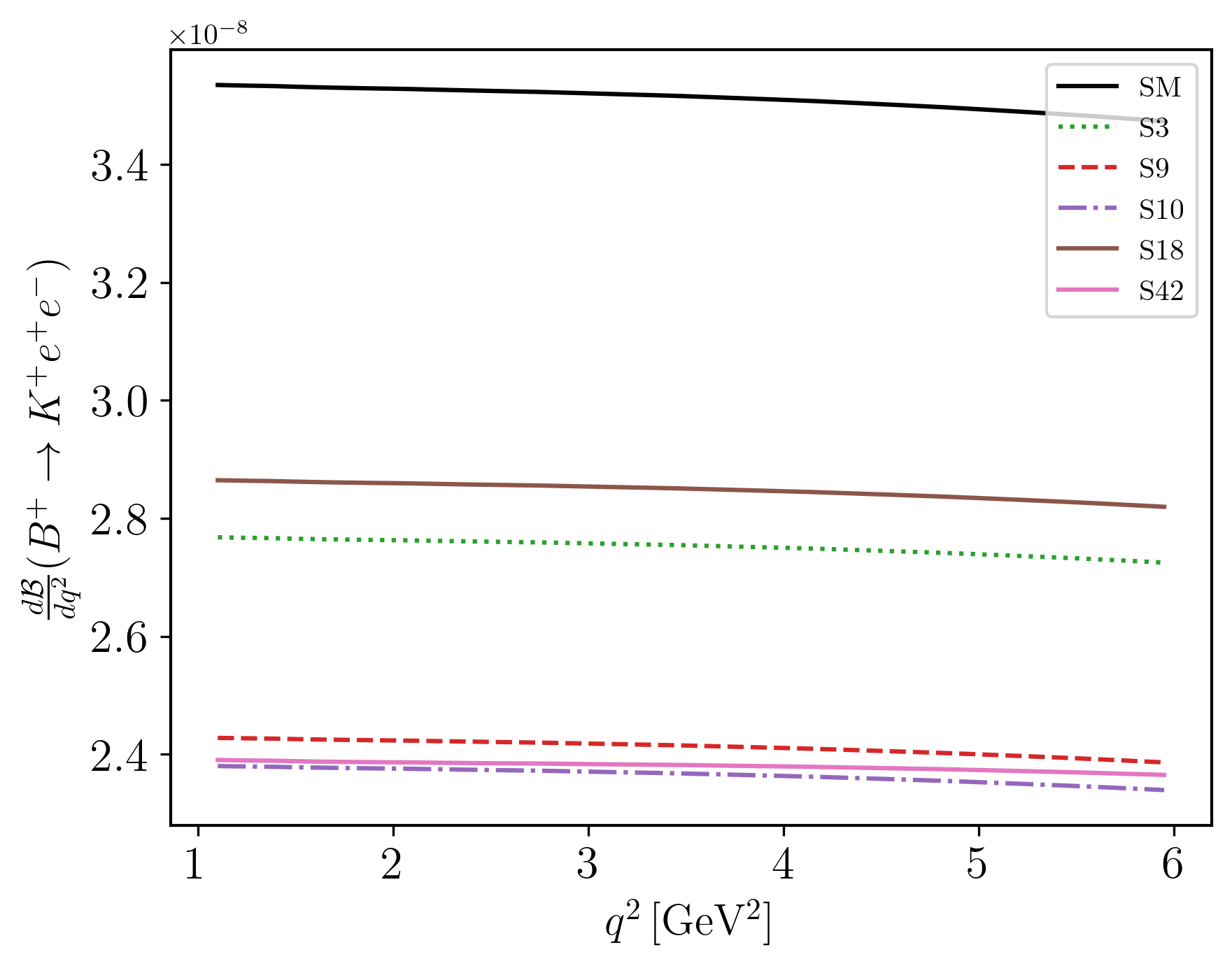}
\includegraphics[width=5.9cm,height=4.3cm]{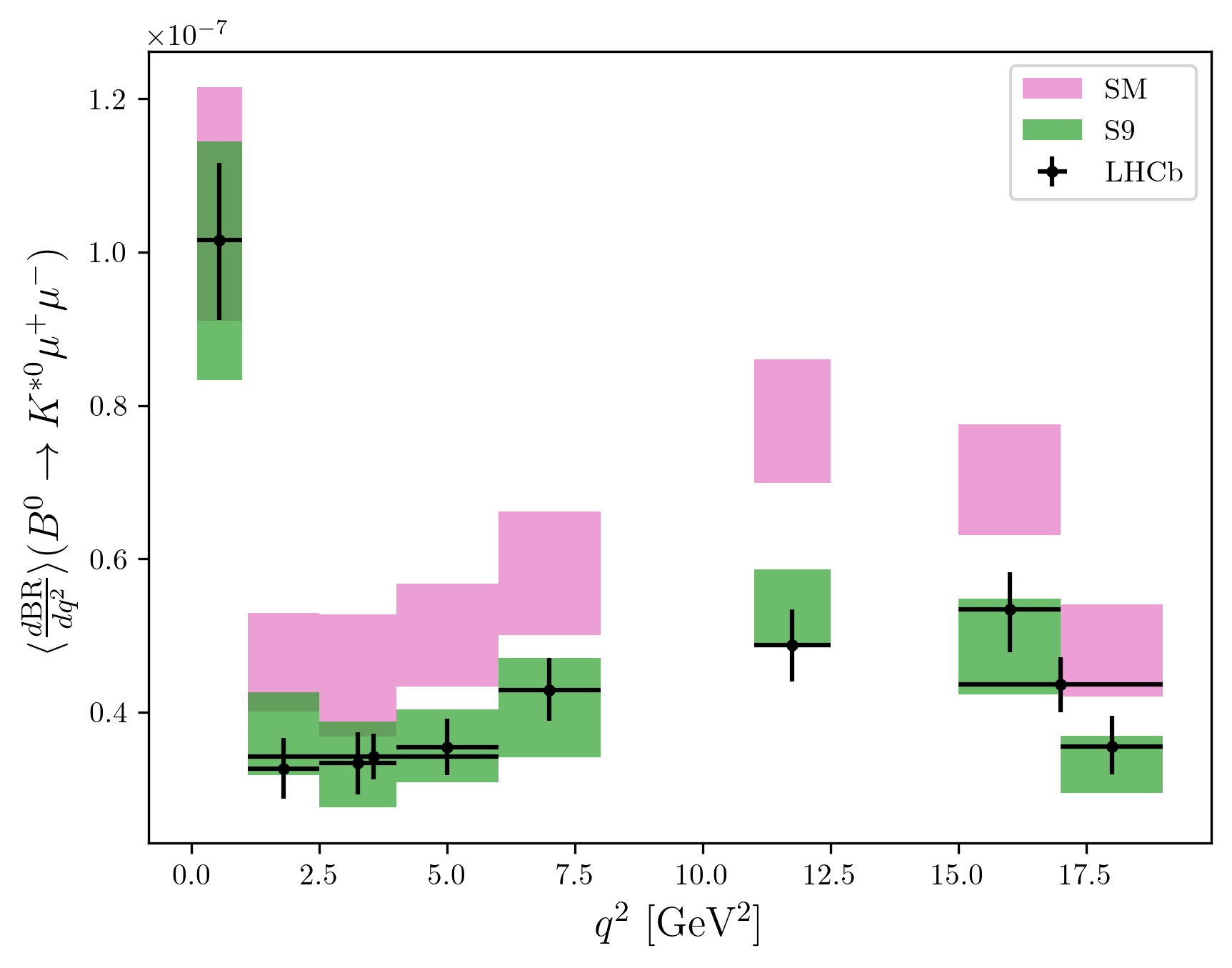}
\includegraphics[width=5.9cm,height=4.3cm]{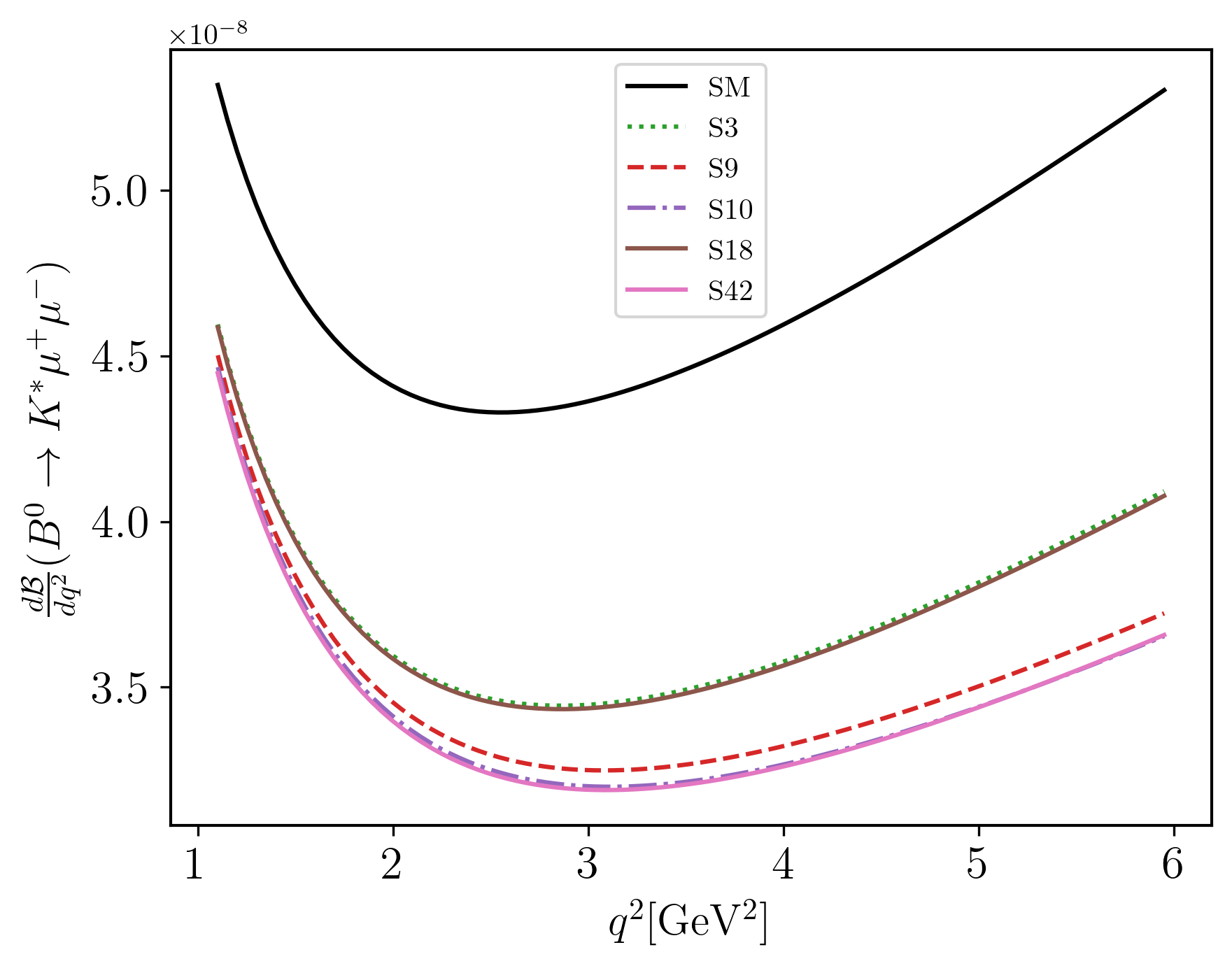}
\includegraphics[width=5.9cm,height=4.3cm]{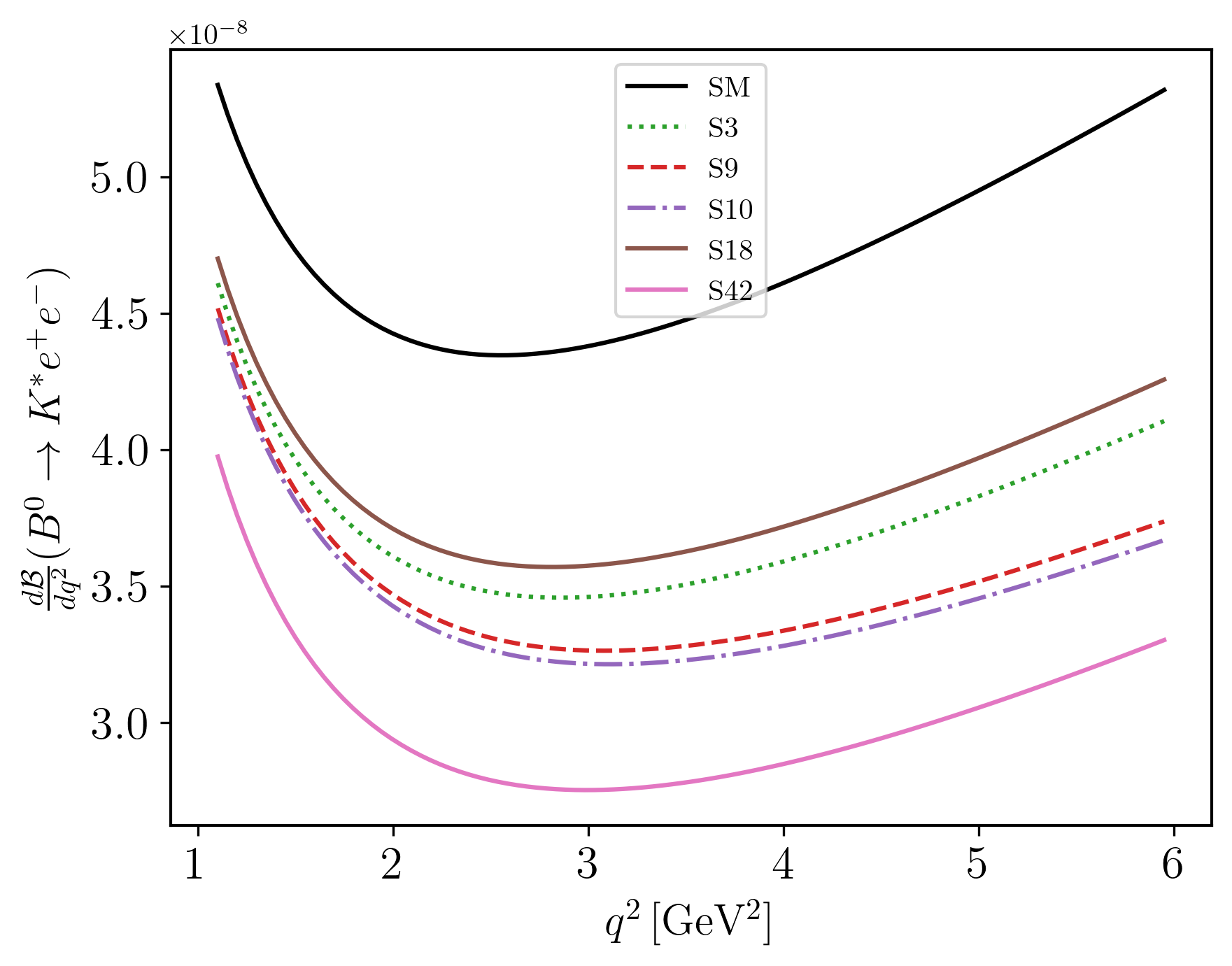}
\caption{Left panel: bin-wise distribution of the branching fractions of $B \to K^{(*)} \mu^+ \mu^-$ decays in SM and in the presence of the NP scenario S9. Both SM and S9 results are compared with the LHCb data. 
Middle and the right panel: $q^2$-dependency curves for the same observable in the presence of SM and various NP scenarios for both muon and electron final states.}
\label{fig_q2_br}
\end{figure}

\item Forward-backward asymmetry (AFB): The AFB for $B^+ \to K^+ \ell^+ \ell^-$ ($\ell \in e, \mu$) is zero in SM (Table~\ref{tab_ap3}) however, non-zero contributions come from NP scenarios at the order of $\mathcal{O}(10^{-13})$ in the electron mode and $\mathcal{O}(10^{-8})$ in the muon mode.    
AFB would be more interesting as of vector final states are concerned such as $B^0 \to K^{0*} \ell^+ \ell^-$ ($\ell \in e, \mu$) modes. Looking into Table~\ref{tab_ap4} for the integrated values at different bins, it is clear that the NP effects are distinguishable from SM in both the final muon and electron states. However, there is an exception in the electron mode as of scenario S42 is concerned which is almost SM-like.
Deviations up to $2\sigma$ are found in S9, S10 and S42 (only in the muon final state), in contrast to the same of lesser significance in S3 and S18. Additionally, the contributions from S9, S10 and S42 (only in the muon final state) are almost similar. 

AFB distribution results for the $B^0 \to K^{0*} \mu^+ \mu^-$ while considering the theoretical error estimates of the same for the SM (pink boxes) and the NP scenario S9 (green boxes) are shown in the bottom left panel of Fig~\ref{fig_q2_afb}. Both the SM and NP predictions are compared with the experimental data from LHCb. We observe that except for the lowest bin at [0.1, 0.98] ${\rm GeV}^2$, our global fit is consistent with the data and can be clearly distinguished from SM at least in the lower bins including [1.1, 2.5] ${\rm GeV}^2$, [2.5, 4] ${\rm GeV}^2$ and [4, 6] ${\rm GeV}^2$. Additionally, AFBs for the final electron and muon states have zero crossings in the $q^2$ distribution curves from [1.1, 6] ${\rm GeV}^2$. These distributions are shown in the bottom middle and right panel of Fig~\ref{fig_q2_afb}. At the leading order, the positions of these zeros depend on the combinations of WCs $C_9$ and $C_7$ which are independent of hadronic form factors. The zero crossing point for SM is located at $q^2 \sim 3.4$ $\rm GeV^2$ for both electron and muon final states of $B^0 \to K^{0*} \ell^+ \ell^-$ decays. However, the locations of the same point for the NP scenarios are hardly the same. First,  the zero-crossing point for S3 and S18 is found at $q^2 \sim 3.8$ $\rm GeV^2$, for S9 and S10 the same is at $q^2 \sim 4.7$ $\rm GeV^2$ in the case of both $B^0 \to K^{0*} \mu^+ \mu^-$ and $B^0 \to K^{0*} e^+ e^-$ modes. However, the zero-crossing point for S42 is the same as SM for $B^0 \to K^{0*} e^+ e^-$ mode and the same is in $q^2 \sim 4.6$ $\rm GeV^2$ for $B^0 \to K^{0*} \mu^+ \mu^-$ mode. Interestingly, the zero crossing point of S42 can distinguish between the nature of AFB associated with the electron and muon modes.

\begin{figure}[h]
\centering
\includegraphics[width=5.9cm,height=4.3cm]{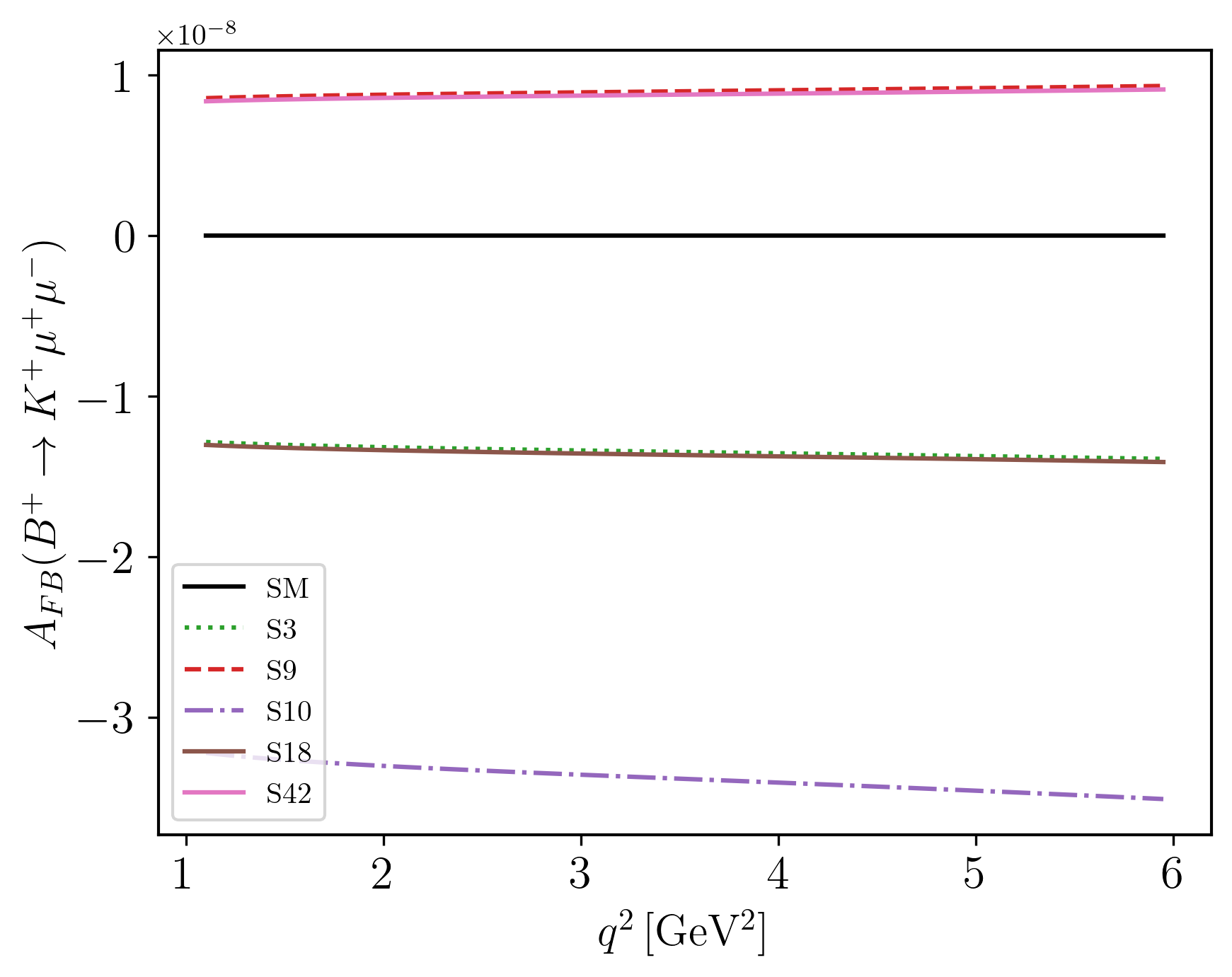}
\includegraphics[width=5.9cm,height=4.3cm]{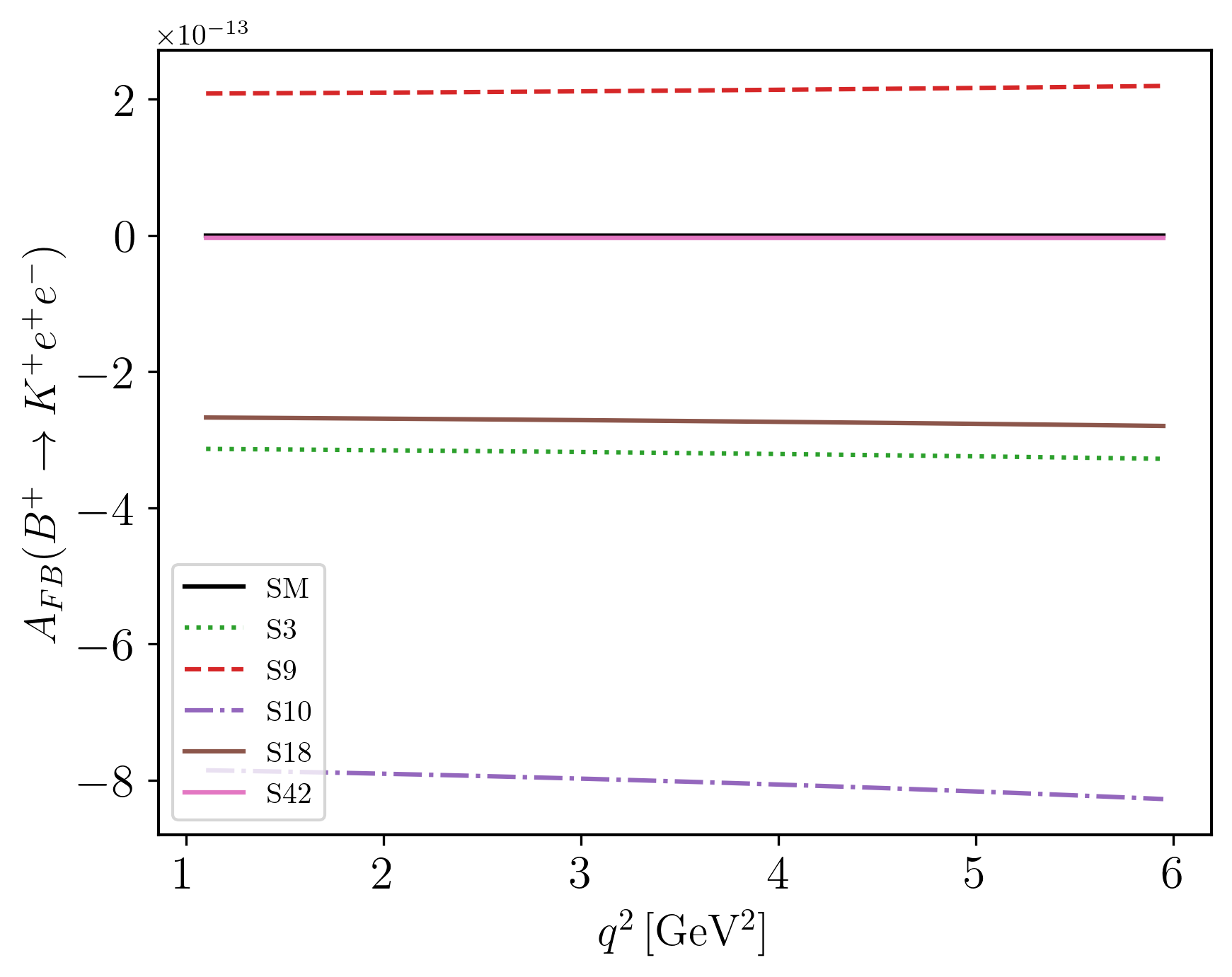}

\includegraphics[width=5.9cm,height=4.3cm]{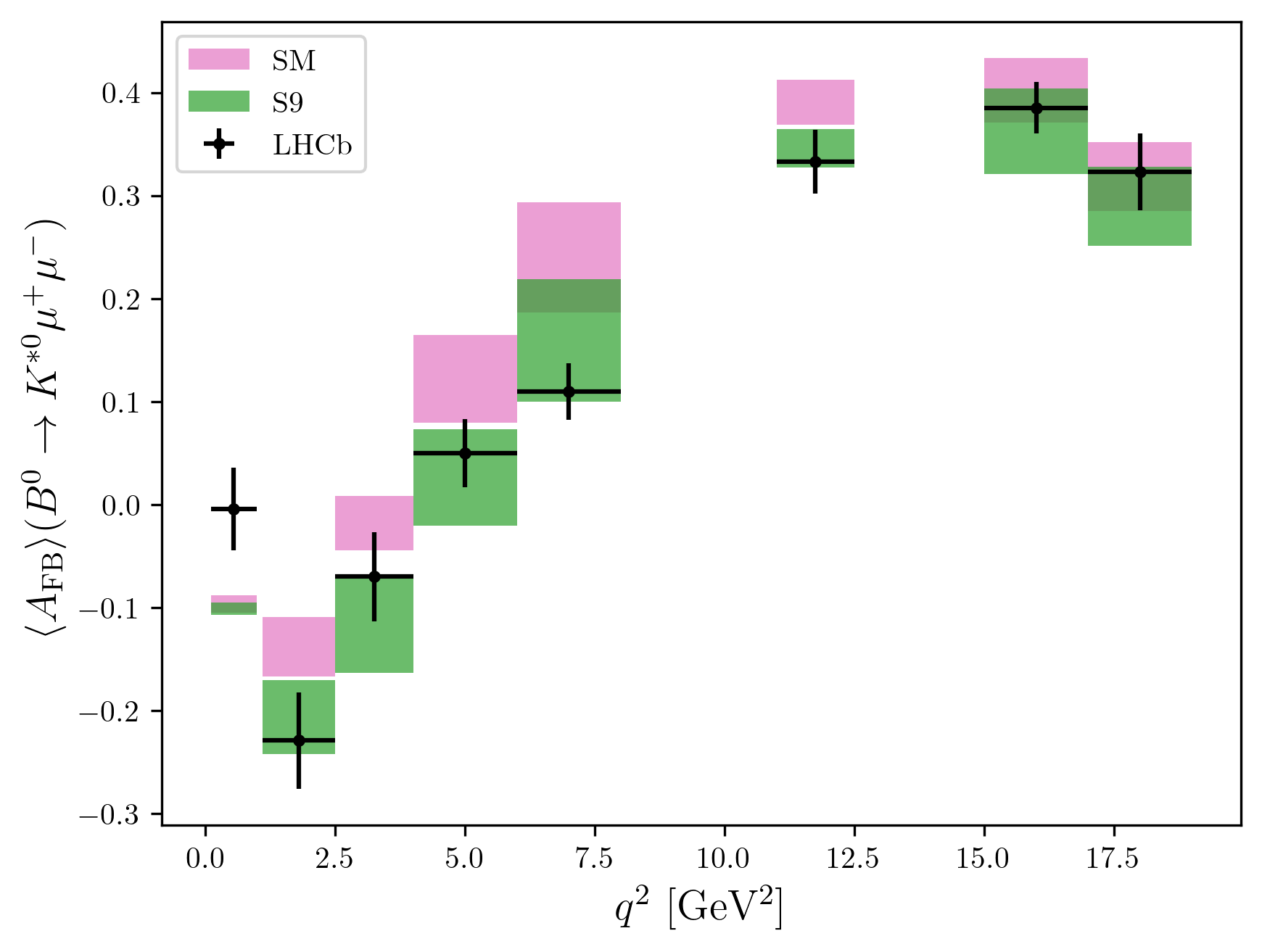}
\includegraphics[width=5.9cm,height=4.3cm]{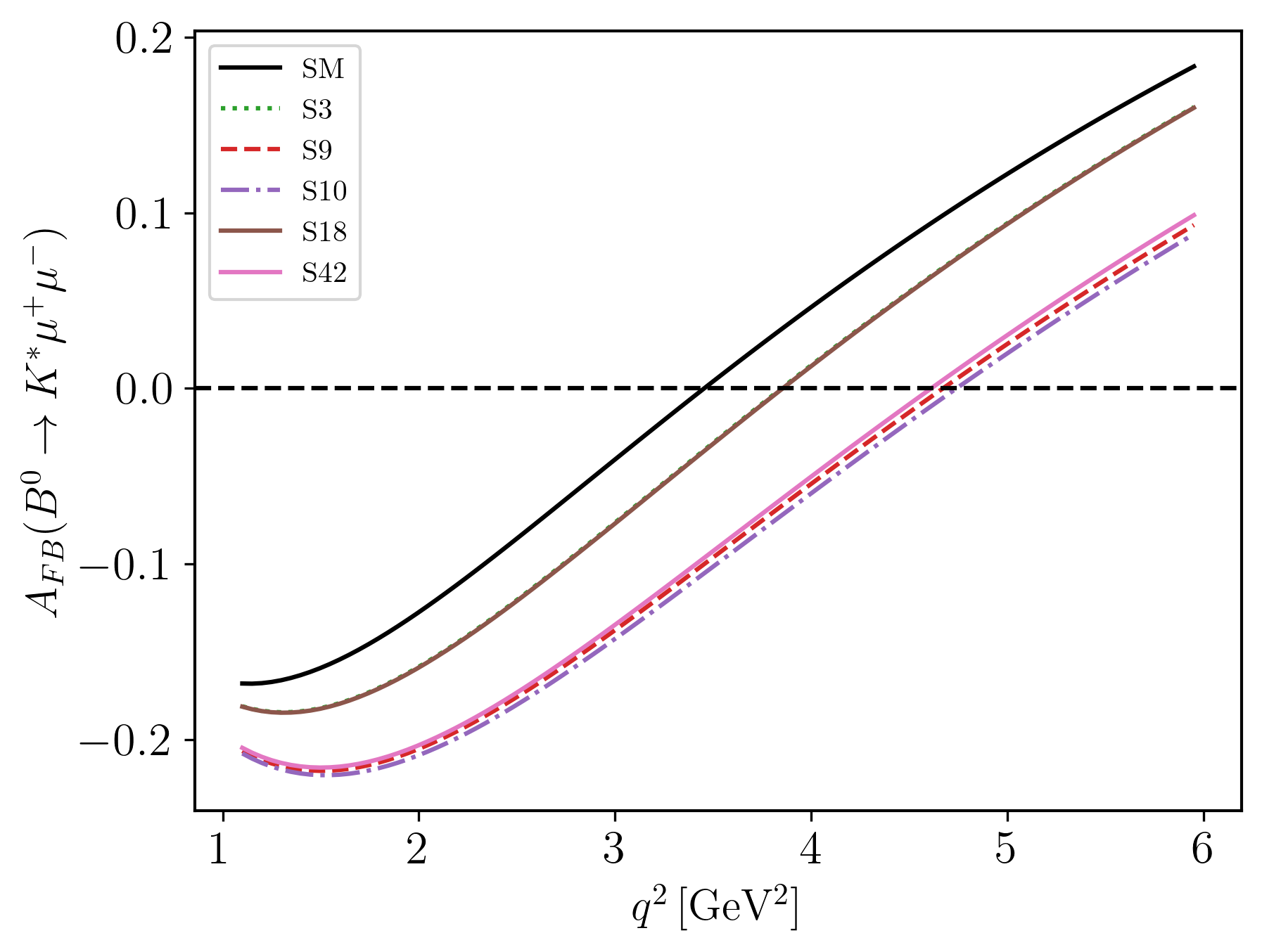}
\includegraphics[width=5.9cm,height=4.3cm]{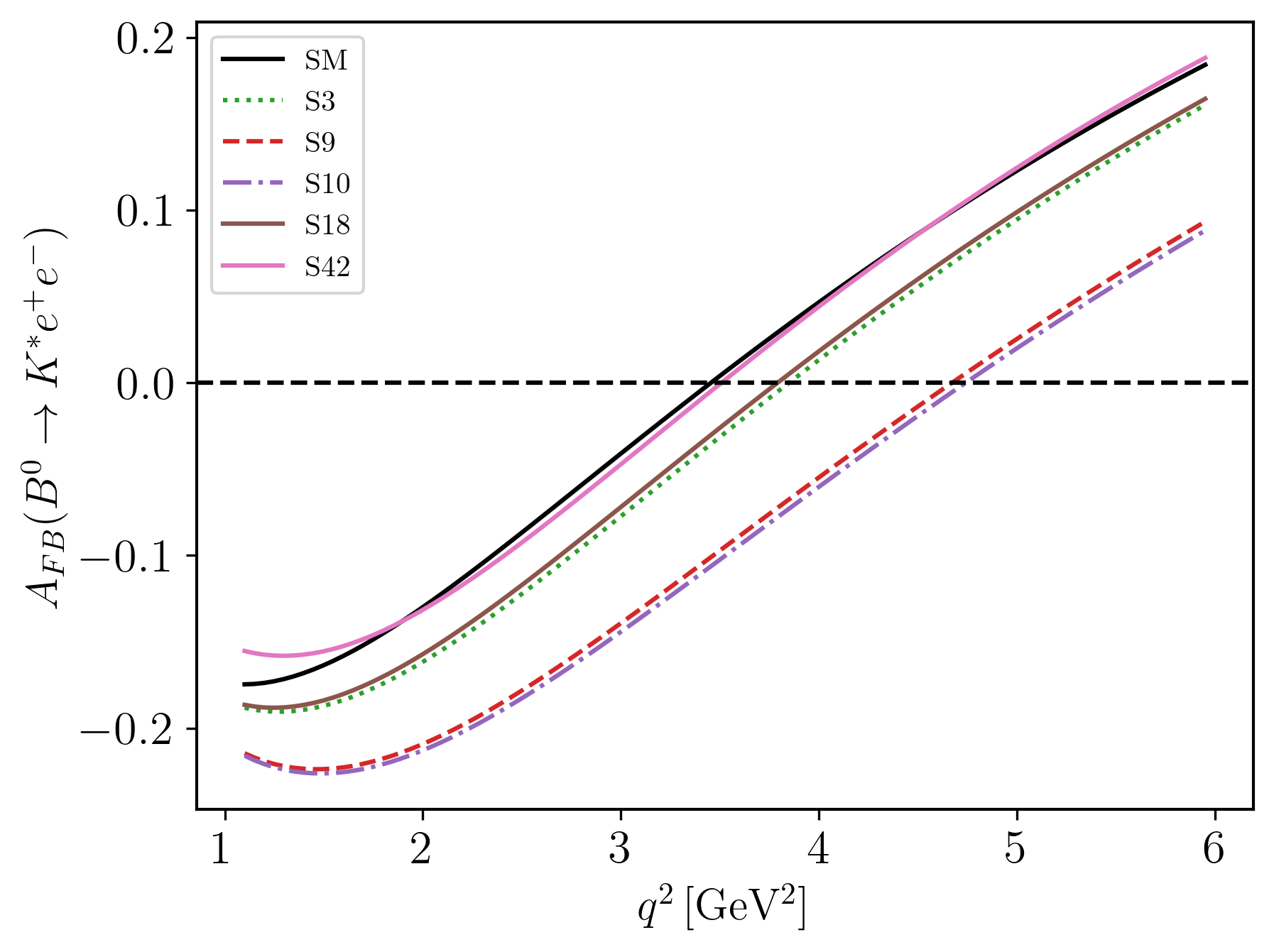}
\caption{Top panel: $q^2$ dependency curves of forward-backward asymmetry for $B \to K \ell^+ \ell^-$ ($\ell \in e,\mu$) decays in SM and in the presence of various NP scenarios. Bottom left: bin-wise distribution of forward-backward asymmetry for $B \to K^* \mu^+ \mu^-$ decay in SM and in the presence of NP S9. SM and S9 are compared with the LHCb data. Bottom middle and right panel: $q^2$ dependency curves of the same observable in SM and in the presence of various NP scenarios for both muon and electron final states.}
\label{fig_q2_afb}
 \end{figure}

\item Longitudinal polarization fraction (FL): We now explore the longitudinal polarization fraction of $K^*$ in $B^0 \to K^{0*} \ell^+ \ell^-$ ($\ell \in e, \mu$).
It is clear from Table ~\ref{tab_ap5} that the contributions from the NP scenarios S3 and S18 are not so different from that of SM especially in the bins [2.5, 4] ${\rm GeV}^2$ and [4, 6] ${\rm GeV}^2$ for both the electron and muon final states. However, a limited degree of deviations ($1\sigma$-$1.5\sigma$) from SM is apparent for S9, S10, and S42 in the muon mode. Similar signatures from S9 and S10 are also found for the electron mode.  
We further investigate the FL distribution results for $B^0 \to K^{0*} \mu^+ \mu^-$ while considering theoretical error estimates of the same for SM (pink boxes) and NP scenario S9 (green boxes) in the left panel plot of Fig~\ref{fig_q2_fl}. With the experimental data from LHCb, one finds that although the global fit is consistent with the data, the NP contribution from S9 is hardly so different from SM in most of the bins except for [1.1, 2.5] ${\rm GeV}^2$. Even in the high $q^2$ regime, the SM result overlaps with the data. Therefore, only the bin [1.1, 2.5] ${\rm GeV}^2$ might be interesting as of the current data are concerned. In the future with increasing data on FL, other bins may also show some interesting NP signatures. In the same Fig.~\ref{fig_q2_fl}, the $q^2$ distribution curves from [1.1, 6] ${\rm GeV}^2$ are shown in the middle and right panel plots respectively for the muon and electron final states. Here we notice that scenarios S3 and S18 remain close to SM and overlap more with the SM curve with an increase in $q^2$. Similarly, S9 and S10 also move closer to SM as $q^2$ increases. However, scenario S42 exhibits different behaviour for the electron mode and it crosses the SM curve approximately at $\sim 2\, \rm GeV^2$. Hence S42 can distinguish the nature of FL for the moun mode from that of the electron mode. 

\begin{figure}[h]
 \centering
\includegraphics[width=5.9cm,height=4.3cm]{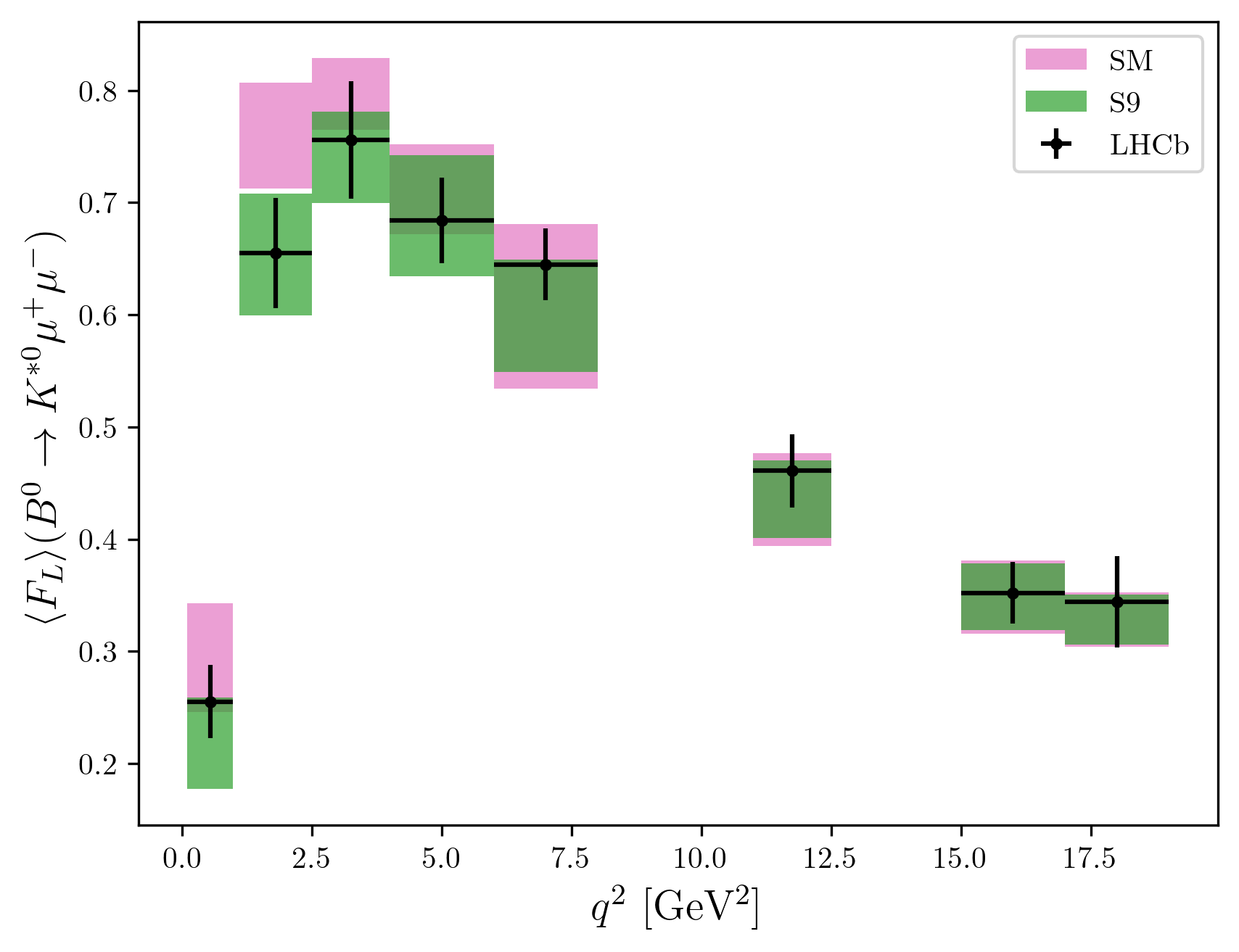}
\includegraphics[width=5.9cm,height=4.3cm]{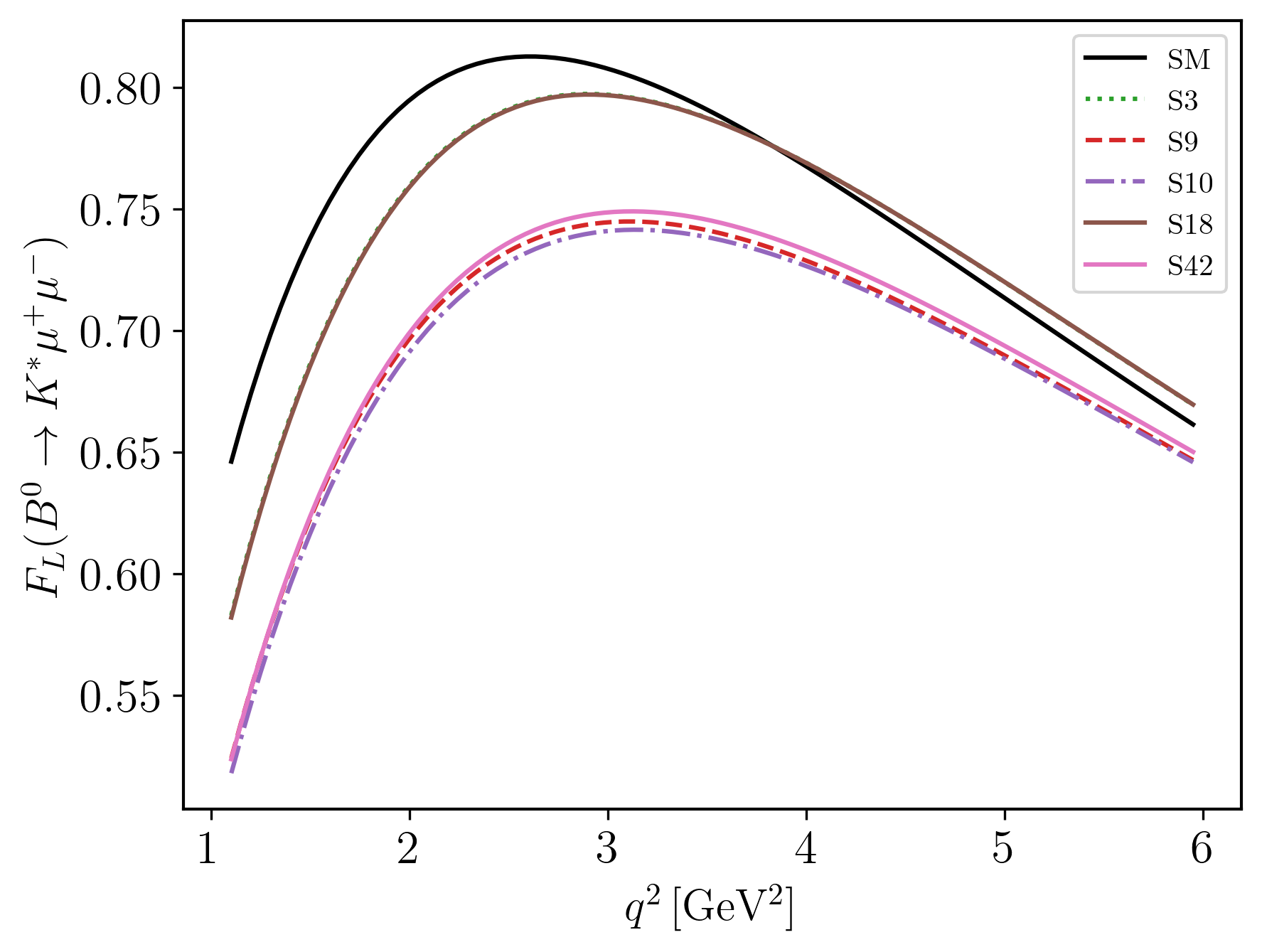}
\includegraphics[width=5.9cm,height=4.3cm]{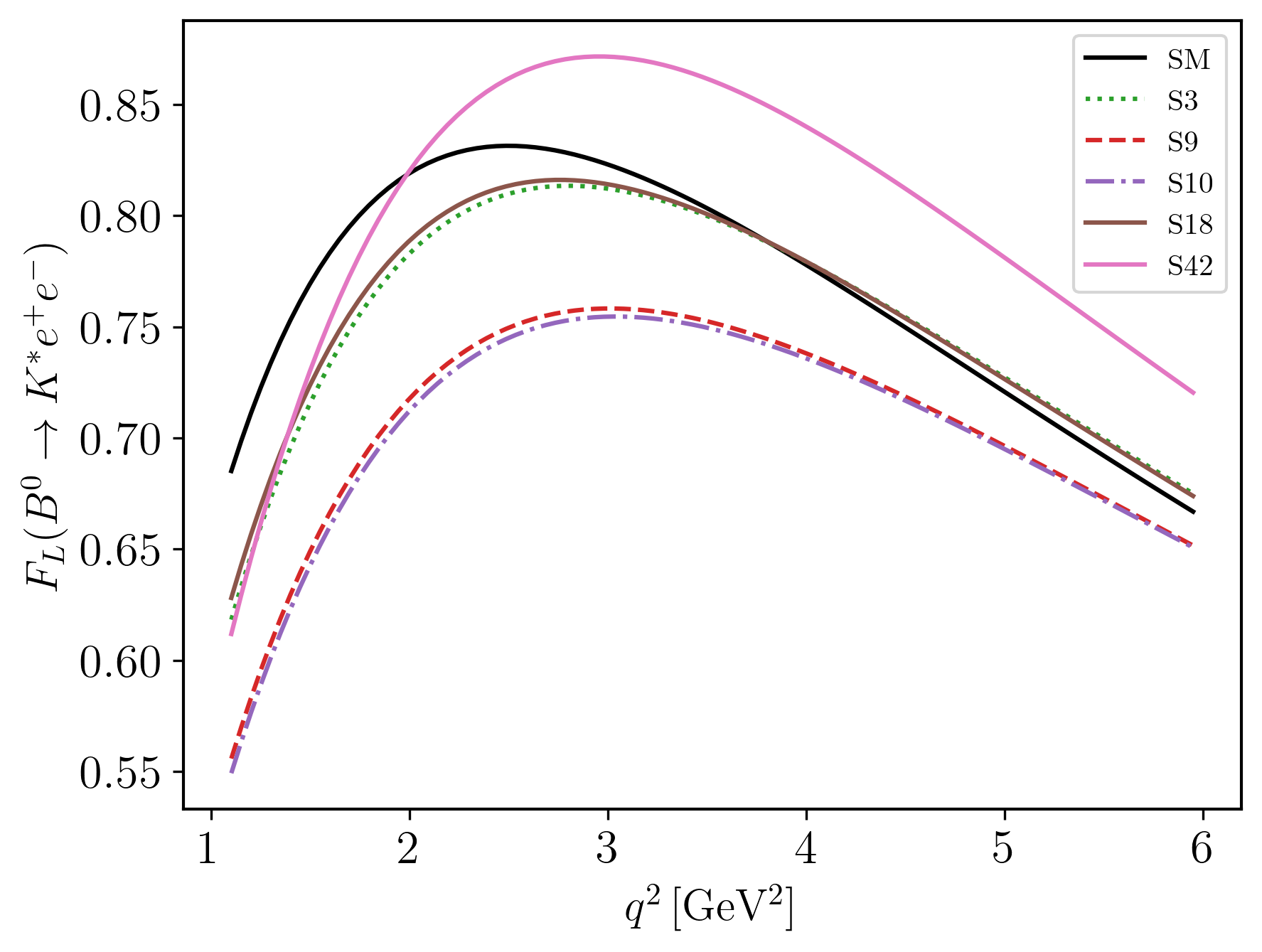}
 \caption{Left panel: bin-wise distribution of longitudinal polarization fraction for $B \to K^* \mu^+ \mu^-$ decay in the SM and NP S9. SM and S9 are compared with the LHCb data. Middle and right panels: $q^2$ dependency curves of the same observable in the SM and various NP scenarios for both muon and electron final states.}
 \label{fig_q2_fl}
 \end{figure}

\item $P_1$: The NP contribution in $P_1$ (see Table~\ref{tab_ap6}) is not significant enough to distinguish itself from SM beyond $1\sigma$. In the bin-wise distribution plot in Fig~\ref{fig_q2_p1} (left panel), we notice that the experimental values of $P_1$ in each bin are associated with very large error bars. Thus, distinguishing NP scenarios from SM becomes a difficult task. The global fit more or less coincides with the SM expectation rather than the experimental data. Moreover, looking at the $q^2$ distribution plots, although there are zero crossing points, none of the NP zero crossings can claim any interesting distinction from SM, especially in the case of $B^0 \to K^{0*} \mu^+ \mu^-$ mode. The same is true for $B^0 \to K^{0*} e^+ e^-$ mode except for the case of S42 which has a zero crossing point at $q^2 \sim 2.6$ GeV$^2$ different from the SM zero crossing of $q^2 \sim 2.0$ GeV$^2$.

\begin{figure}[h]
 \centering
\includegraphics[width=5.9cm,height=4.3cm]{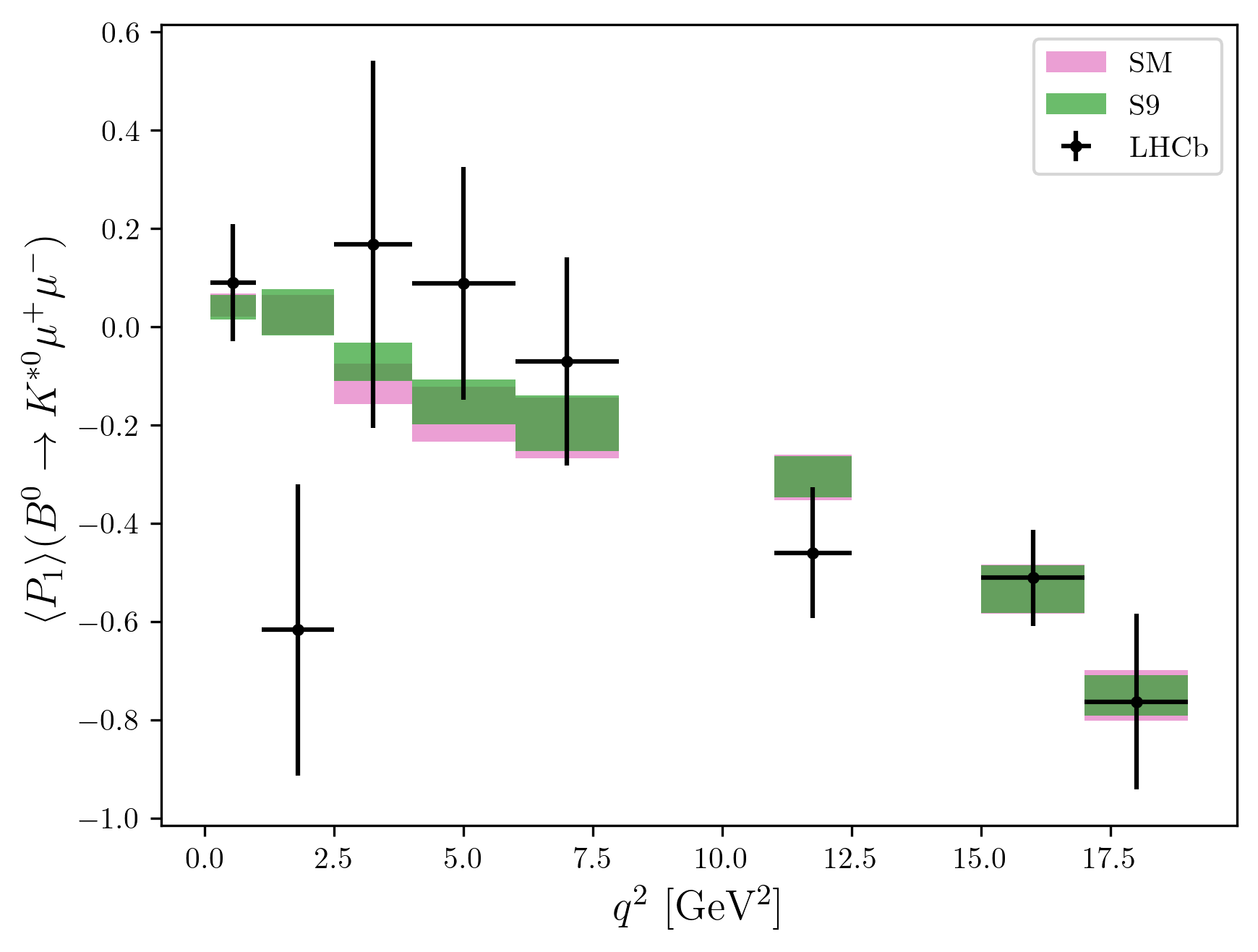}
\includegraphics[width=5.9cm,height=4.3cm]{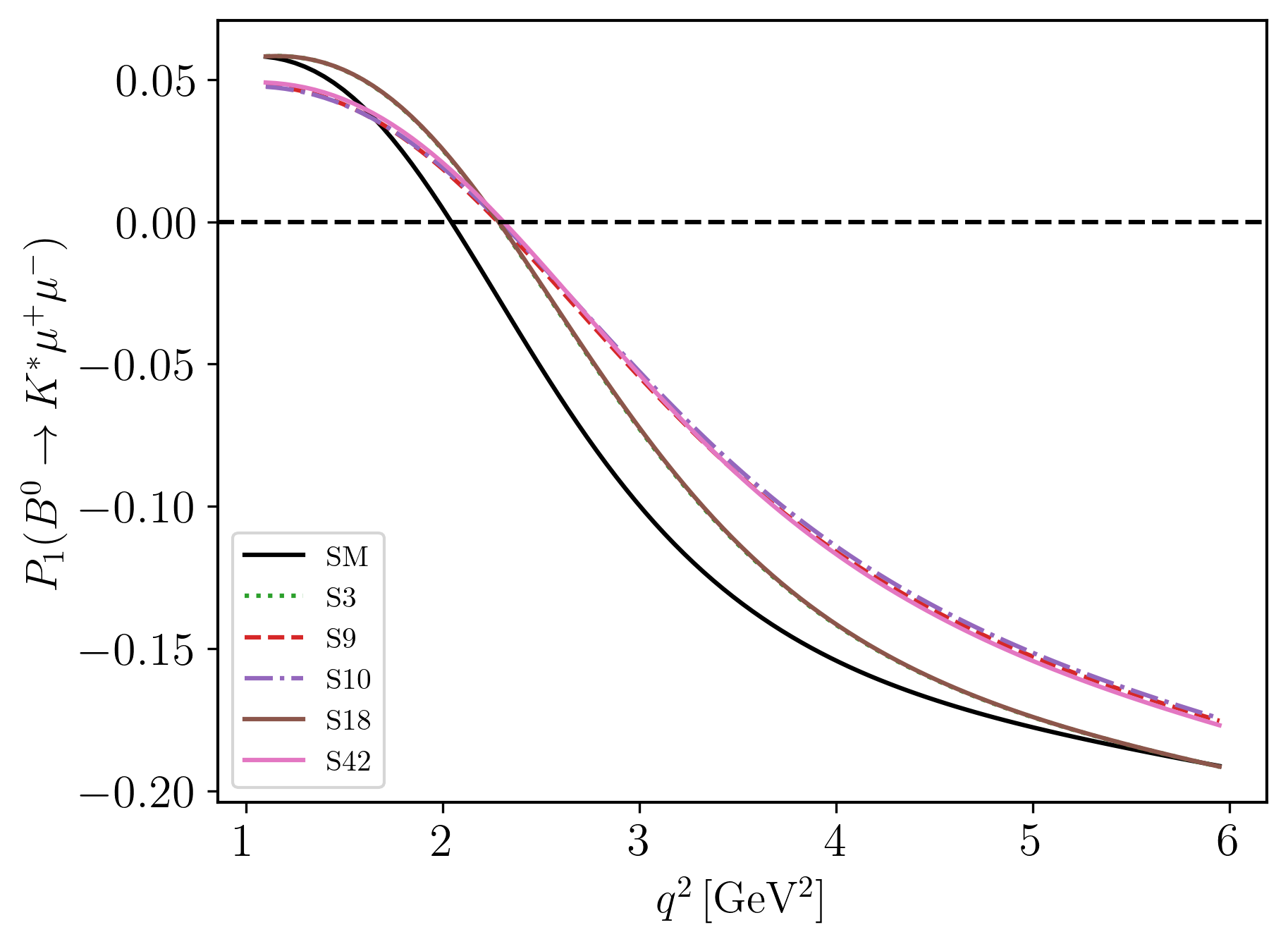}
\includegraphics[width=5.9cm,height=4.3cm]{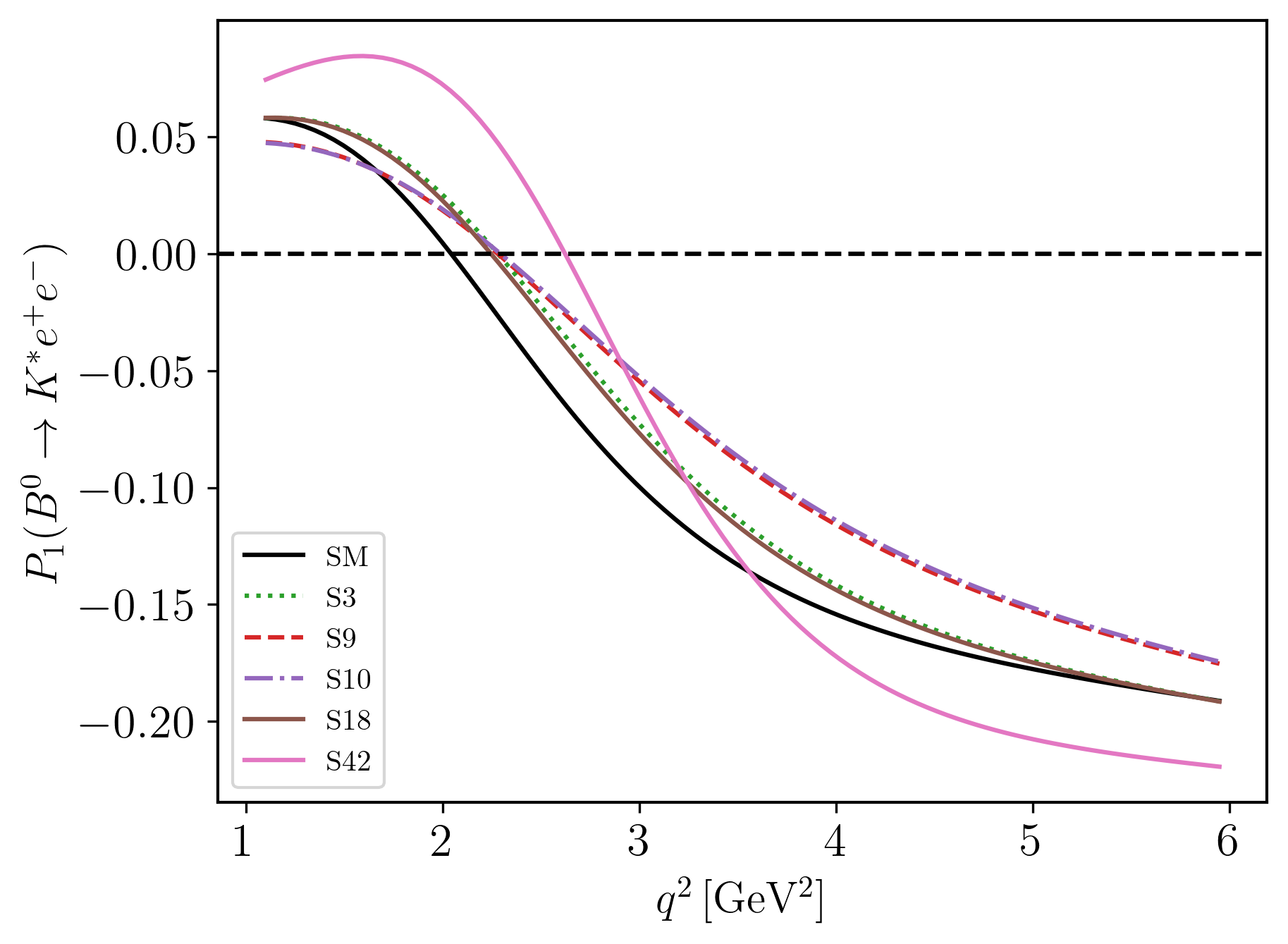}
\caption{Left panel: bin-wise distribution of $P_1$ for $B \to K^* \mu^+ \mu^-$ decay in SM and in the presence of NP S9. SM and S9 are compared with the LHCb data. Middle and right panels: $q^2$ dependency curves of the same observable in SM and in the presence of various NP scenarios for both muon and electron final states.}
 \label{fig_q2_p1}
\end{figure}

\item $P_2$: Although the NP contributions for the scenarios S9, S10, and S42 on $P_2$ in the muon final state look quite similar (Table~\ref{tab_ap7}), a deviation up to $2\sigma$ from SM is apparent in the bins [2.5, 4] ${\rm GeV}^2$, [4, 6] ${\rm GeV}^2$,  and [1.1, 6] ${\rm GeV}^2$. The NP contributions from S9 and S10 in the electron final state are similar to the muon mode. This can be easily verified from the $q^2$ distribution plot in Fig~\ref{fig_q2_p2}. On the other hand, the NP contributions from S3 and S18 have less significance on $P_2$ in both electron and muon final states. In addition, the contribution from S9 at different bins can graphically be interpreted from the bin-wise plot in Fig~\ref{fig_q2_p2}. It is apparent that our fit is consistent with the experimental data, especially in the bins [4, 6]$\rm GeV^2$ and [6, 8]$\rm GeV^2$. With the improved data and less statistical uncertainties, these two bins together with [2.5, 4] $\rm GeV^2$ are interesting as of NP searches are concerned. There are zero crossing points as of the $q^2$ distribution of $P_2$ are concerned. The zero crossing point for SM $q^2$ distribution is observed at $q^2 \sim 3.5$ $\rm GeV^2$ while the contribution of S9 and S10 contribution shifts the zero crossing to $q^2 \sim 4.5$ $\rm GeV^2$ for both muon and electron modes. It is true for S42 as well but only for the muon mode. In the electron mode, S42 crosses the zero line almost close to SM zero crossing.

\begin{figure}[h]
\centering
\includegraphics[width=5.9cm,height=4.3cm]{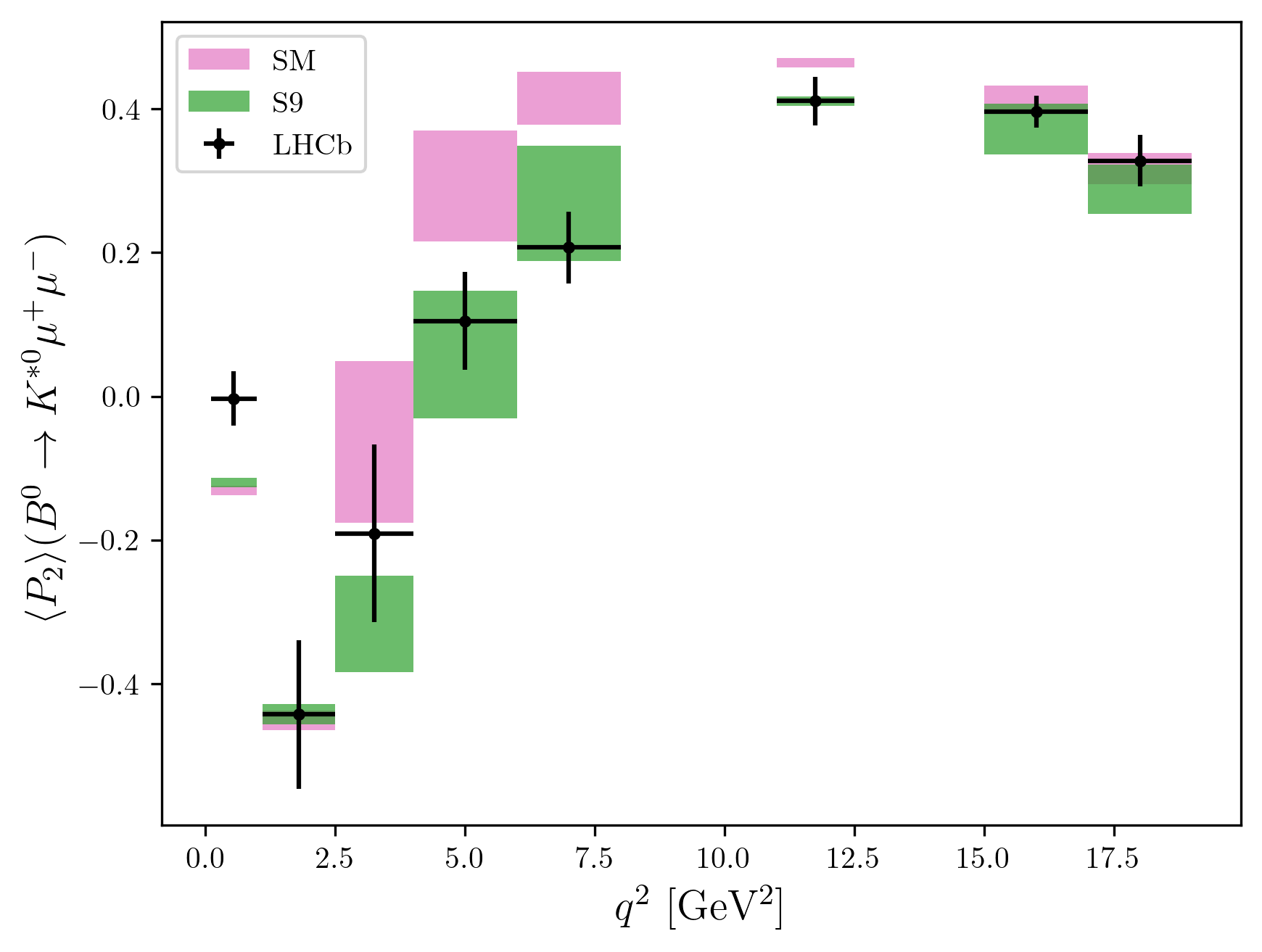}
\includegraphics[width=5.9cm,height=4.3cm]{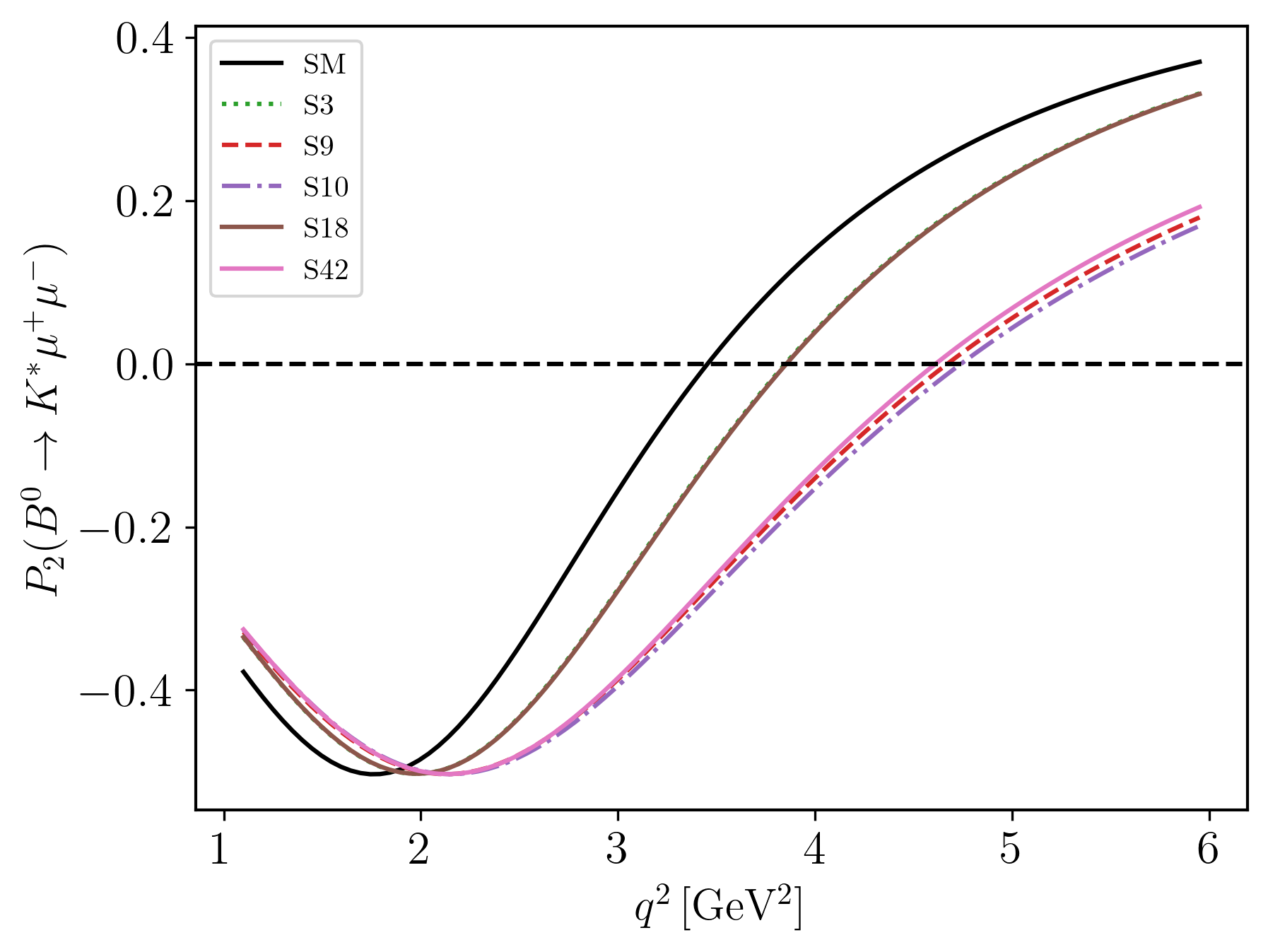}
\includegraphics[width=5.9cm,height=4.3cm]{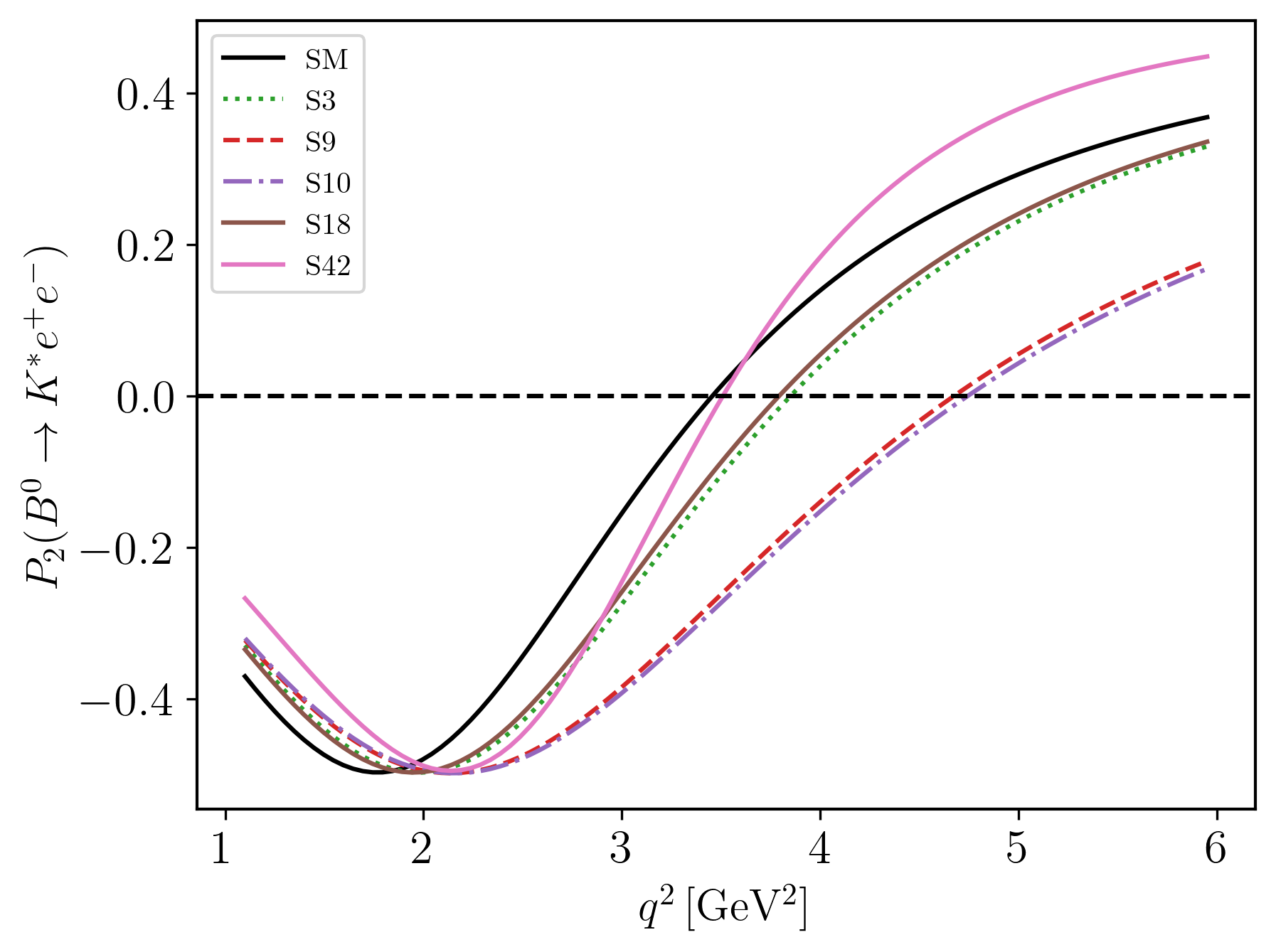}
\caption{Left panel: bin-wise distribution of $P_2$ for $B \to K^* \mu^+ \mu^-$ decay in SM and in the presence of NP S9. SM and S9 are compared with the LHCb data. Middle and right panels: $q^2$ dependency curves of the same observable in SM and in the presence of various NP scenarios for both muon and electron final states.}
\label{fig_q2_p2}
\end{figure}

\item $P'_4$: The NP contributions in $P'_4$ (Fig~\ref{fig_q2_p4p} and Table~\ref{tab_ap9}) are hard to distinguish from one another and also from SM in both the electron and muon final states. At most not more than $1\sigma$ deviation exists in each bin as evident from the bin-wise plot in Fig~\ref{fig_q2_p4p}. The experimental errors of $P'_4$ in the muon final states are very large, thus diminishing the prospect of identifying an NP signature from SM. Although the bin [4, 6] ${\rm GeV}^2$ looks interesting from the rest of the bins as SM can be distinguished beyond the experimental error, the global fit is hardly consistent with the measurement. Indeed, the NP (S9) result overlaps with that of SM. In addition, there is a zero-crossing as of $q^2$ distributions are concerned, however, none of the NP outputs are significant enough to be distinguished from the SM curve. In $B^0 \to K^{0*} e^+ e^-$ mode, the $q^2$ distribution of S42 isolates itself only in the first bin. The zero crossing for S42 occurs at $q^2 \sim 2.5$ $\rm GeV^2$ which is away from the same for SM appearing at $q^2 \sim 1.8$ $\rm GeV^2$.
  
\begin{figure}[htbp]
\centering
\includegraphics[width=5.9cm,height=4.3cm]{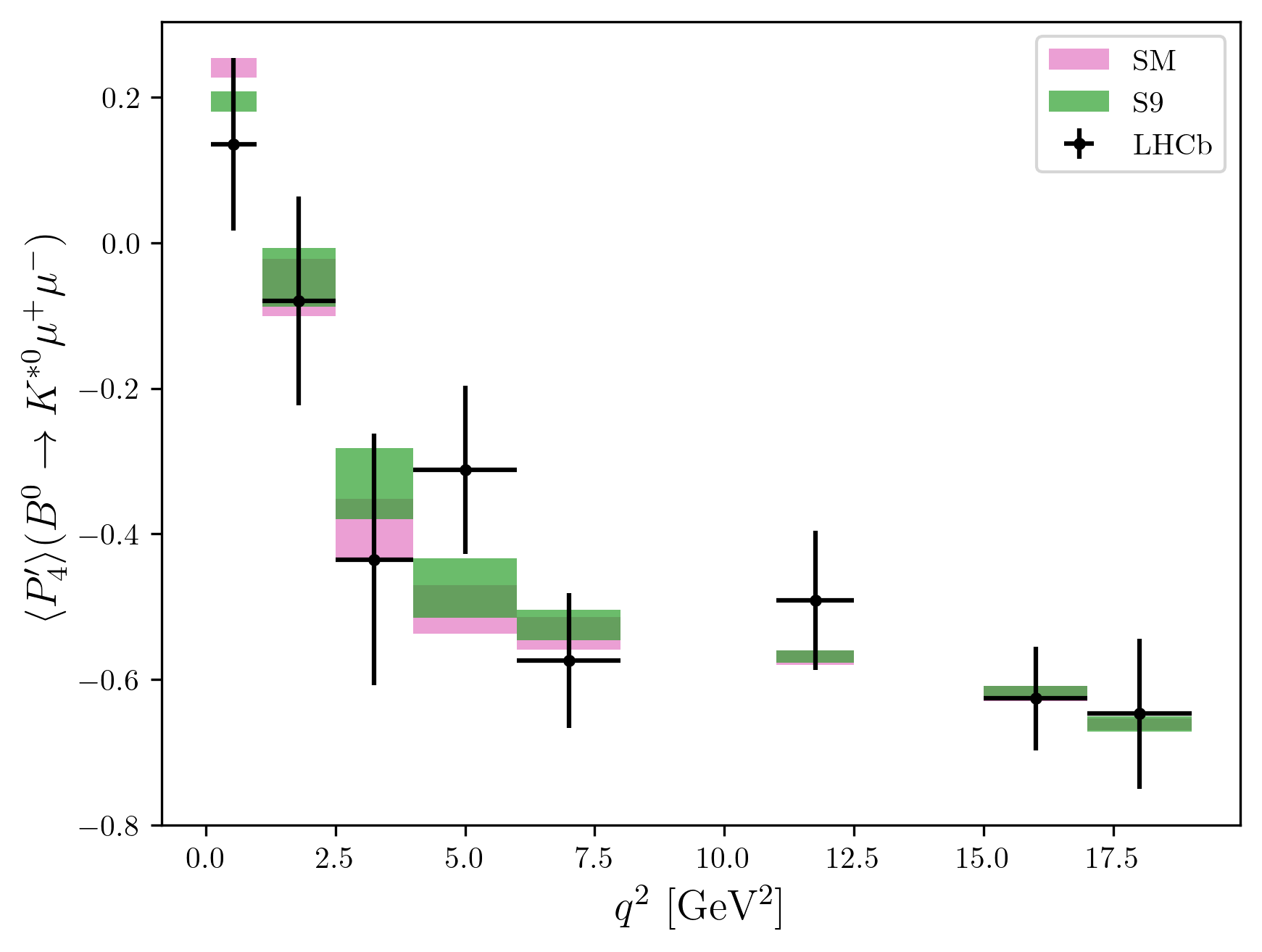}
\includegraphics[width=5.9cm,height=4.3cm]{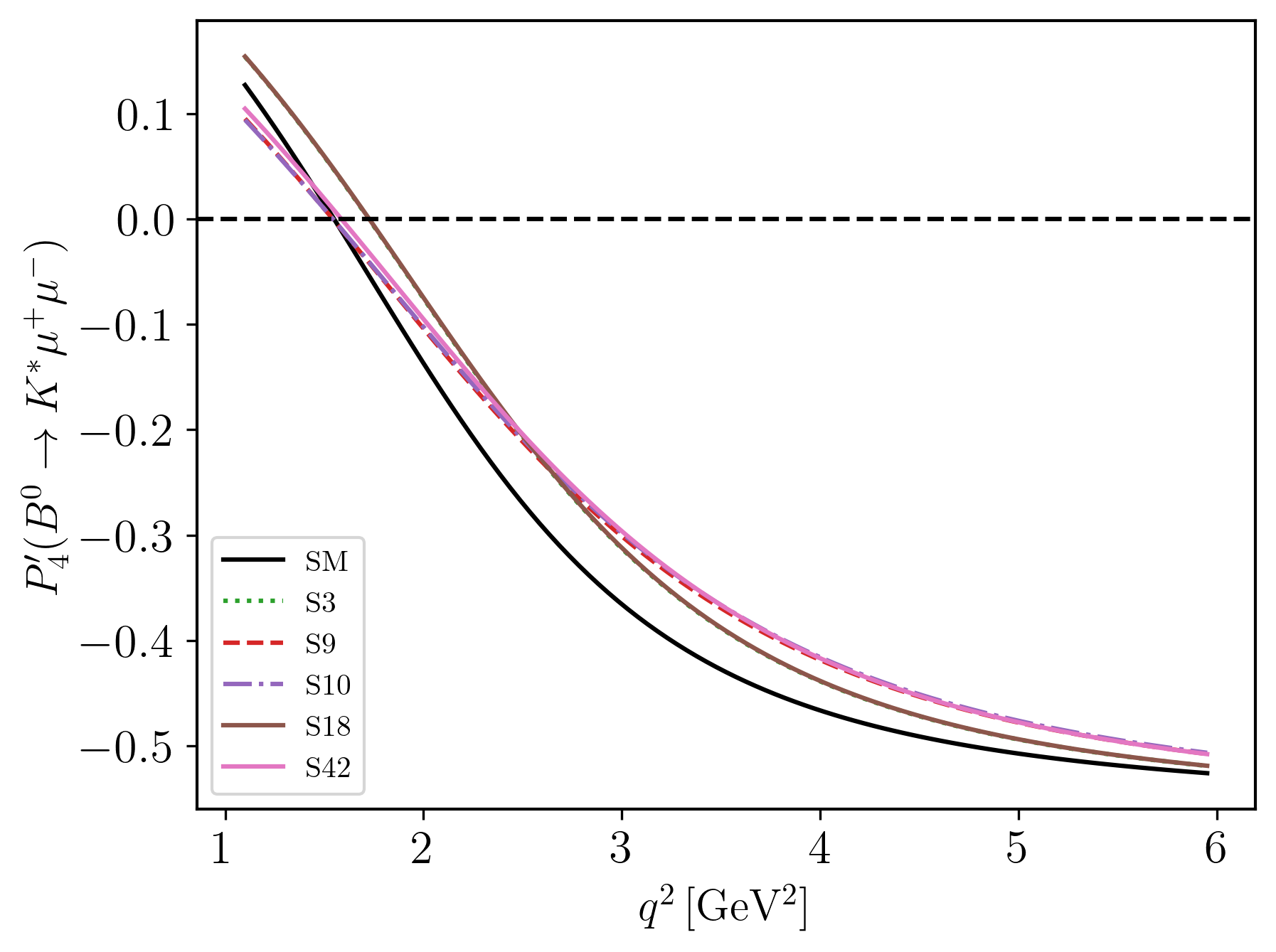}
\includegraphics[width=5.9cm,height=4.3cm]{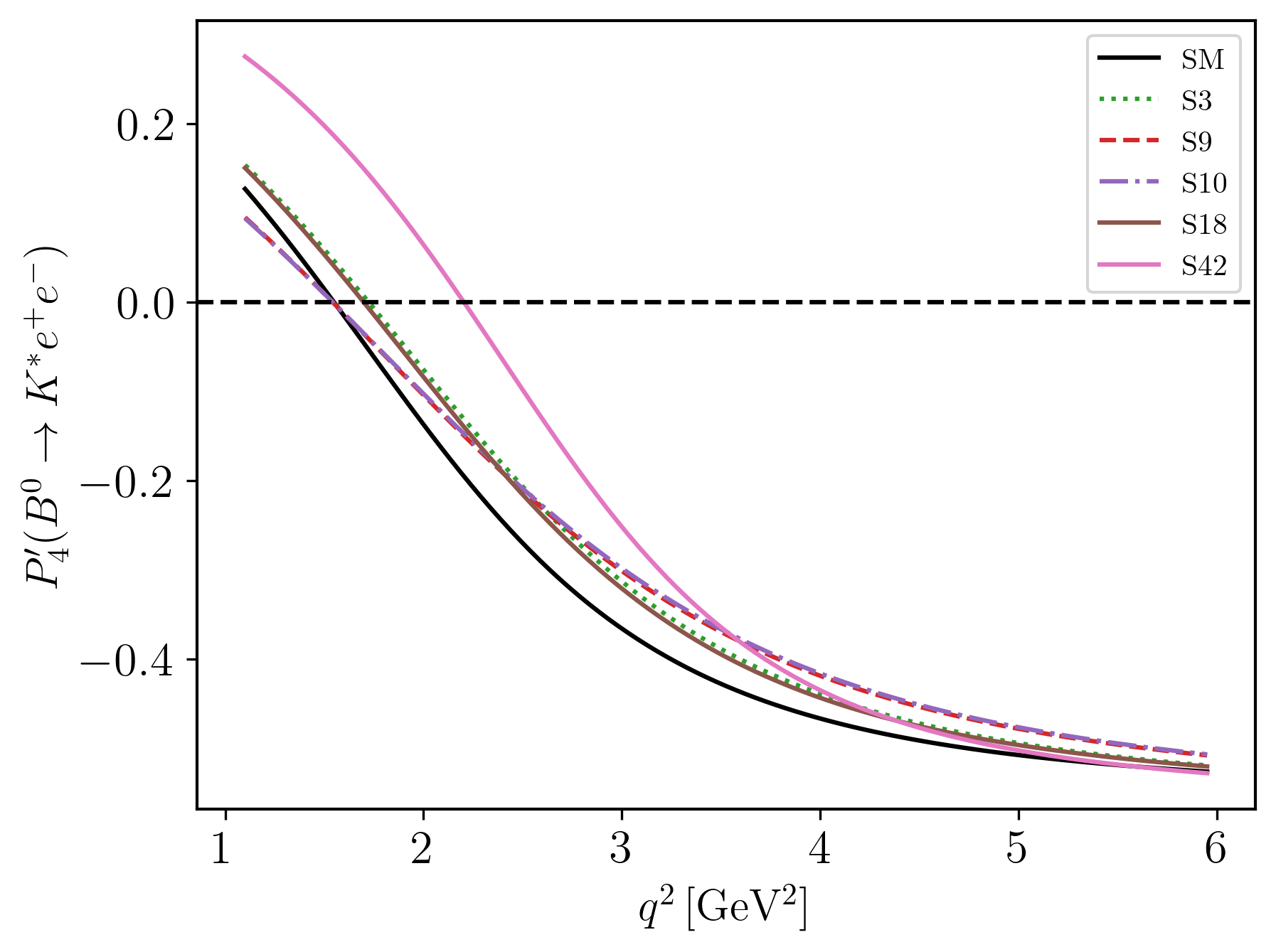}
\caption{Left panel: bin-wise distribution of $P'_4$ for $B \to K^* \mu^+ \mu^-$ decay in SM and in the presence of NP S9. SM and S9 are compared with the LHCb data. Middle and right panels: the $q^2$ dependency curve of the same observable in SM and in the presence of various NP scenarios for both muon and electron final states.}
\label{fig_q2_p4p}
\end{figure}

\item $P'_5$: The anomaly in $P'_5$ is well known in $B^0 \to K^{0*} \mu^+ \mu^-$ decays, where the LHCb measurement disagreed with the SM at $3\sigma$ in the bin $q^2 \in [4, 6]\, \rm GeV^2$ (Fig~\ref{fig_q2_p5p} and Table~\ref{tab_ap10}). In contrast to S3 and S18, the NP contributions from scenarios S9, S10, and S42 fit better with the LHCb data in this $q^2$ bin, showing a significance up to $2\sigma$ from the SM value. This can be verified from the bin-wise distribution plot in the left panel of Fig~\ref{fig_q2_p5p} reported for S9 when compared to the SM and LHCb data for $B^0 \to K^{0*} \mu^+ \mu^-$. In addition to the $q^2$ bin [4, 6] $\rm GeV^2$, the lower energy bins [1.1, 2.5] $\rm GeV^2$ and [2.5, 4] $\rm GeV^2$ also draw attention in regard to similar NP contributions from S9, S10, and S42 that are more than $2\sigma$ away from the SM result. Among the above NP scenarios, S9 and S10 are of LFU-NP type whereas S42 refers to an LFU+LFUV-NP case. Certainly, any measurement of the same observable regarding the electron mode may help us in distinguishing among the mentioned scenarios classified into LFU-NP or LFUV-NP types. Additionally, the $q^2$ distributions of  $P'_5$ for the muon and the electron modes may have interesting zero-crossing points when the same for SM refers to $q^2 \sim 2.1$ $\rm GeV^2$. For the muon mode one has $q^2 \sim 3$ $\rm GeV^2$ as the point for the NP scenarios S9, S10 and S42. The same argument applies to the electron final state except for S42.
  
\begin{figure}[h]
\centering
\includegraphics[width=5.9cm,height=4.3cm]{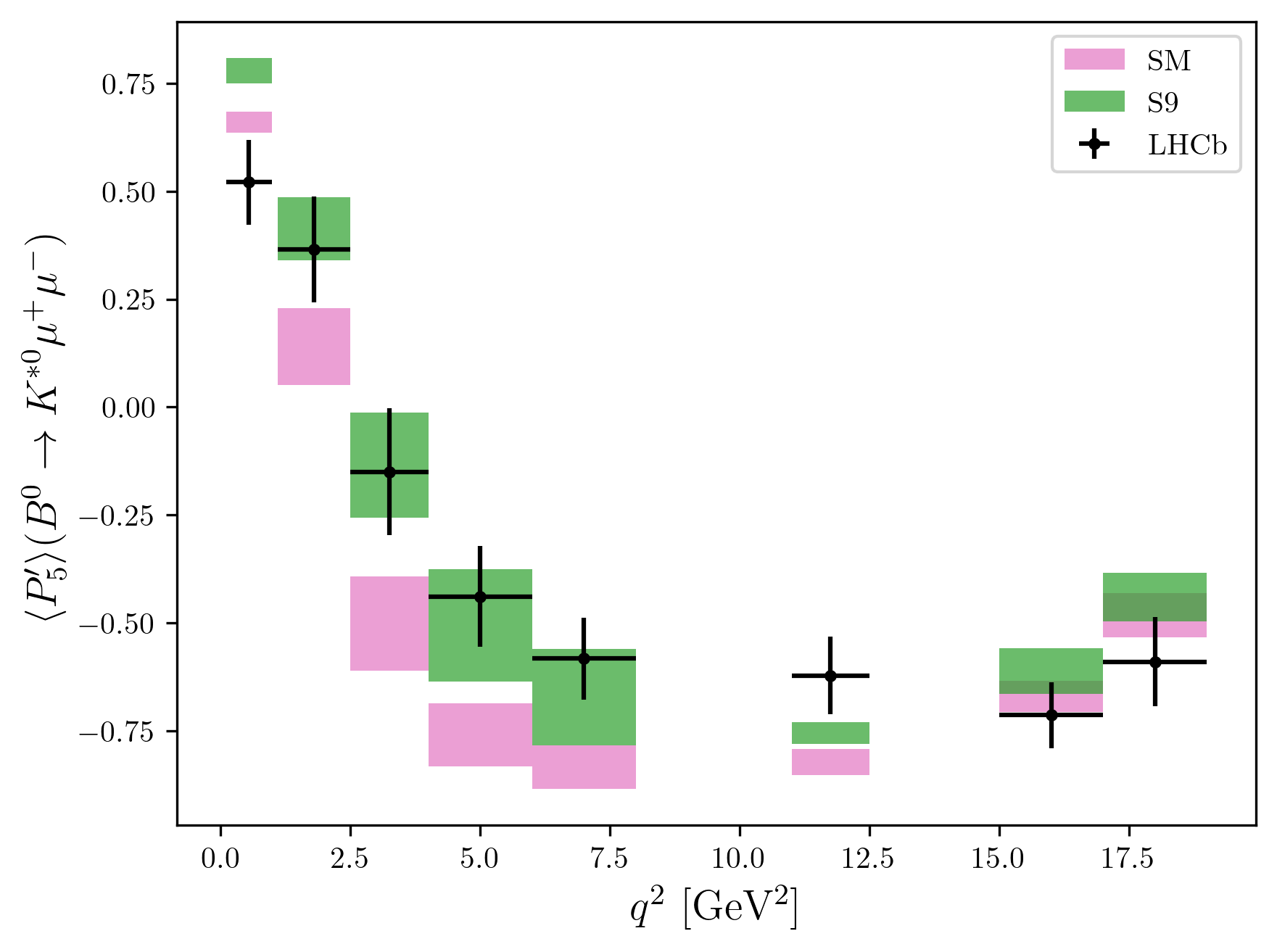}
\includegraphics[width=5.9cm,height=4.3cm]{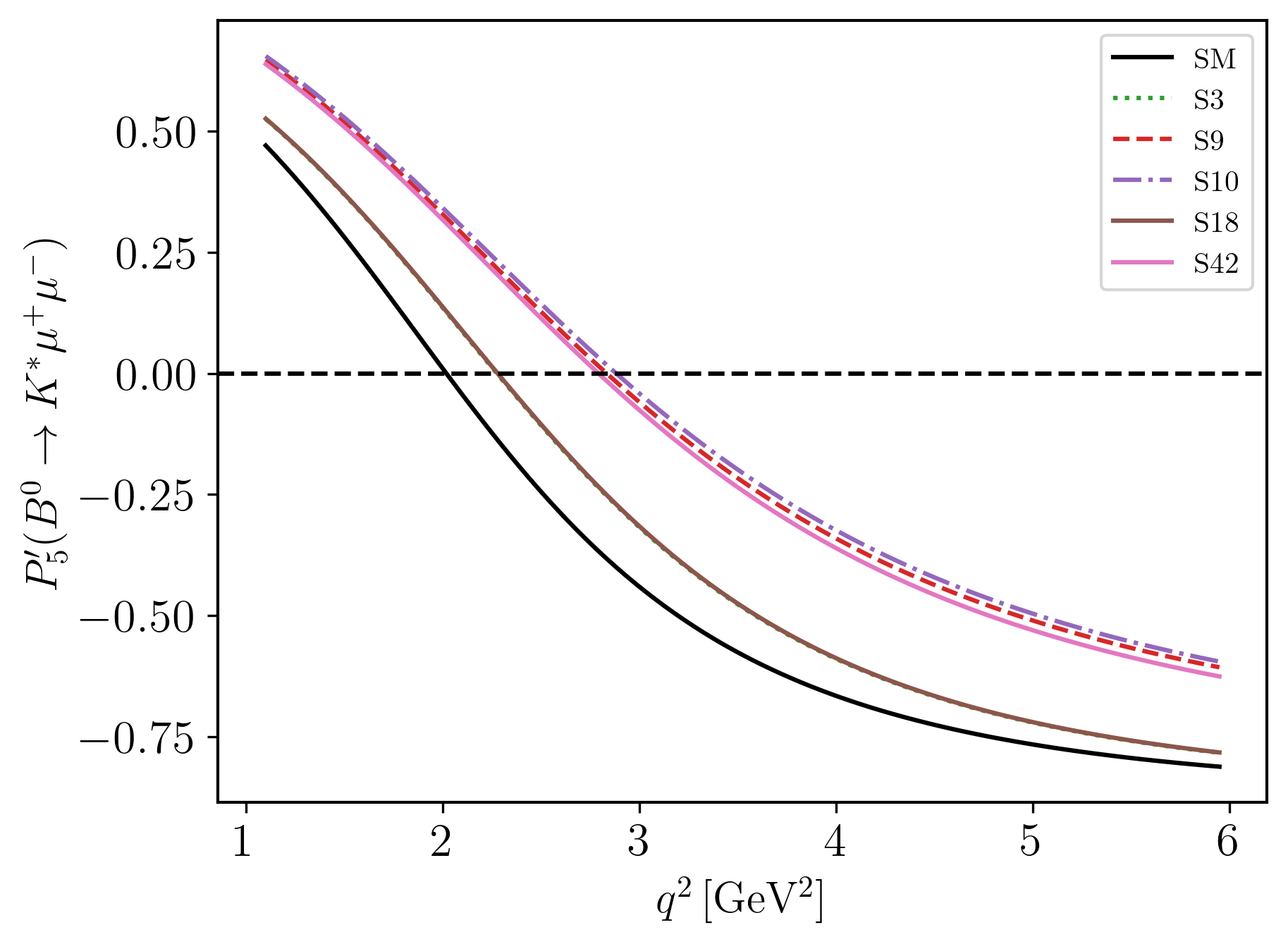}
\includegraphics[width=5.9cm,height=4.3cm]{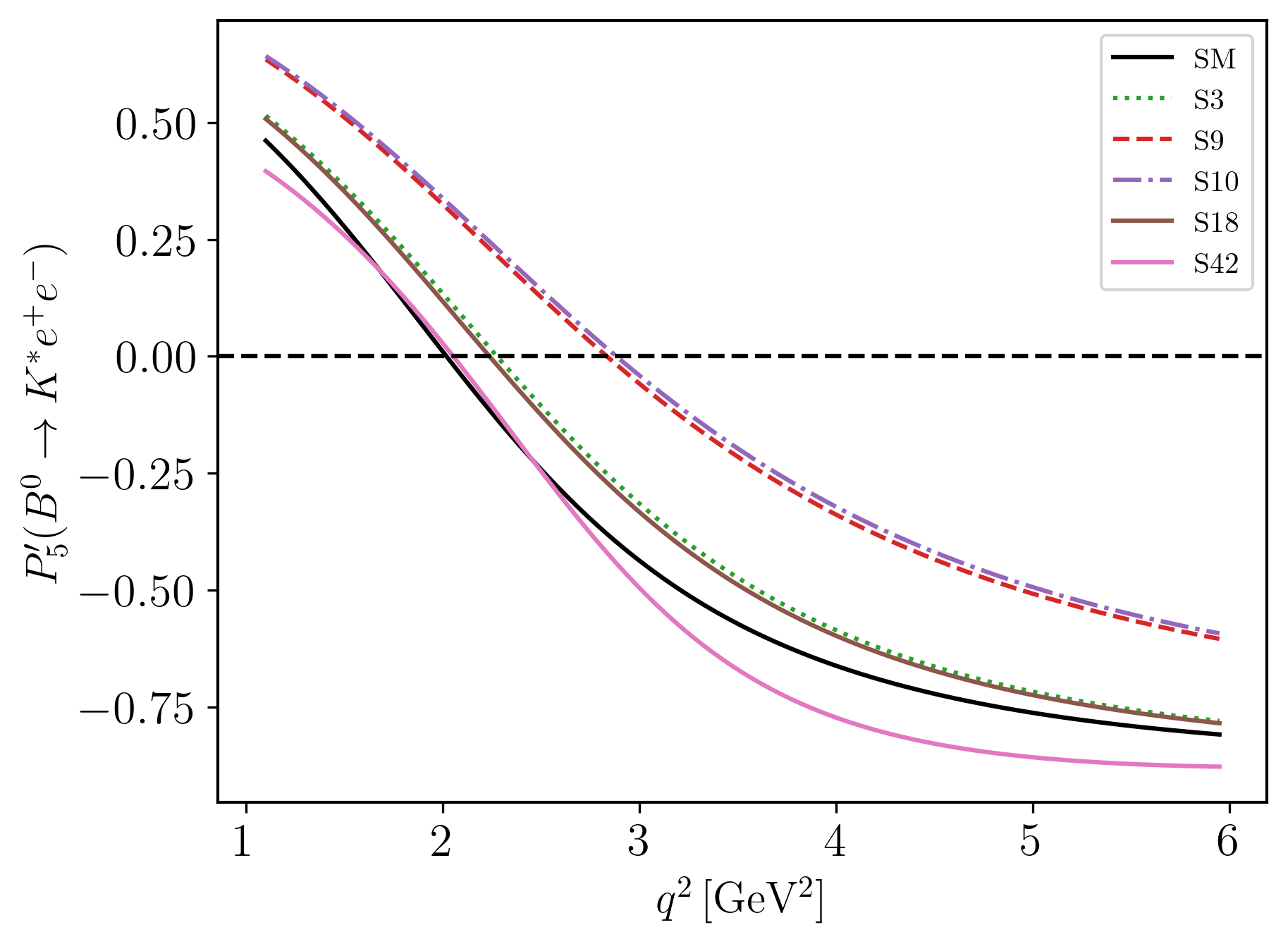}
\caption{Left panel: bin-wise distribution of $P'_5$ for $B \to K^* \mu^+ \mu^-$ decay in SM and the presence of NP S9. SM and S9 are compared with the LHCb data. Middle and right panels: $q^2$ dependency curves of the same observable in SM and the presence of various NP scenarios for both muon and electron final states.}
\label{fig_q2_p5p}
\end{figure}

\item $P_3$, $P'_6$ and $P'_8$: NP contributions in $P_3$, $P'_6$ and $P'_8$ are not significant enough to distinguish themselves from SM in the presence of WCs given as real number inputs. The $q^2$ distribution plots for these observables are reported in Fig~\ref{fig_q2_p3p6p8} and Tables~\ref{tab_ap8}, \ref{tab_ap11}, and \ref{tab_p8p}.

\begin{figure}[h]
\centering
\includegraphics[width=5.9cm,height=4.3cm]{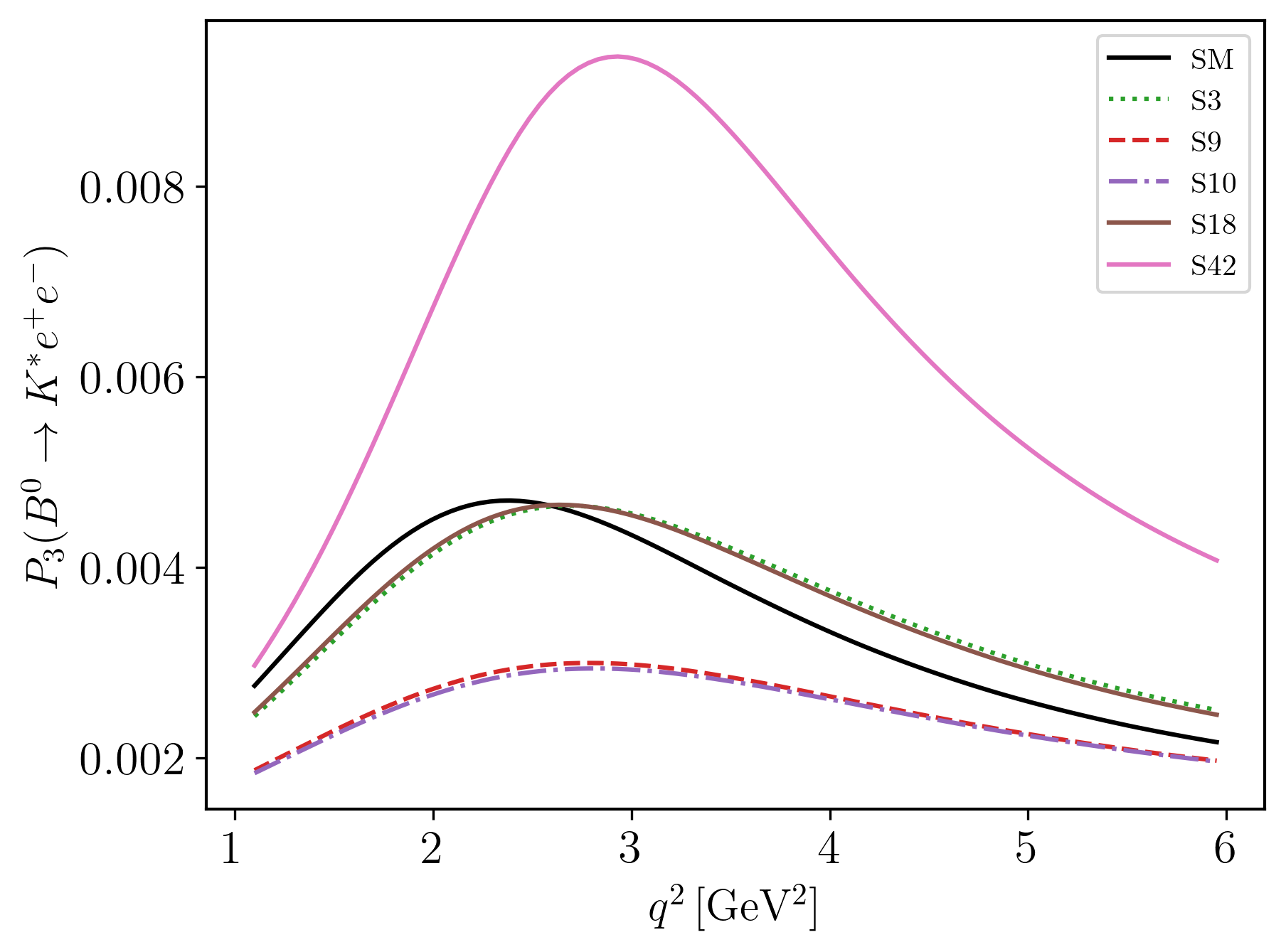}
\includegraphics[width=5.9cm,height=4.3cm]{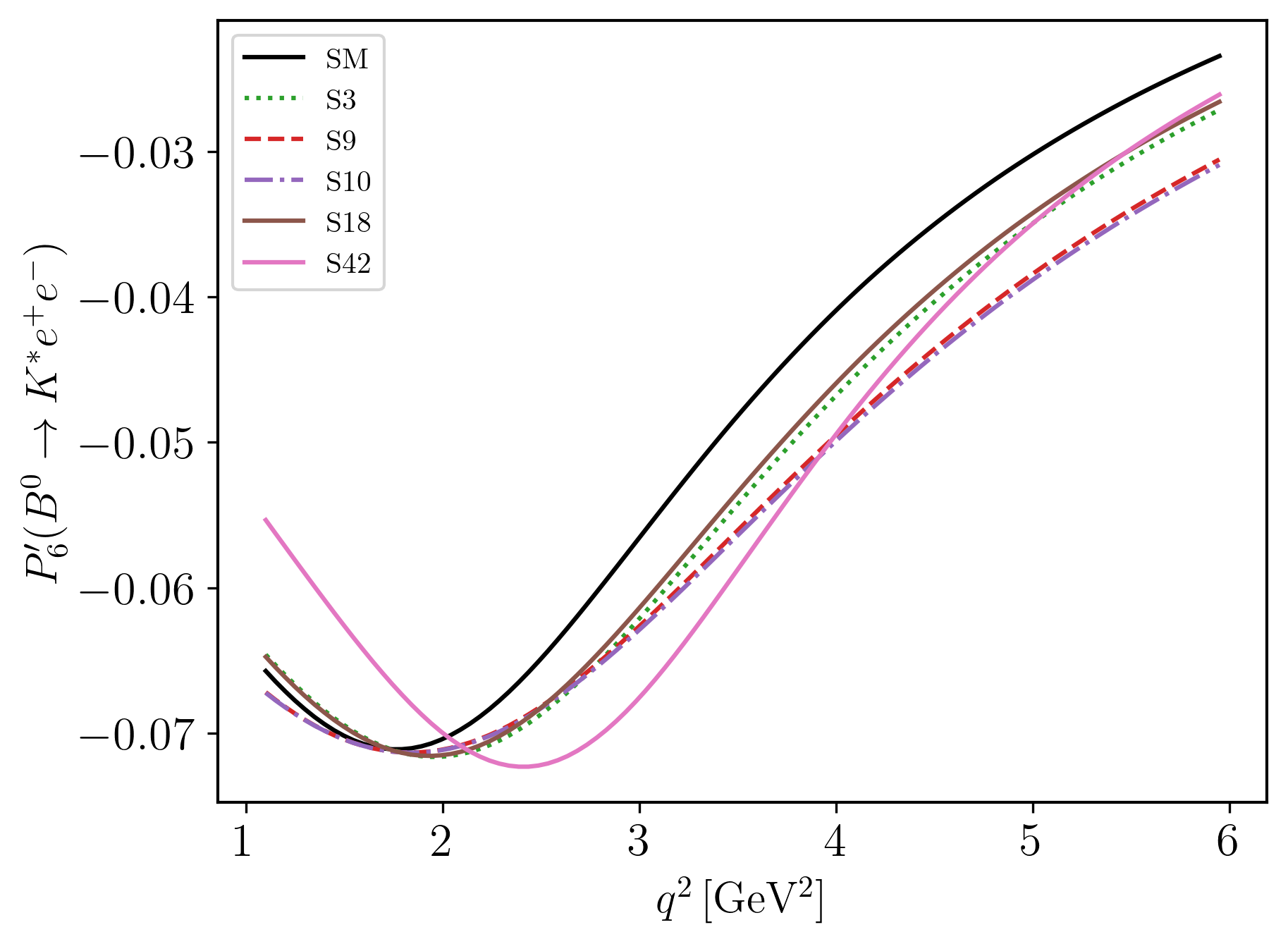}
\includegraphics[width=5.9cm,height=4.3cm]{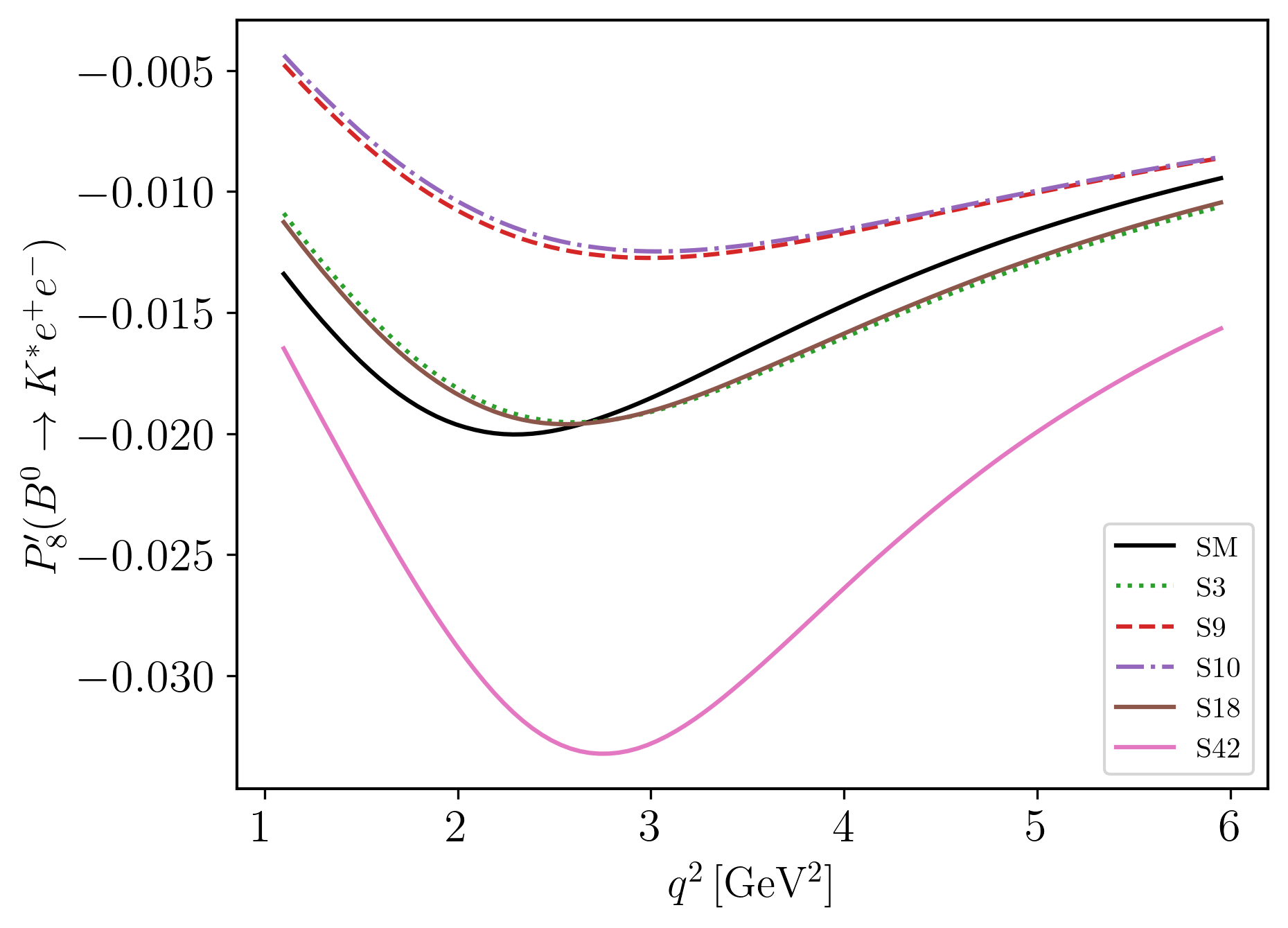}
\includegraphics[width=5.9cm,height=4.3cm]{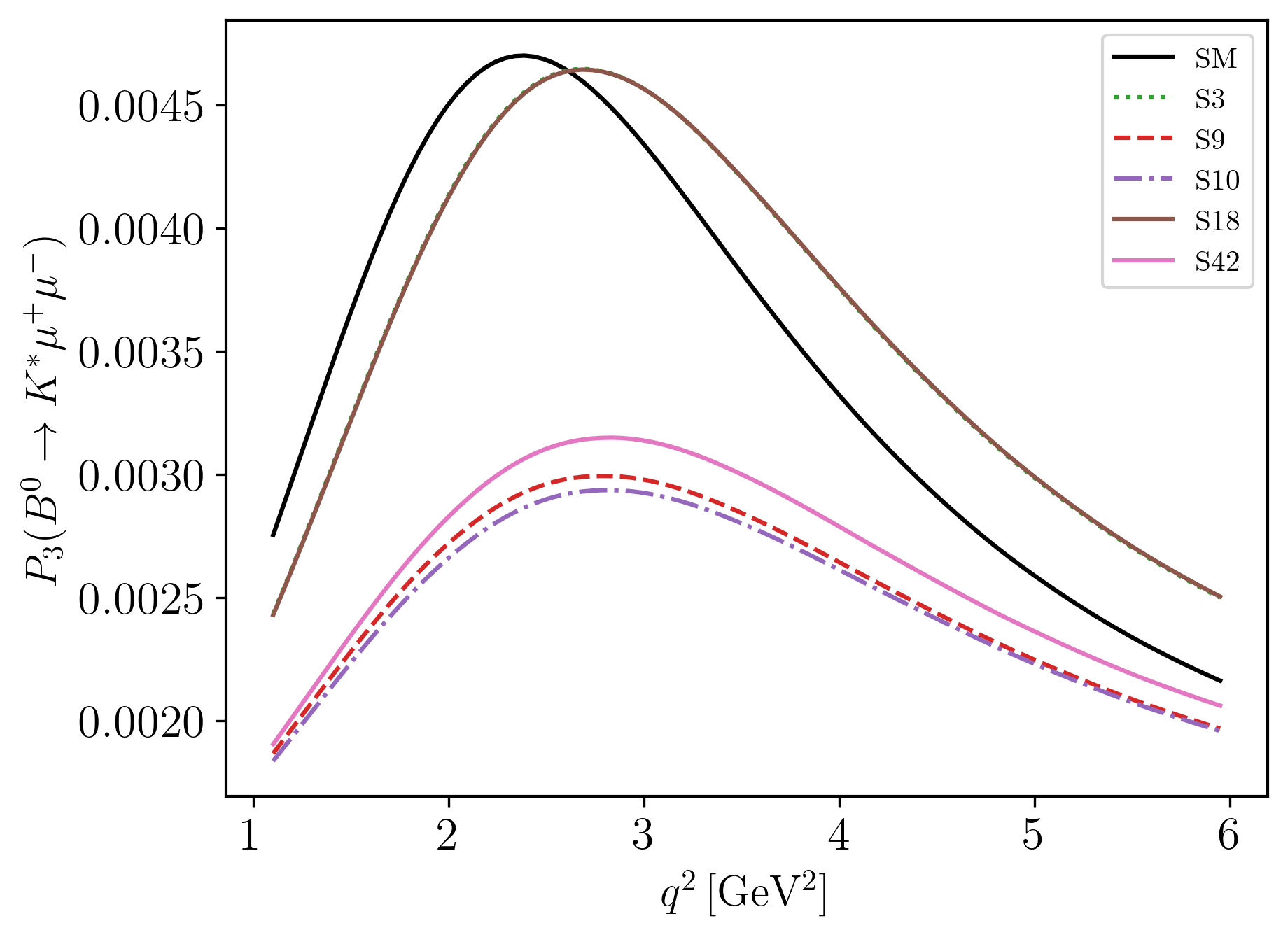}
\includegraphics[width=5.9cm,height=4.3cm]{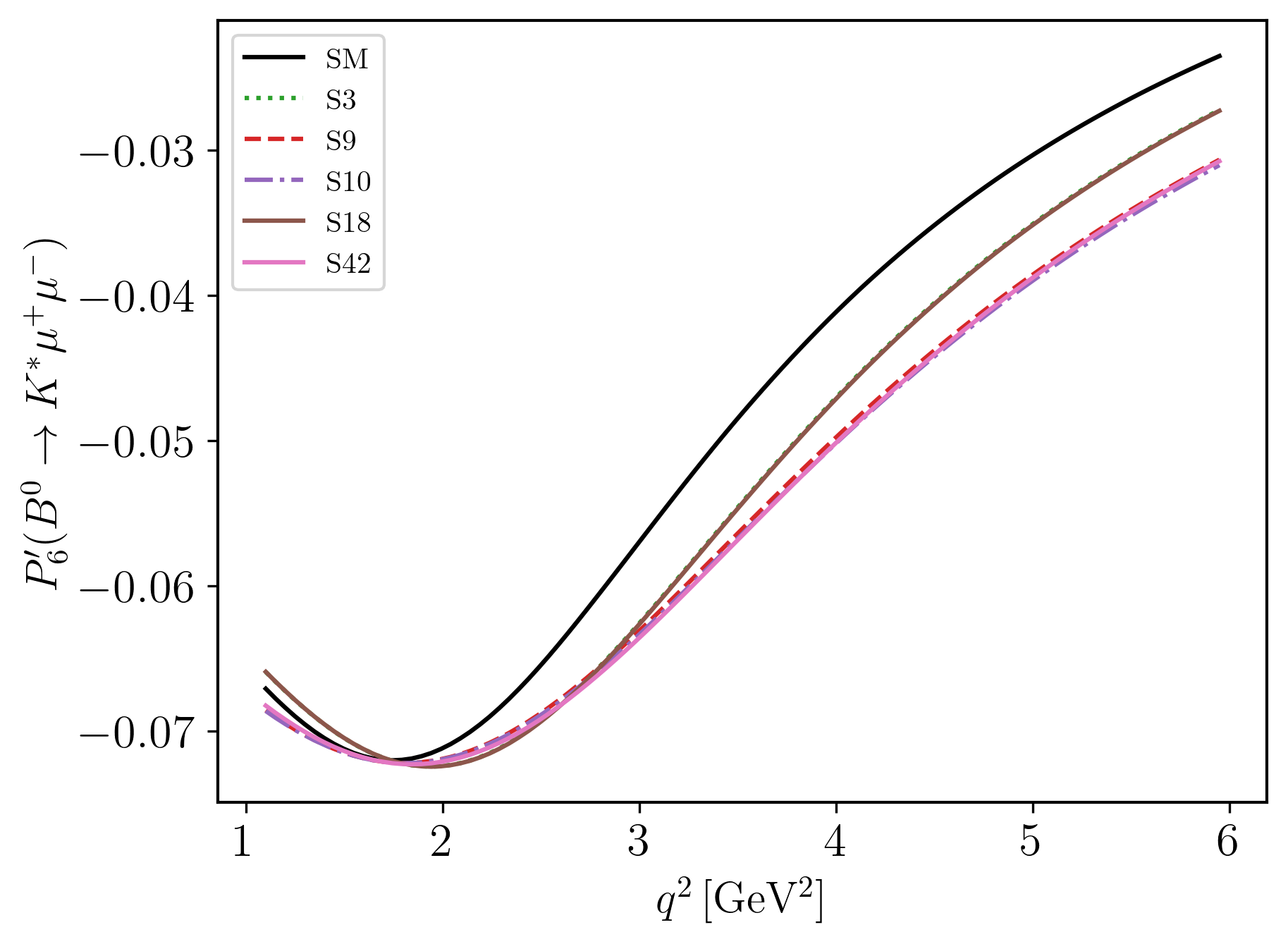}
\includegraphics[width=5.9cm,height=4.3cm]{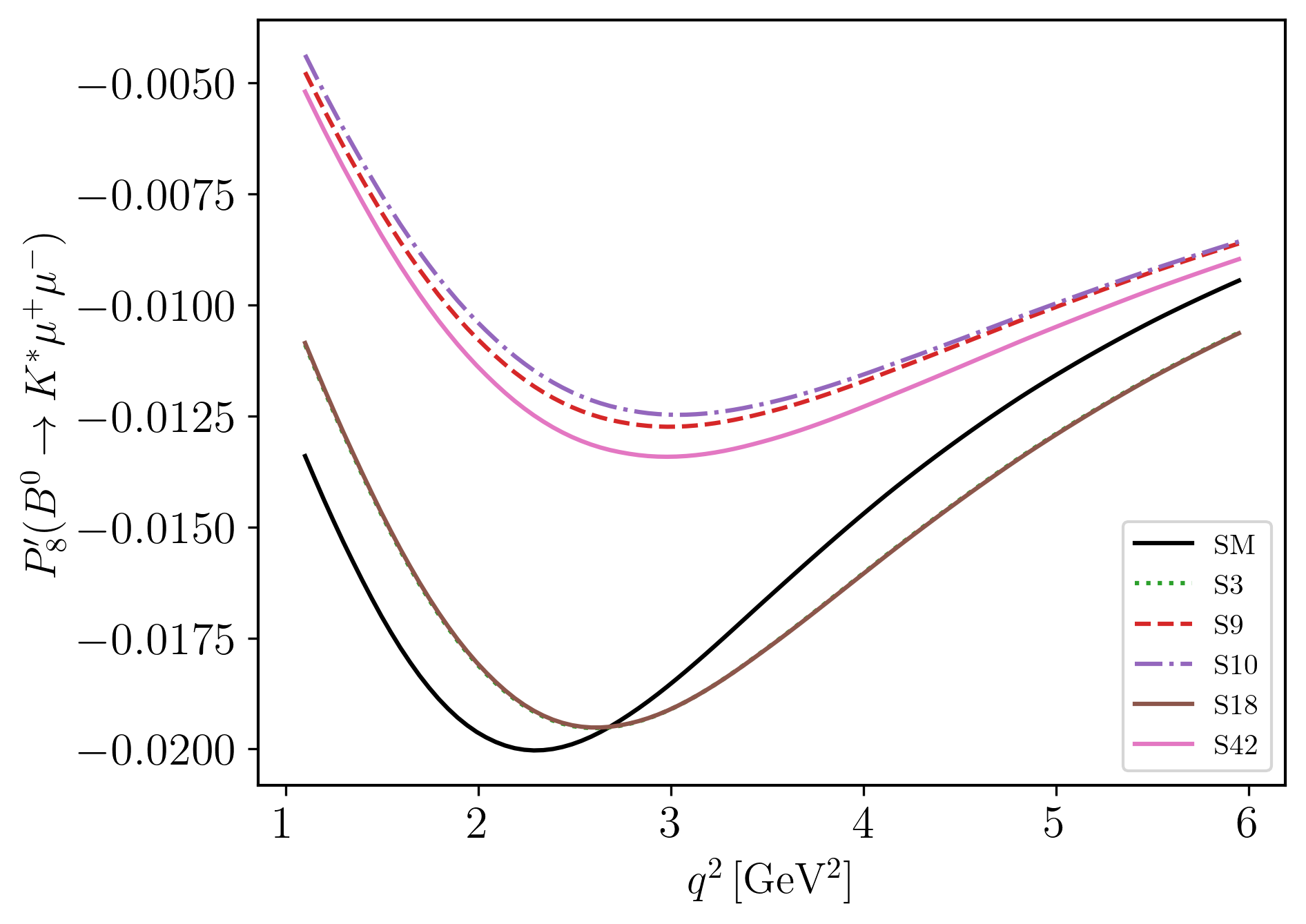}
\caption{$q^2$ dependency curves of $P_3$, $P'_6$ and $P'_8$ observables in SM and in the presence of various NP scenarios for both muon and electron final states.}
 \label{fig_q2_p3p6p8}
 \end{figure}
    
\end{itemize}

\section{Sensitivity of LFUV in $\Delta$-observables}
\label{results2}
One of the primary motivations to study $b \to s \ell^+ \ell^-$ transitions is to probe LFU violation. As already discussed none of the measurements of the LFU-sensitive observables such as $R_{K^{(*)}}$ can establish the presence of any NP. Moreover, the updated values of $R_{K^{(*)}}$ tend to align with SM expectations as of the recent LHCb results are concerned. The above in turn indicates the need to probe further observables that are sensitive to the LFUV. 
In this connection, being inspired by observables defined by the ratio of branching fractions such as that used for $R_{K^{(*)}}$, a few earlier analyses~\cite{Alguero:2019pjc,Alguero:2018nvb} probed some observables involving the differences between two quantities of different lepton flavors. These observables often denoted as $"Q"$ or "$\Delta$" observables may potentially have low degrees of sensitivity toward hadronic uncertainties. In addition, they are protected from long-distance charm-loop contribution in SM. Any deviation in $\Delta$ from zero would point out NP very distinctly, confirming the LFU violation. The $\Delta$-observables are defined involving several angular observables such as AFB, FL, and $P_i$'s such as,
\begin{equation}
     \Delta A_{FB} = \langle A_{FB}^{\mu} \rangle - \langle A_{FB}^{e} \rangle,\hspace{0.2cm} \Delta F_{L} = \langle F_{L}^{\mu} \rangle - \langle F_{L}^{e} \rangle, \hspace{0.2cm} \Delta P_{i} = \langle P_{i}^{\mu} \rangle - \langle P_{i}^{e} \rangle.
\end{equation}

Among the best NP scenarios mentioned in Table~\ref{tab_fit3}, S3, S9, and S10 contribute as LFU-NP scenarios, whereas S18 and S42 do the same as LFUV-NP and LFU+LFUV-NP respectively. Non-vanishing values for $\Delta$-observables are possible only from the LFUV-NP cases mentioned above. With S18 showing hardly any deviation from SM, one finds only S42 to be important. Hence, we report in Table~\ref{tab_delta1} and \ref{tab_delta2} the binned values for each $\Delta$-observables in SM and S42. The contribution from S42 on $\Delta A_{FB}$ deviates from SM up to $1\sigma$-$1.3\sigma$ in all the three bins. Similarly, there is  $1.3\sigma$ level of deviation in $\Delta F_{L}$ in the bin [2.5, 4] ${\rm GeV}^2$ and $<1\sigma$ in the other two bins. While $\Delta P_{1}$ cannot be distinguished beyond $1\sigma$ from SM, $\Delta P_{2}$ in contrast, shows a very significant deviation, particularly in the bins [2.5, 4] ${\rm GeV}^2$ and [4, 6] ${\rm GeV}^2$. On the other hand, $\Delta P'_{4}$ is sensitive to NP in the bin [1.1, 2.5] ${\rm GeV}^2$. Besides, interestingly $\Delta P'_{5}$ shows significant deviations from SM in all the three bins including [1.1, 6] ${\rm GeV}^2$. In particular, one can expect $>2\sigma$ deviation in [2.5, 4] ${\rm GeV}^2$ and [4, 6] ${\rm GeV}^2$ bins. The uncertainty associated with $\Delta P'_{5}$ is of the order of 10\% which is one order more than the same associated with the rest of the observables. Figure~\ref{fig_delta} display the binned plots for $\Delta P_{2}$ and $\Delta P'_{5}$ in SM and the contribution from the NP scenario S42.

\begin{figure}[htbp]
\centering
\includegraphics[width=0.40\textwidth]{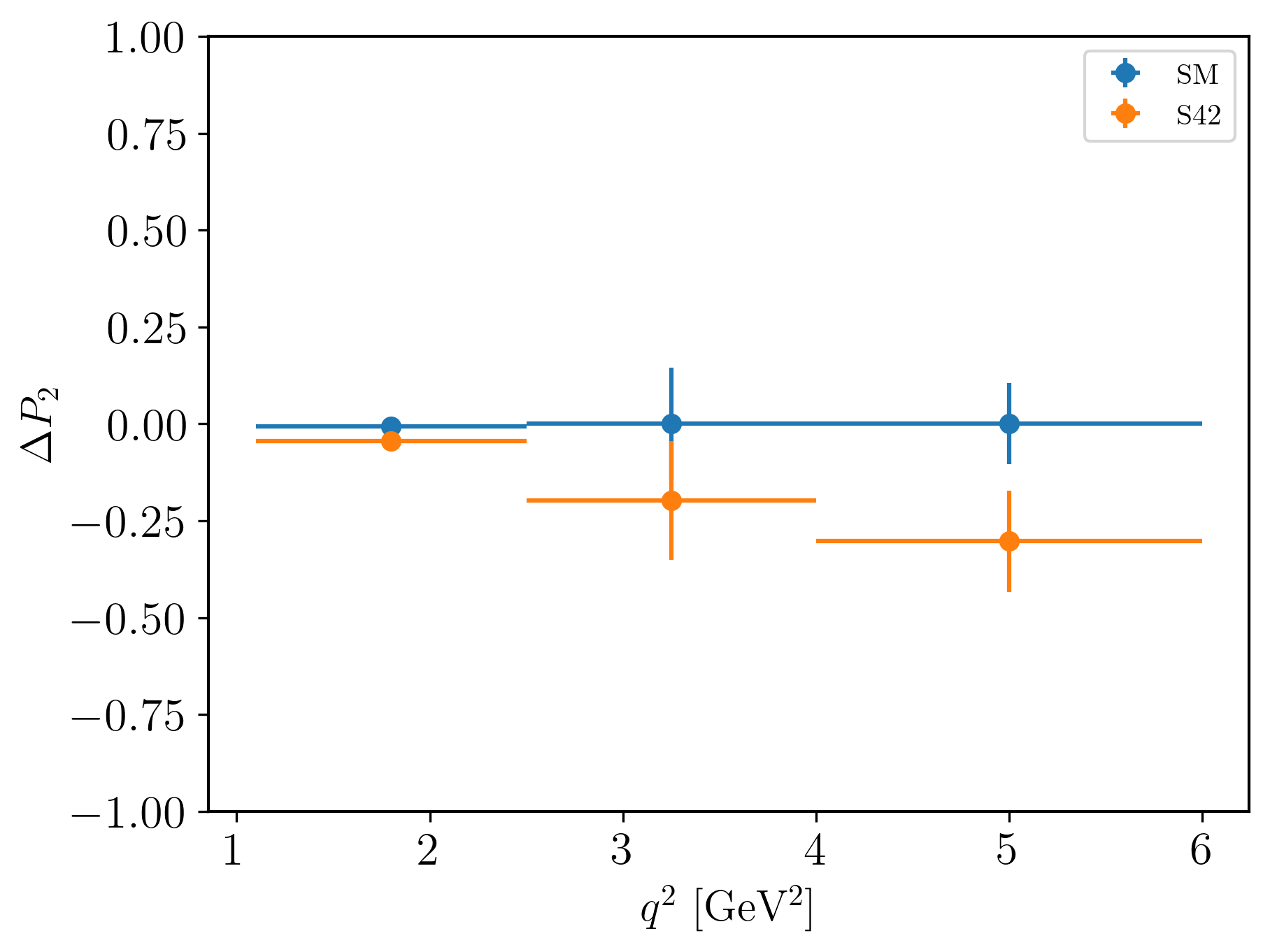}
\includegraphics[width=0.40\textwidth]{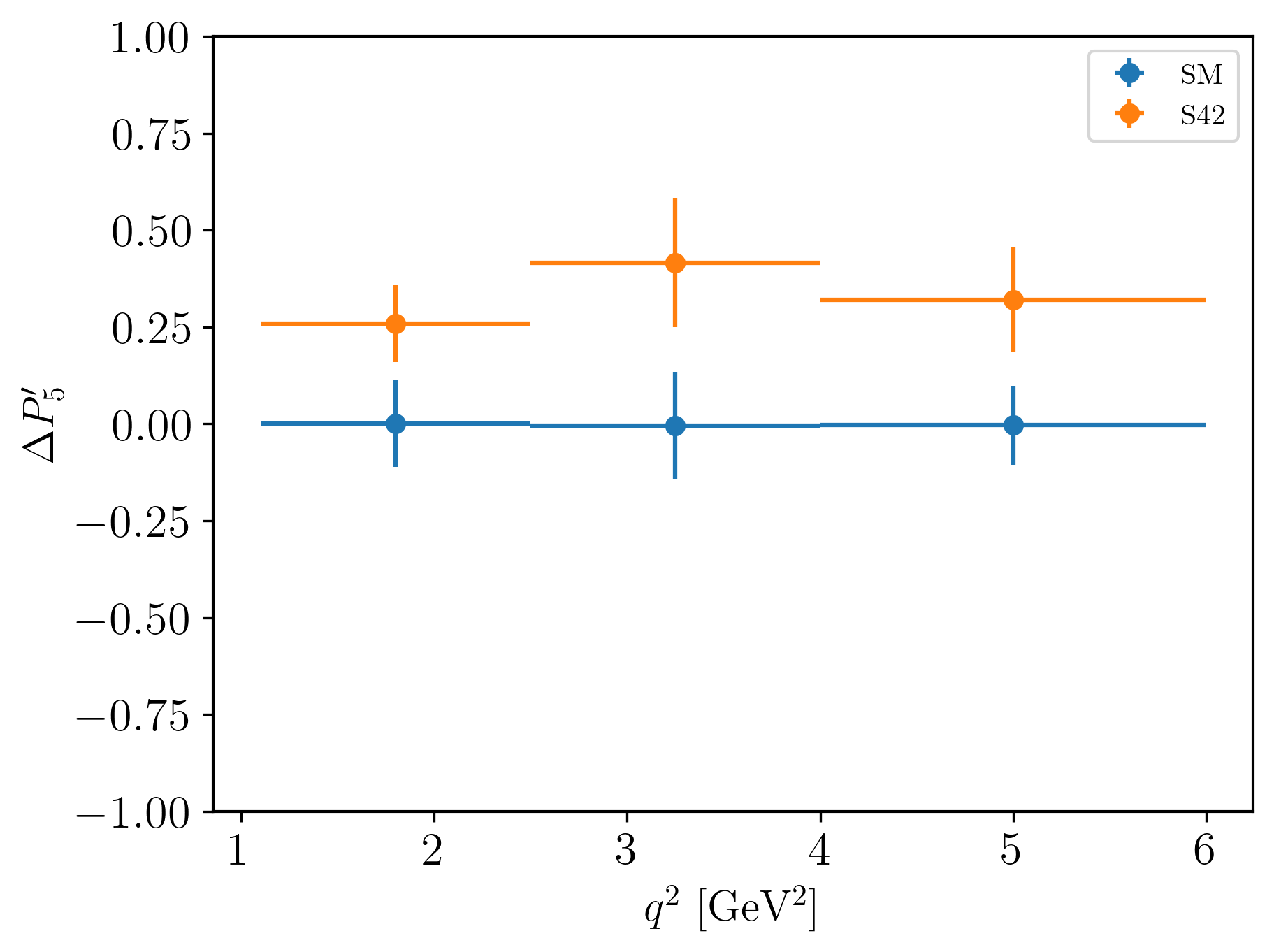}
\caption{Binned $\Delta P_{2}$ and $\Delta P'_{5}$ in SM and in the presence of the NP scenario S42.}
\label{fig_delta}
\end{figure}

\section{Summary and Conclusion}
\label{summary}
Motivated by the recent $R_{K^{(*)}}$ measurements, which have moved closer to the SM expectation, we revisit the global fit to the subset of dimension-6 SMEFT operators contributing at tree level to $b \to s\, \ell^+ \ell^-$ transitions. Although the LFU ratios appear SM-like, persistent tensions in branching fractions and angular observables remain. Under moderate hadronic assumptions, such tensions can still be interpreted as possible hints of NP, which motivates a systematic analysis in the SMEFT framework. To this end, we adopt a balanced treatment of hadronic uncertainties, ensuring that our results are not overly biased towards either a purely hadronic or a strongly NP-driven explanation.

The subset of dimension-6 SMEFT operators under consideration specifically generate vector and axial-vector semileptonic operators after matching onto LEFT. We explore their impact on various decay observables, including the LFU-sensitive observables in $B \to K^{(*)} \ell^+ \ell^-$ decays. Several NP scenarios are constructed by switching on one, two, three, or four SMEFT coefficients at a time with distinct lepton indices for the electron and muon. The relevant SMEFT operators can be broadly classified into two categories based on their structures: (i) the LFUV-NP scenarios arising from $2q2\ell$ operators involving quark and lepton fields, and (ii) the LFU-NP branch originating from Higgs-quark operators. Both categories contribute to the low-energy Wilson coefficients $C_{9,10}^{(\prime)}$ upon tree-level matching at the $M_W$ scale. Our analysis makes use of the packages \texttt{flavio} and \texttt{wilson}, which consistently implement the RGE running and operator mixing effects.

Since we assume that NP couples to both electron and muon simultaneously, the various NP scenarios are constructed out of SMEFT coefficients that involve both the leptons. When the global fit is performed with one operator at a time, we assume the LFU condition to hold good among the $2q2\ell$ operators. A large $\rm pull$ value among the 1D scenarios is found for $[C_{\ell q}^{(1),(3)}]_{ii23}$. This coefficient is evolved and matched appropriately to the left-handed coupling $C_{9,10}$. The SMEFT coefficients involving right chiral currents are suppressed by the global data. Once we consider two SMEFT coefficients at a time, $\rm pull$ is increased marginally. Our 2D scenarios include NP contributions of all the types, namely, those involving only LFU-NP, only LFUV-NP} as well as LFU+LFUV NP. Large values of $\rm pull$ are found in S9: $[C_{qe}]_{2311}=[C_{qe}]_{2322}$, $[C_{\phi q}]_{23}$; S10: $[C_{\ell q}]_{1123}=[C_{\ell q}]_{2223}$, $[C_{\phi q}]_{23}$; S18: $[C_{\ell q}]_{1123}$, $[C_{\ell q}]_{2223}$ and S42: $[C_{qe}]_{2322}$, $[C_{\phi q}]_{23}$ NP scenarios. In particular, the scenarios S9, S10, S42 and thereafter S18 have 
large values for $\rm pull$.
Interestingly, S9 and S10 contribute as LFU-NP, S42 as LFU+LFUV NP and S18 as LFUV-NP. Among the different types of NP contributions which are not distinguished very prominently via global data, the later mildly favor the LFU-NP and LFU+LFUV NP cases in contrast to LFUV-NP.
When considering the three and four nonzero SMEFT coefficients at a time, although not exhaustive, some scenarios have $\rm pull_{SM}$ relatively higher than 1D and 2D scenarios. However, the most dominant coefficients are observed to be $[C_{qe}]_{23ii}$, $[C_{\ell q}]_{ii23}$ and $[C_{\phi q}]_{23}$, all these fall among the best choices in the 2D combinations as well. 

When analysing the contour plots generated in the plane of two SMEFT coefficients, we come across NP parameter space allowed by the data but may stand away from the SM point. This has been analyzed for four 2D NP scenarios respectively for S9, S10, S18, and S42. For scenarios S9 and S10, the global fit can simultaneously explain the SM-like behaviour of average of $R_K$ \& $R_{K^*}$, $R_\phi$, $\mathcal{B}(B_s \to \mu^+ \mu^-)$ and accommodate the measurements of $P'_5$, $\mathcal{B}(B \to K^{(*)} \mu^+ \mu^-)$, and $\mathcal{B}(B \to K^{(*)} e^+ e^-)$ at $1\sigma$ of their respective errors. However, $\mathcal{B}(B_s \to \phi \mu^+ \mu^-)$ and $\mathcal{B}(B_s \to \phi e^+ e^-)$ can hardly be accommodated within $1\sigma$. For scenario S18, we found that the best fit cannot fully encompass both $P'_5$, $\mathcal{B}(B_s \to \phi \mu^+ \mu^-)$ and $\mathcal{B}(B_s \to \phi e^+ e^-)$ within $1\sigma$.  For scenario S42, again $\mathcal{B}(B_s \to \phi \mu^+ \mu^-)$ can hardly be accommodated within $1\sigma$. Similarly, this is also true for the average of $R_K$ and $R_{K^*}$ at the level of $1\sigma$, though individually they can be accommodated well.

Beyond the observables that generally agree with SM predictions, we also focused on those such as branching fractions, $A_{FB}$, $P_2$, and $P'_5$, which remain particularly sensitive to NP effects. Under moderate hadronic assumptions, these observables still exhibit residual tensions that can be consistently addressed within selected SMEFT scenarios. For instance, scenario S42, which incorporates both LFU-preserving and violating NP, can significantly affect $\Delta P_{2}$ and $\Delta P'_{5}$. Among the scenarios that satisfy the present experimental limits, 
we identify four representative benchmark cases classified according to their LFU or LFUV nature. The LFU candidates are (i) S9 ($[C_{qe}]_{2311}=[C_{qe}]_{2322}$, $[C_{\phi q}^{(3)}]_{23}$) and (ii) S10 ($[C_{\ell q}^{(3)}]_{1123}=[C_{\ell q}^{(3)}]_{2223}$, $[C_{\phi q}^{(3)}]_{23}$), while the LFUV candidates are (i) S18 ($[C_{\ell q}^{(3)}]_{1123}$, $[C_{\ell q}^{(3)}]_{2223}$) and (ii) S42 ($[C_{qe}]_{2322}$, $[C_{\phi q}^{(3)}]_{23}$). Future high-precision measurements in both the electron and muon sectors, together with continued theoretical developments aimed at more precise determinations of hadronic quantities, will be crucial in disentangling these scenarios prominently.

\section*{Acknowledgments}
NR thanks Rupak Dutta and Niladri Sahoo for the useful discussions. NR acknowledges the Department of Space (DOS) and the Physical Research Laboratory, Ahmedabad, India for financial support. The authors thank Namit Mahajan for carefully reading the draft and for valuable comments. 

\appendix

\section{Hadronic matrix elements, differential decay distribution and $q^2$ observables in $B \to K^{(*)} \ell^+ \ell^-$ decays}
\label{ap1}
The hadronic matrix elements for the exclusive $B \to K$ transition in terms of form factors are given by \cite{Dutta:2019wxo}
\begin{eqnarray}
\langle K|\bar{s}\gamma^{\mu}b|B \rangle = f_{+}(q^2)\Big[p_{B}^{\mu}+p_{K}^{\mu}-\frac{M_{B}^2-M_{K}^2}{q^2}\,q^{\mu}\Big] + f_0(q^2) \frac{M_{B}^2-M_{K}^2}{q^2}\,q^{\mu}\,,\nonumber \\
\langle K|\bar{s}\sigma^{\mu\nu}\,q_{\nu}b|B \rangle =\frac{i\,f_T(q^2)}{M_{B}+M_{K}}\Big[q^2(p_{B}^{\mu}+p_{K}^{\mu} - (M_{B}^2-M_{K}^2)q^{\mu}\Big]\,,
\end{eqnarray}

Similarly, for $B \to K^{\ast}$ transitions, the hadronic matrix elements in terms of the form factors can be written as
\begin{eqnarray}
\langle K^{\ast}|\bar{s}\gamma^{\mu}b|B \rangle &=& \frac{2\,i\,V(q^2)}{M_{B}+M_{K^{\ast}}}\,\epsilon^{\mu\nu\rho\sigma}\epsilon^{\ast}_{\nu}\,
p_{B_{{\rho}}}\,p_{{K^{\ast}}_{\sigma}}\,, \nonumber \\
\langle K^{\ast}|\bar{s}\gamma^{\mu}\gamma_5\,b|B \rangle &=& 2\,M_{K^{\ast}}\,A_0(q^2)\frac{\epsilon^{\ast}\cdot q}{q^2}\,q^{\mu} + 
(M_{B} + M_{K^{\ast}})\,A_1(q^2)\Big(\epsilon^{{\ast}^{\mu}}-\frac{\epsilon^{\ast}\cdot q}{q^2}\,q^{\mu}\Big)\, \nonumber \\
&&-
A_2(q^2)\frac{\epsilon^{\ast}\cdot q}{M_{B} + M_{K^{\ast}}}\,\Big[p_{B}^{\mu}+p_{K^{\ast}}^{\mu}-\frac{M_{B}^2-M_{K^{\ast}}^2}
{q^2}\,q^{\mu}\Big]\,, \nonumber \\
\langle K^{\ast}|\bar{s}\,i\,\sigma^{\mu\nu}\,q_{\nu}b|B \rangle &=& 2\,T_1(q^2)\,\epsilon^{\mu\nu\rho\sigma}\epsilon^{\ast}_{\nu}\,p_{B_{{\rho}}}\,
p_{{K^{\ast}}_{\sigma}}\,, \nonumber \\
\langle K^{\ast}|\bar{s}\,i\,\sigma^{\mu\nu}\,\gamma_5\,q_{\nu}b|B \rangle &=& T_2(q^2)\,\Big[(M_{B}^2-M_{K^{\ast}}^2)\epsilon^{{\ast}^{\mu}} - 
(\epsilon^{\ast}\cdot q)(p_{B}^{\mu}+p_{K^{\ast}}^{\mu})\Big] \nonumber \\
&&+ 
T_3(q^2)\,(\epsilon^{\ast}\cdot q)\Big[q^{\mu}-\frac{q^2}
{M_{B}^2-M_{K^{\ast}}^2}(p_{B}^{\mu}+p_{K^{\ast}}^{\mu})\Big]\,,
\end{eqnarray}
where $q^{\mu}=(p_B ^ \mu -p_{K^*}^\mu)$ is the four momentum transfer and $\epsilon_{\mu}$ is polarization vector of the $K^{\ast}$ meson. \newline

The $q^2$ dependent differential branching ratio for $B\to K\ell^+\ell^-$ decays is given by\cite{Bouchard:2013eph}:
\begin{equation}
\frac{d\mathcal{BR}}{dq^2} = \frac{\tau_{B}}{\hbar}(2a_\ell + \frac{2}{3}c_\ell) ,
\end{equation}
where the parameters $a_\ell$ and $c_\ell$ are given by
\begin{eqnarray}
a_\ell &=& \frac{G_F^2 \alpha_{EW}^2 |V_{tb} V_{ts}^*|^2}{2^9\pi^5 m_{B}^3}\beta_\ell \sqrt{\lambda} \Big[ q^2 |F_P|^2 + \frac{\lambda}{4}(|F_A|^2 + |F_V|^2) + 4m_\ell^2 m_{B}^2 |F_A|^2 \nonumber \\
&+& 2m_\ell(m_{B}^2 - m_{D_s}^2 + q^2){\rm Re}(F_PF_A^*) \Big], \\
c_\ell &=& -\frac{G_F^2 \alpha_{EW}^2 |V_{tb} V_{ts}^*|^2}{2^9\pi^5 m_{B}^3}\beta_\ell \sqrt{\lambda}\frac{ \lambda \beta_\ell^2}{4}(|F_A|^2 + |F_V|^2).
\end{eqnarray}
Here the kinematical factor $\lambda$ and the mass correction factor $\beta_\ell$ given in the above equations are given by
\begin{eqnarray}
\lambda &=& q^4 + m_{B}^4 + m_{K}^4 - 2(m_{B}^2m_{K}^2 + m_{B}^2q^2 + m_{K}^2q^2), \nonumber\\
\beta_\ell &=& \sqrt{1-4m_\ell^2/q^2}.
\end{eqnarray}
However, the explicit expressions of the form factors such as $F_P$, $F_V$, and $F_A$ are given as follows:
\begin{eqnarray}
F_P &=& -m_\ell C_{10} \Big[ f_+ - \frac{M_B^2-M_K^2}{q^2}(f_0-f_+) \Big], \\
F_V &=& C_9^{\rm eff} f_+ + \frac{2m_b}{M_B+M_K}C_7^{\rm eff}f_T, \\
F_A &=& C_{10}f_+. \\
\end{eqnarray}

Similarly, the $q^2$ dependent differential branching ratio for $B \to K^* \ell^+\ell^-$ decays is given as~\cite{Rajeev:2020aut}
\begin{equation}
d\mathcal{BR}/dq^2=\frac{d\Gamma / dq^2}{\Gamma _{Total}}=\frac{\tau _{B}}{\hbar} \frac{1}{4}\bigg[3I_{1}^c + 6I_{1}^s - I_{2}^c -2I_{2}^s\bigg],
\end{equation}
where the angular coefficients are given by,
\begin{eqnarray}
I_{1}^{c} &=& \bigg(|A_{L0}|^2 + |A_{R0}|^2\bigg) + 8\frac{m_{l}^2}{q^2} Re\bigg[A_{L0}A_{R0}^{*}\bigg] + 4\frac{m_{l}^2}{q^2}|A_{t}|^2, \nonumber \\
I_{2}^{c} &=& -\beta_{l}^2 \bigg(|A_{L0}|^2 + |A_{R0}|^2\bigg), \nonumber \\
 I_{1}^{s} &=& \frac{3}{4} \bigg[|A_{L\perp}|^2 + |A_{L\parallel}|^2 + |A_{R\perp}|^2 + |A_{R\parallel}|^2\bigg] \bigg(1-\frac{4m_{l}^2}{3q^2}\bigg) +
 \frac{4m_{l}^2}{q^2} Re\bigg[A_{L\perp} A_{R\perp}^{*} + A_{L\parallel} A_{R\parallel}^{*}\bigg], \nonumber \\
 I_{2}^{s} &=& \frac{1}{4} \beta_{l}^2 \bigg[|A_{L\perp}|^2 + |A_{L\parallel}|^2 + |A_{R\perp}|^2 + |A_{R\parallel}|^2\bigg],   \nonumber \\
 I_{3} &=& \frac{1}{2} \beta_{l}^2 \bigg[|A_{L\perp}|^2 - |A_{L\parallel}|^2 + |A_{R\perp}|^2 - |A_{R\parallel}|^2\bigg], \nonumber \\
 I_{4} &=& \frac{1}{\sqrt{2}} \beta_{l}^2 \bigg[Re\bigg(A_{L0}A_{L\parallel}^{*}\bigg) + Re\bigg(A_{R0}A_{R\parallel}^{*}\bigg)\bigg],  \nonumber \\
 I_{5} &=& \sqrt{2} \beta_l \bigg[Re\bigg(A_{L0}A_{L\perp}^{*}\bigg) - Re\bigg(A_{R0}A_{R\perp}^{*}\bigg)\bigg], \nonumber \\
 I_{6} &=& 2 \beta_l \bigg[Re\bigg(A_{L\parallel}A_{L\perp}^{*}\bigg) - Re\bigg(A_{R\parallel}A_{R\perp}^{*}\bigg)\bigg], \nonumber \\
 I_{7} &=& \sqrt{2} \beta_l \bigg[Im\bigg(A_{L0}A_{L\parallel}^{*}\bigg) - Im\bigg(A_{R0}A_{R\parallel}^{*}\bigg)\bigg], \nonumber \\
 I_{8} &=& \frac{1}{\sqrt{2}} \beta_{l}^2 \bigg[Im\bigg(A_{L0}A_{L\perp}^{*}\bigg) + Im\bigg(A_{R0}A_{R\perp}^{*}\bigg)\bigg], \nonumber \\
 I_{9} &=& \beta_{l}^2 \bigg[Im\bigg(A_{L\parallel}A_{L\perp}^{*}\bigg) + Im\bigg(A_{R\parallel}A_{R\perp}^{*}\bigg)\bigg]\,,
\end{eqnarray}
where $\beta _ \ell = \sqrt{1-4m_ \ell ^2/q^2}$. The transversity amplitudes are expressed in terms of the corresponding WCs and form factors as
\begin{eqnarray}
 A_{L0} &=& N \frac{1}{2m_{{{K^*}}}\sqrt{q^2}}
 \bigg\{(C_{9} - C_{10}) \bigg[(m_{B}^2 - m_{{K^*}}^2 - q^2)(m_{B} + m_{{K^*}})A_1 - \frac{\lambda}{m_{B} + m_{{K^*}}}A_2\bigg]+ \nonumber \\  
 && 2\,m_b\, C_{7}\, \bigg[(m_{B}^2 + 3m_{{K^*}}^2 - q^2)T_2 - \frac{\lambda}{m_{B}^2 -m_{{K^*}}^2}T_3 \bigg] \bigg\}\,, \nonumber \\
 A_{L\perp} &=& - N\sqrt{2} \bigg[(C_{9} - C_{10})
 \frac{\sqrt{\lambda}}{m_{B} + m_{{K^*}}} V + \frac{\sqrt{\lambda}\,2\,m_b\,C_{7}^{eff}}{q^2} T_1 \bigg]\,, \nonumber \\
 A_{L\parallel} &=& N\sqrt{2} \bigg[(C_{9} - C_{10})
 (m_{B} + m_{{K^*}}) A_1 + \frac{2\,m_b\,C_{7}(m_{B}^2 - m_{{K^*}}^2)}{q^2} T_2 \bigg]\,, \nonumber \\
 A_{Lt} &=& N (C_{9} - C_{10}) \frac{\sqrt{\lambda}}{\sqrt{q^2}}A_0\,,
\end{eqnarray}
where $\lambda=(m_{B}^4\,+m_{K^*}^4\,+q^4\,-\,2\,(m_{B}^2m_{K^*}^2+m_{k^*}^2 q^2+q^2 m_{B}^2)$ and $N$, the normalization constant which is defined as 
\begin{equation}
 N= \bigg[\frac{G_{F}^2\alpha_{em}^2}{3\cdot 2^{10}\pi^5\,m_{B}^3}|V_{tb}V_{ts}^{*}|^2 q^2 \sqrt{\lambda}\bigg(1-\frac{4m_{\ell}^2}{q^2}\bigg)^{1/2}\bigg]^{1/2}\,.
\end{equation}
The right chiral component $A_{Ri}$ of the transversity amplitudes can be obtained by replacing $A_{Li}$ by $A_{Li}|_{C_{10} \to -C_{10}} (i= 0, \parallel, \perp, t)$.\newline

The definition of prominent observables such as the forward-backward asymmetry $( A_{FB} )$, the longitudinal polarization fraction $(F_L)$ and the angular observable $\langle P_5 ' \rangle$ which are given by
\begin{equation}
 F_L(q^2)=\frac{3I_{1}^c - I_{2}^c}{3I_{1}^c + 6I_{1}^s - I_{2}^c -2I_{2}^s}, \hspace{0.1cm}
 A_{FB}(q^2)=\frac{3I_{6}}{3I_{1}^c + 6I_{1}^s - I_{2}^c -2I_{2}^s},
\end{equation}
\begin{equation}
\langle P_1 \rangle = \frac{\int_{bin}dq^2 I_{3}}{2 \int_{bin}dq^2 I_{2s}}, \hspace{0.1cm}
\langle P_2 \rangle = \frac{\int_{bin}dq^2 I_{6}}{8 \int_{bin}dq^2 I_{2s}}, \hspace{0.1cm}
\langle P_3 \rangle = -\frac{\int_{bin}dq^2 I_{9}}{4 \int_{bin}dq^2 I_{2s}},
\end{equation}
\begin{equation}
\langle P^{\prime}_4 \rangle = \frac{\int_{bin}dq^2 I_{4}}{2 \sqrt{-\int_{bin}dq^2 I_{2}^c \int_{bin}dq^2 I_{2}^s}}, \hspace{0.1cm}
 \langle P^{\prime}_5 \rangle = \frac{\int_{bin}dq^2 I_{5}}{2 \sqrt{-\int_{bin}dq^2 I_{2}^c \int_{bin}dq^2 I_{2}^s}}, 
\end{equation}
\begin{equation}
\hspace{0.1cm}
 \langle P^{\prime}_6 \rangle = -\frac{\int_{bin}dq^2 I_{7}}{2 \sqrt{-\int_{bin}dq^2 I_{2}^c \int_{bin}dq^2 I_{2}^s}},\hspace{0.1cm}
 \langle P^{\prime}_8 \rangle = -\frac{\int_{bin}dq^2 I_{8}}{2 \sqrt{-\int_{bin}dq^2 I_{2}^c \int_{bin}dq^2 I_{2}^s}}.
\end{equation}

the ratio of the branching ratios of $\mu$ to $e$ transition in $B \to K^{(*)} \ell^ + \ell^-$ decays are defined as
\begin{eqnarray}
R_{K^{(*)}} = \frac{\mathcal{B}\,\Big(B \to K^{(*)}\, \mu^+\mu^-\Big)}{\mathcal{B}\,\Big(B \to K^{(*)}\, e^+\,e^-\Big)}\,.
\end{eqnarray}


\section{Fast likelihood approach~\cite{Straub:2018kue}}
\label{fastfit}
In general, we consider a set of $N$ measurements $\{x_i^{\rm exp}\}$ corresponding to $M$ observables, where $i=1,…,N$. For each observable, the theoretical prediction $x_i^{\rm th}$ depends on a set of theory parameters $\theta_j$, which may include Wilson coefficients as well as other model parameters. The likelihood function can then be constructed by comparing the experimental measurements with their corresponding theoretical predictions, such that
\begin{equation}
    \mathcal{L}_{\rm exp}(\hat{\theta})= \prod_{i=1}^{N} f_i \bigg(x_i^{\rm exp}, x_i^{\rm th}(\vec{\theta}) \bigg),
\end{equation}
where, $f_i$ is the probability distribution function of measurement $i$. Since we are interested in determining the value of $\vec{\theta}$ preferred by the data by direct measurements or theoretical considerations, we can split $\vec{\theta}$ into fit-parameters $(\vec{C})$ and nuisance parameters. \newline

There exists a rich set of experimental data on $b \to s\, \ell^+ \ell^-$ transitions, particularly for $\ell \in \mu$, with many observables measured in finely binned distributions. To properly constrain the WCs in a way that accurately reflects the experimental measurements, it is crucial to account for the correlations of theoretical uncertainties — both between different observables and across different bins of the same observable.

A rigorous way to incorporate these correlations is through global Bayesian analyses, where all uncertainties are parameterized by nuisance parameters and marginalized over using advanced numerical methods such as Markov Chain Monte Carlo (MCMC) techniques. However, a major drawback of this approach is its computational intensity: the required computing time grows significantly with the number of parameters involved.

In contrast, the \texttt{flavio}~\cite{Straub:2018kue} package adopts a slightly different approach which introduces slightly different concept called a "fast fit". It is based on the approximation of assuming the likelihood to be of the form 
\begin{equation}
    \mathcal{L}=e^{-\chi^2 (\vec{C}^{\rm NP})/2},
\end{equation}
where, the $\chi^2$ function is defined as, 
\begin{equation}
    \chi^2 (\vec{C}^{\rm NP})= \bigg[ \vec{O}_{\rm exp} - \vec{O}_{\rm th} (\vec{C}^{\rm NP}) \bigg]^T \big[ C_{\rm exp} + C_{\rm th} \big]^{-1} \bigg[ \vec{O}_{\rm exp} - \vec{O}_{\rm th} (\vec{C}^{\rm NP}) \bigg].
\end{equation}

The $\chi^2$ function depends directly on the WCs, while the theoretical uncertainties are incorporated via precomputed covariance matrices. This significantly simplifies the numerical evaluation while still retaining the essential correlations among observables.
Here, $\vec{O}_{\rm exp}$ represents the experimentally measured central values of all observables, while $\vec{O}_{\rm th}$ denotes the corresponding theoretical predictions, which depend on the NP WCs. $C_{\rm exp}$ is the covariance matrix incorporating the experimental uncertainties and their correlations, and $C_{\rm th}$ is the covariance matrix for the theoretical predictions, containing theoretical uncertainties and their correlations.

In constructing the above $\chi^2$, two key approximations are made. First, both experimental and theoretical uncertainties are assumed to follow Gaussian distributions. Second, the dependence of the theoretical uncertainties on the NP contributions is neglected. This means that the theoretical uncertainties and their correlations are evaluated for the WCs fixed at their Standard Model (SM) values. This approximation is well justified, as no significant deviations from the SM predictions have been observed so far.

The theoretical covariance matrix $C_{\rm th}$ is determined by computing all relevant observables for a large set of randomly generated theory parameters. These parameters are sampled from normal distributions, using the uncertainties and correlations specified earlier. This procedure allows us to consistently include both correlations between different observables and correlations between different bins of the same observable. We find that these correlations significantly affect our numerical results.

Similarly, the experimental covariance matrix $C_{\rm exp}$ symmetrizes the experimental error bars and incorporates the experimental correlations provided in the LHCb results on the $B \to K^* \mu^+ \mu^-$ analysis. For branching ratio measurements where no official error correlations are available, an approximate treatment is assumed such that the statistical uncertainties are uncorrelated, while systematic uncertainties are fully correlated across measurements of the same observable by a given experiment.


\FloatBarrier

 \section{Prediction of $B^+ \to K^+\, \ell^+\,\ell^-$ and $B^0 \to K^*\, \ell^+\,\ell^-$ observables in SM and in the presence of various NP scenarios at different $q^2$ bins}
\label{ap3}

\begin{table}[h]
 \centering
 \setlength{\tabcolsep}{6pt} 
 \renewcommand{\arraystretch}{1.5} 
 \resizebox{\columnwidth}{!}{
 \begin{tabular}{|c|c|c|c|c|c|c|c|}
 \hline
Observable & $q^2$ bin (GeV$^2$) & SM & S3 & S9 & S10 & S18 & S42 \\
\hline \hline
\multirow{5}{*}{$\langle \rm BR \rangle(B^+ \to K^+\, e^+\,e^-)\times 10^{8}$}
& [1.1, 2.5] & $3.52\pm0.55$ & $2.71\pm 0.55$ & $2.42\pm0.43$ & $2.37\pm0.46$ & $2.86\pm0.54$ & $2.39\pm0.40$\\
& [2.5, 4] & $3.49\pm0.58$ & $2.69\pm0.44$ & $2.41\pm0.41$ & $2.36\pm0.41$ & $2.84\pm0.44$ & $2.38\pm0.41$\\
& [4, 6] & $3.45\pm0.63$ &$2.66\pm0.43$& $2.38\pm0.38$ & $2.33\pm0.40$ & $2.81\pm0.49$ & $2.36\pm0.37$\\
& [1.1, 6] & $3.48 \pm 0.59$ & $2.68 \pm 0.44$ & $2.40 \pm 0.41$ & $2.35 \pm 0.42$ & $2.83 \pm 0.53$ & $2.38 \pm 0.43$  \\
& [15, 22]  & $1.50 \pm 0.18$ & $1.16 \pm 0.14$ & $1.04 \pm 0.11$ & $1.02 \pm 0.11$ & $1.22 \pm 0.16$ & $1.03 \pm 0.12$  \\ \hline
\multirow{5}{*}{$\langle \rm BR \rangle(B^+ \to K^+\, \mu^+\,\mu^-)\times 10^{8}$}
& [1.1, 2.5] &  $3.52\pm0.61$ & $2.71\pm 0.47 $ & $2.43\pm0.44$ & $2.38\pm0.44$ &$2.75\pm0.44$&$2.40\pm0.43$\\
& [2.5, 4] & $3.50\pm0.58$ &$2.69\pm0.43$ & $2.41\pm0.38$ & $2.36\pm0.34$ &$2.74\pm0.41$&$2.38\pm0.42$\\
& [4, 6] & $3.46\pm0.57$ &$2.67\pm0.42$ & $2.38\pm0.36$ & $2.33\pm0.44$ &$2.71\pm0.0.42$&$2.35\pm0.40$\\
& [1.1, 6] & $3.48 \pm 0.57$ & $2.68 \pm 0.43$ & $2.40 \pm 0.44$ & $2.35 \pm 0.37$ & $2.73 \pm 0.46$ & $2.37 \pm 0.41$  \\
& [15, 22]  & $1.50 \pm 0.16$ & $1.16 \pm 0.15$ & $1.04 \pm 0.11$ & $1.02 \pm 0.10$ & $1.18 \pm 0.14$ & $1.03 \pm 0.13$  \\ 
 \hline
 \end{tabular}}
\caption{The branching ratios of $B^+ \to K^+\, e^+\,e^-$ and $B^+ \to K^+\, \mu^+\,\mu^-$ in SM and the presence of various NP scenarios at different $q^2$ bins. Each entry denotes the central value and the corresponding $1\sigma$ error.}  
 \label{tab_ap1}
 \end{table}

\begin{table}[ht]
 \centering
 \setlength{\tabcolsep}{6pt} 
 \renewcommand{\arraystretch}{1.5} 
 \resizebox{\columnwidth}{!}{
 \begin{tabular}{|c|c|c|c|c|c|c|c|}
 \hline
Observable & $q^2$ bin (GeV$^2$) & SM & S3 & S9 & S10 & S18 & S42 \\
\hline \hline
\multirow{5}{*}{$\langle \rm BR \rangle(B^0 \to K^\star\, e^+\,e^-)\times 10^{8}$}
& [1.1, 2.5] & $4.67\pm0.66$   & $3.86\pm 0.59$  & $3.73\pm0.46$   & $3.69\pm0.48$  & $3.96\pm0.57$   & $3.20\pm0.49$\\
& [2.5, 4]   & $4.49\pm0.71$   & $3.54\pm0.60$   & $3.33\pm0.52$   & $3.28\pm0.51$  & $3.66\pm0.52$   & $2.82\pm0.55$\\
& [4, 6]     & $5.02\pm0.78$   & $3.89\pm0.60$   & $3.57\pm0.52$   & $3.51\pm0.48$  & $4.03\pm0.61$   & $3.11\pm0.54$\\
& [1.1, 6]   & $4.76 \pm 0.76$ & $3.78 \pm 0.57$ & $3.54 \pm 0.49$ & $3.49\pm0.52$  & $3.90\pm0.58$   & $3.05 \pm 0.60$  \\
& [15, 22]    & $5.93 \pm 0.71$ & $4.63 \pm 0.53$ & $4.10 \pm 0.49$ & $4.02\pm0.49$  & $4.80\pm0.54$   & $3.95 \pm 0.56$  \\ \hline
\multirow{5}{*}{$\langle \rm BR \rangle(B^0 \to K^\star\, \mu^+\,\mu^-)\times 10^{8}$}
& [1.1, 2.5] &  $4.65\pm0.66$  & $3.85\pm 0.67$  & $3.71\pm0.51$   & $3.68\pm0.51$  & $3.84\pm0.59$   & $3.66\pm0.50$\\
& [2.5, 4]   & $4.48\pm0.65$   & $3.53\pm 0.58$  & $3.31\pm0.43$   & $3.27\pm0.41$  & $3.52\pm0.53$   & $3.26\pm0.48$\\
& [4, 6]     & $5.00\pm0.70$   & $3.88\pm 0.67$  & $3.56\pm0.48$   & $3.50\pm0.60$  & $3.87\pm0.59$   & $3.50\pm0.48$\\
& [1.1, 6]   & $4.74 \pm 0.68$ & $3.77 \pm 0.55$ & $3.53 \pm 0.51$ & $3.48\pm0.52$  & $3.75\pm0.54$   & $3.47 \pm 0.51$  \\
& [15, 22]    & $5.92 \pm 0.72$ & $4.63 \pm 0.60$ & $4.09 \pm 0.53$ & $4.01\pm0.49$  & $4.61\pm0.57$   & $4.03 \pm 0.45$  \\ 
 \hline
 \end{tabular}}
\caption{The branching ratios of $B^0 \to K^\star\, e^+\,e^-$ and $B^0 \to K^\star\, \mu^+\,\mu^-$ in SM and the presence of various NP scenarios at different $q^2$ bins. Each entry denotes the central value and the corresponding $1\sigma$ error.}  
 \label{tab_ap2}
 \end{table}

 \begin{table}[ht]
 \centering
 \setlength{\tabcolsep}{6pt} 
 \renewcommand{\arraystretch}{1.5} 
 \resizebox{\columnwidth}{!}{
 \begin{tabular}{|c|c|c|c|c|c|c|c|}
 \hline
Observable & $q^2$ bin (GeV$^2$) & SM & S3 & S9 & S10 & S18 & S42 \\
\hline \hline
\multirow{5}{*}{$\langle \rm A_{FB} \rangle(B^+ \to K^+\, e^+\,e^-)\times 10^{13}$}
& [1.1, 2.5] & $0$ & $-3.42\pm0.07$ & $2.11\pm0.06$ & $-7.96\pm0.18$ & $-2.69\pm0.07$ & $-0.03\pm0.001$\\
& [2.5, 4] & $0$ & $-3.48\pm0.09$ & $2.15\pm0.06$ & $-8.12\pm0.24$ & $-2.74\pm0.08$ & $-0.03\pm0.001$\\
& [4, 6] & $0$ & $-3.57\pm0.09$ & $2.22\pm0.07$ & $-8.35\pm0.28$ & $-2.81\pm0.08$ & $-0.03\pm0.001$\\
& [1.1, 6] & 0 & $-3.49 \pm 0.09$ & $2.17 \pm 0.06$ & $-8.17 \pm 0.22$ & $-2.76 \pm 0.06$ & $-0.03 \pm 0.001$  \\
& [15, 22]  & 0 & $-5.32 \pm 0.29$ & $3.32 \pm 0.19$ & $-12.49 \pm 0.70$ & $-4.20 \pm 0.20$ & $-0.05 \pm 0.003$  \\ \hline
\multirow{5}{*}{$\langle \rm A_{FB} \rangle(B^+ \to K^+\, \mu^+\,\mu^-)\times 10^{8}$}
& [1.1, 2.5] & $0$ &$-1.42\pm0.03$ & $0.88\pm0.02$ & $-3.31\pm0.08$ & $-1.33\pm0.03$ & $0.86\pm0.02$\\
& [2.5, 4] & $0$ &$-1.47\pm0.04$ & $0.91\pm0.03$ & $-3.42\pm0.10$ & $-1.37\pm0.03$ & $0.89\pm0.02$\\
& [4, 6] & $0$ &$-1.51\pm0.04$ & $0.94\pm0.03$ & $-3.54\pm0.11$ & $-1.42\pm0.05$ & $0.92\pm0.03$\\
& [1.1, 6] & 0 & $-1.47 \pm 0.04$ & $0.91 \pm 0.03$ & $-3.43 \pm 0.09$ & $-1.38 \pm 0.04$ & $0.81 \pm 0.02$ \\
& [15, 22]  & 0 & $-2.26 \pm 0.13$ & $1.41 \pm 0.08$ & $-5.31 \pm 0.31$ & $-2.12 \pm 0.11$ & $1.37 \pm 0.07$ \\
 \hline
\end{tabular}}
\caption{The forward-backward asymmetry of $B^+ \to K^+\, e^+\,e^-$ and $B^+ \to K^+\, \mu^+\,\mu^-$ in SM and in the presence of various NP scenarios at different $q^2$ bins. Each entry denotes the central value and the corresponding $1\sigma$ error.}  
 \label{tab_ap3}
 \end{table}

 \begin{table}[ht]
 \centering
 \setlength{\tabcolsep}{6pt} 
 \renewcommand{\arraystretch}{1.5} 
 \resizebox{\columnwidth}{!}{
 \begin{tabular}{|c|c|c|c|c|c|c|c|}
 \hline
Observable & $q^2$ bin (GeV$^2$) & SM & S3 & S9 & S10 & S18 & S42 \\
\hline \hline
\multirow{5}{*}{$\langle \rm A_{FB} \rangle(B^0 \to K^\star\, e^+\,e^-) $}
& [1.1, 2.5] & $-0.142\pm0.030$ & $-0.17\pm0.03$ & $-0.21\pm0.04$   & $-0.21\pm0.04$   & $-0.17\pm0.03$   & $-0.14\pm0.03$\\
& [2.5, 4]   & $-0.018\pm0.031$ & $-0.05\pm0.03$ & $-0.12\pm0.04$   & $-0.12\pm0.04$   & $-0.05\pm0.03$   & $-0.02\pm0.04$\\
& [4, 6]     & $0.123\pm0.045$ & $0.10\pm0.05$  & $0.03\pm0.05$   & $0.02\pm0.06$   & $0.10\pm0.05$   & $0.12\pm0.05$\\
& [1.1, 6]   & $0.008\pm0.027$  & $-0.02\pm0.04$ & $-0.09\pm0.04$ & $-0.09\pm0.05$  & $-0.02\pm0.04$   & $0.00 \pm 0.04$  \\
& [15, 22]    & $0.365\pm0.032$   & $0.37\pm 0.03$ & $0.33\pm0.04$ & $0.33\pm0.04$  & $0.37\pm0.03$   & $0.36 \pm 0.03$  \\ \hline
\multirow{5}{*}{$\langle \rm A_{FB} \rangle(B^0 \to K^\star\, \mu^+\,\mu^-) $}
& [1.1, 2.5] & $-0.138\pm0.032$ &$-0.17\pm0.03$ & $-0.21\pm0.03$   & $-0.21\pm0.04$   & $-0.17\pm0.03$   & $-0.20\pm0.04$\\
& [2.5, 4]   & $-0.018\pm0.031$ &$-0.05\pm0.03$ & $-0.11\pm0.04$   & $-0.12\pm0.04$   & $-0.05\pm0.03$   & $-0.11\pm0.04$\\
& [4, 6]     & $0.122\pm0.040$ &$0.09\pm0.05$  & $0.03\pm0.04$   & $0.02\pm0.05$   & $0.09\pm0.05$   & $0.03\pm0.04$\\
& [1.1, 6]   & $0.009\pm0.027$ &$-0.02\pm0.03$ & $-0.08\pm0.04$ & $-0.09\pm0.04$   & $-0.02\pm0.03$   & $-0.08 \pm 0.04$ \\
& [15, 22]    & $0.365\pm0.031$ & $0.36\pm0.03$ & $0.33\pm0.04$ & $0.33\pm0.04$   & $0.37\pm0.03$   & $0.34 \pm 0.04$ \\
 \hline
\end{tabular}}
\caption{The forward-backward asymmetry of $B^0 \to K^\star\, e^+\,e^-$ and $B^0 \to K^\star\, \mu^+\,\mu^-$ in SM and in the presence of various NP scenarios. Each entry denotes the central value and the corresponding $1\sigma$ error.}  
 \label{tab_ap4}
 \end{table}

\begin{table}[ht]
 \centering
 \setlength{\tabcolsep}{6pt} 
 \renewcommand{\arraystretch}{1.5} 
 \resizebox{\columnwidth}{!}{
 \begin{tabular}{|c|c|c|c|c|c|c|c|}
 \hline
Observable & $q^2$ bin (GeV$^2$) & SM & S3 & S9 & S10 & S18 & S42 \\
\hline \hline
\multirow{5}{*}{$\langle \rm F_L \rangle(B^0 \to K^\star\, e^+\,e^-)$}
& [1.1, 2.5] & $0.788\pm0.046 $  & $0.743\pm0.055 $  &  $0.678\pm0.056 $  & $0.671\pm0.050 $  & $0.750\pm0.054 $   & $0.766\pm0.058 $\\
& [2.5, 4]   & $0.810\pm0.032 $  & $0.803\pm0.042 $  &  $0.753\pm0.046 $  & $0.750\pm0.041 $  & $0.805\pm0.036 $   & $0.862\pm0.037 $\\
& [4, 6]     & $0.719\pm0.051 $  & $0.726\pm0.051 $  &  $0.695\pm0.052 $  & $0.694\pm0.052 $  & $0.725\pm0.041 $   & $0.778\pm0.052 $\\
& [1.1, 6]   & $0.765\pm0.038 $  & $0.753\pm0.050 $  &  $0.706\pm0.051 $  & $0.703\pm0.048 $  & $0.755\pm0.040 $   & $0.799\pm0.045 $\\
& [15, 19]    & $0.340\pm0.033 $  & $0.342\pm0.038 $  &  $0.341\pm0.030 $  & $0.341\pm0.029 $  & $0.342\pm0.030 $   & $0.345\pm0.041 $\\ \hline
\multirow{5}{*}{$\langle \rm F_L \rangle(B^0 \to K^\star\, \mu^+\,\mu^-)$}
& [1.1, 2.5] & $0.760\pm0.040 $  & $0.716\pm0.049 $  &  $0.654\pm0.056 $  & $0.648\pm0.060 $  & $0.715\pm0.054 $   & $0.655\pm0.062 $\\
& [2.5, 4]   & $0.797\pm0.032 $  & $0.789\pm0.038 $  &  $0.741\pm0.044 $  & $0.737\pm0.047 $  & $0.789\pm0.039 $   & $0.745\pm0.040 $\\
& [4, 6]     & $0.712\pm0.055 $  & $0.718\pm0.047 $  &  $0.688\pm0.048 $  & $0.687\pm0.046 $  & $0.718\pm0.057 $   & $0.692\pm0.050 $\\
& [1.1, 6]   & $0.750\pm0.041 $  & $0.738\pm0.040 $  &  $0.693\pm0.046 $  & $0.690\pm0.045 $  & $0.738\pm0.049 $   & $0.696\pm0.046 $\\
& [15, 19]    & $0.340\pm0.029 $  & $0.341\pm0.037 $  &  $0.340\pm0.030 $  & $0.340\pm0.029 $  & $0.341\pm0.033 $   & $0.341\pm0.031 $\\ 
 \hline
\end{tabular}}
\caption{The longitudinal polarization fraction of $B^0 \to K^\star\, e^+\,e^-$ and $B^0 \to K^\star\, \mu^+\,\mu^-$ in SM and in the presence of various NP scenarios. Each entry denotes the central value and the corresponding $1\sigma$ error.}  
 \label{tab_ap5}
 \end{table}

\begin{table}[ht]
 \centering
 \setlength{\tabcolsep}{6pt} 
 \renewcommand{\arraystretch}{1.5} 
 \resizebox{\columnwidth}{!}{
 \begin{tabular}{|c|c|c|c|c|c|c|c|}
 \hline
Observable & $q^2$ bin (GeV$^2$) & SM & S3 & S9 & S10 & S18 & S42 \\
\hline \hline
\multirow{5}{*}{$\langle \rm P_1 \rangle(B^0 \to K^\star\, e^+\,e^-) $}
& [1.1, 2.5] & $ 0.024\pm0.054 $  & $ 0.038\pm0.052 $  &  $ 0.029\pm0.052 $  & $ 0.030\pm0.051 $  & $ 0.036\pm0.047 $   & $ 0.074\pm0.059 $\\
& [2.5, 4]   & $-0.116\pm0.037 $  & $-0.092\pm0.040 $  &  $-0.071\pm0.046 $  & $-0.069\pm0.043 $  & $-0.096\pm0.041 $   & $-0.094\pm0.052 $\\
& [4, 6]     & $-0.178\pm0.045 $  & $-0.174\pm0.049 $  &  $-0.153\pm0.045 $  & $-0.152\pm0.038 $  & $-0.175\pm0.048 $   & $-0.206\pm0.063 $\\
& [1.1, 6]   & $-0.112\pm0.036 $  & $-0.091\pm0.036 $  &  $-0.073\pm0.034 $  & $-0.071\pm0.034 $  & $-0.094\pm0.034 $   & $-0.087\pm0.040 $\\
& [15, 19]    & $-0.624\pm0.044 $  & $-0.625\pm0.054 $  &  $-0.623\pm0.055 $  & $-0.623\pm0.049 $  & $-0.625\pm0.045 $   & $-0.629\pm0.061 $\\ \hline
\multirow{5}{*}{$\langle \rm P_1 \rangle(B^0 \to K^\star\, \mu^+\,\mu^-) $}
& [1.1, 2.5] & $ 0.024\pm0.046 $  & $ 0.038\pm0.053 $  &  $ 0.029\pm0.048 $  & $ 0.029\pm0.048 $  & $ 0.038\pm0.058 $   & $ 0.031\pm0.052 $\\
& [2.5, 4]   & $-0.116\pm0.042 $  & $-0.093\pm0.040 $  &  $-0.071\pm0.042 $  & $-0.069\pm0.048 $  & $-0.092\pm0.041 $   & $-0.071\pm0.043 $\\
& [4, 6]     & $-0.178\pm0.049 $  & $-0.174\pm0.048 $  &  $-0.153\pm0.041 $  & $-0.152\pm0.036 $  & $-0.174\pm0.045 $   & $-0.154\pm0.039 $\\
& [1.1, 6]   & $-0.113\pm0.032 $  & $-0.092\pm0.034 $  &  $-0.074\pm0.037 $  & $-0.072\pm0.034 $  & $-0.092\pm0.031 $   & $-0.074\pm0.034 $\\
& [15, 19]    & $-0.624\pm0.047 $  & $-0.625\pm0.048 $  &  $-0.623\pm0.053 $  & $-0.623\pm0.046 $  & $-0.625\pm0.051 $   & $-0.624\pm0.045 $\\ 
 \hline
\end{tabular}}
\caption{The angular observable $P_1$ in $B^0 \to K^\star\, e^+\,e^-$ and $B^0 \to K^\star\, \mu^+\,\mu^-$ in SM and in the presence of various NP scenarios. Each entry denotes the central value and the corresponding $1\sigma$ error.}  
 \label{tab_ap6}
 \end{table}

\begin{table}[ht]
 \centering
 \setlength{\tabcolsep}{6pt} 
 \renewcommand{\arraystretch}{1.5} 
 \resizebox{\columnwidth}{!}{
 \begin{tabular}{|c|c|c|c|c|c|c|c|}
 \hline
Observable & $q^2$ bin (GeV$^2$) & SM & S3 & S9 & S10 & S18 & S42 \\
\hline \hline
\multirow{5}{*}{$\langle \rm P_2 \rangle(B^0 \to K^\star\, e^+\,e^-) $}
& [1.1, 2.5] & $-0.445\pm0.013 $  & $-0.438\pm0.012 $  &  $-0.436\pm0.012 $  & $-0.435\pm0.014 $  & $-0.440\pm0.012 $   & $-0.396\pm0.018 $\\
& [2.5, 4]   & $-0.063\pm0.111 $  & $-0.181\pm0.097 $  &  $-0.315\pm0.078 $  & $-0.324\pm0.071 $  & $-0.165\pm0.106 $   & $-0.114\pm0.139 $\\
& [4, 6]     & $ 0.292\pm0.080 $  & $ 0.230\pm0.088 $  &  $ 0.059\pm0.079 $  & $ 0.046\pm0.102 $  & $ 0.240\pm0.083 $   & $ 0.070\pm0.075 $\\
& [1.1, 6]   & $ 0.023\pm0.085 $  & $-0.067\pm0.083 $  &  $-0.196\pm0.067 $  & $-0.204\pm0.068 $  & $-0.055\pm0.087 $   & $-0.190\pm0.095 $\\
& [15, 19]    & $ 0.373\pm0.021 $  & $ 0.370\pm0.028 $  &  $ 0.337\pm0.032 $  & $ 0.334\pm0.034 $  & $ 0.370\pm0.024 $   & $ 0.364\pm0.024 $\\ \hline
\multirow{5}{*}{$\langle \rm P_2 \rangle(B^0 \to K^\star\, \mu^+\,\mu^-) $}
& [1.1, 2.5] & $-0.451\pm0.013 $  & $-0.444\pm0.011 $  &  $-0.443\pm0.014 $  & $-0.441\pm0.014 $  & $-0.444\pm0.011 $   & $-0.441\pm0.014 $\\
& [2.5, 4]   & $-0.063\pm0.101 $  & $-0.182\pm0.116 $  &  $-0.317\pm0.077 $  & $-0.327\pm0.075 $  & $-0.184\pm0.111 $   & $-0.313\pm0.094 $\\
& [4, 6]     & $ 0.293\pm0.063 $  & $ 0.231\pm0.103 $  &  $ 0.058\pm0.110 $  & $ 0.046\pm0.107 $  & $ 0.230\pm0.100 $   & $ 0.070\pm0.104 $\\
& [1.1, 6]   & $ 0.025\pm0.083 $  & $-0.065\pm0.091 $  &  $-0.196\pm0.075 $  & $-0.204\pm0.079 $  & $-0.067\pm0.079 $   & $-0.190\pm0.088 $\\
& [15, 19]    & $ 0.373\pm0.023 $  & $ 0.371\pm0.023 $  &  $ 0.337\pm0.032 $  & $ 0.335\pm0.032 $  & $ 0.370\pm0.025 $   & $ 0.342\pm0.030 $\\ 
 \hline
\end{tabular}}
\caption{The angular observable $P_2$ in $B^0 \to K^\star\, e^+\,e^-$ and $B^0 \to K^\star\, \mu^+\,\mu^-$ in SM and in the presence of various NP scenarios. Each entry denotes the central value and the corresponding $1\sigma$ error.}  
 \label{tab_ap7}
 \end{table}

\begin{table}[ht]
 \centering
 \setlength{\tabcolsep}{6pt} 
 \renewcommand{\arraystretch}{1.5} 
 \resizebox{\columnwidth}{!}{
 \begin{tabular}{|c|c|c|c|c|c|c|c|}
 \hline
Observable & $q^2$ bin (GeV$^2$) & SM & S3 & S9 & S10 & S18 & S42 \\
\hline \hline
\multirow{5}{*}{$\langle \rm P'_4 \rangle(B^0 \to K^\star\, e^+\,e^-)$}
& [1.1, 2.5] & $-0.060\pm0.046$  & $-0.008\pm0.044 $ &  $-0.046\pm0.037 $  & $-0.046\pm0.036$  & $-0.015\pm0.038 $   & $0.128\pm0.040 $\\
& [2.5, 4]   & $ -0.392\pm0.043$ & $-0.346\pm0.052$  &  $-0.331\pm0.057$   & $-0.328\pm0.054 $ & $-0.352\pm0.058 $   & $-0.300\pm0.081 $\\
& [4, 6]     & $-0.503\pm0.027$  & $-0.490 \pm0.036$ &  $-0.474\pm0.042 $  & $-0.472\pm0.039 $ & $-0.492\pm0.039 $   & $-0.495\pm0.039 $\\
& [1.1, 6]   & $ -0.351\pm0.040$ & $-0.307\pm0.044$  &  $-0.302\pm0.047 $  & $-0.299\pm0.048$  & $-0.313\pm0.049 $   & $-0.247\pm0.066 $\\
& [15, 19]    & $-0.636\pm0.010 $ & $-0.636\pm0.011$  &  $-0.636\pm0.011 $  & $-0.636\pm0.011 $ & $-0.636\pm0.009 $   & $-0.637\pm0.011 $\\ \hline
\multirow{5}{*}{$\langle \rm P'_4 \rangle(B^0 \to K^\star\, \mu^+\,\mu^-)$}
& [1.1, 2.5] & $-0.061\pm0.045 $  & $ -0.009\pm0.042$  &  $-0.047\pm0.039 $  & $-0.047\pm0.036 $  & $-0.008\pm0.043 $   & $-0.038\pm0.038 $\\
& [2.5, 4]   & $-0.392\pm0.046 $  & $ -0.346\pm0.048$  &  $-0.331\pm0.041 $  & $-0.328\pm0.055 $  & $-0.345\pm0.054 $   & $-0.327\pm0.050 $\\
& [4, 6]     & $-0.503\pm0.029 $  & $ -0.490\pm0.037$  &  $-0.474\pm0.040 $  & $-0.472\pm0.035 $  & $-0.489\pm0.038 $   & $-0.474\pm0.039 $\\
& [1.1, 6]   & $-0.353\pm0.039 $  & $ -0.310\pm0.040$  &  $-0.304\pm0.043 $  & $-0.301\pm0.044 $  & $-0.309\pm0.047 $   & $-0.300\pm0.042 $\\
& [15, 19]    & $-0.636\pm0.010 $  & $ -0.636\pm0.010$  &  $-0.636\pm0.011 $  & $-0.636\pm0.010 $  & $-0.636\pm0.012 $   & $-0.636\pm0.010 $\\ 
 \hline
\end{tabular}}
\caption{The angular observable $P'_4$ in $B^0 \to K^\star\, e^+\,e^-$ and $B^0 \to K^\star\, \mu^+\,\mu^-$ in SM and in the presence of various NP scenarios. Each entry denotes the central value and the corresponding $1\sigma$ error.}  
 \label{tab_ap9}
 \end{table}

\begin{table}[ht]
 \centering
 \setlength{\tabcolsep}{6pt} 
 \renewcommand{\arraystretch}{1.5} 
 \resizebox{\columnwidth}{!}{
 \begin{tabular}{|c|c|c|c|c|c|c|c|}
 \hline
Observable & $q^2$ bin (GeV$^2$) & SM & S3 & S9 & S10 & S18 & S42 \\
\hline \hline
\multirow{5}{*}{$\langle \rm P'_5 \rangle(B^0 \to K^\star\, e^+\,e^-)$}
& [1.1, 2.5] & $ 0.139\pm0.089 $  & $ 0.243\pm0.075 $  &  $ 0.408\pm0.070 $  & $ 0.420\pm0.069 $  & $ 0.229\pm0.085 $   & $ 0.143\pm0.081 $\\
& [2.5, 4]   & $-0.498\pm0.102 $  & $-0.387\pm0.126 $  &  $-0.134\pm0.129 $  & $-0.116\pm0.126 $  & $-0.404\pm0.118 $   & $-0.568\pm0.120 $\\
& [4, 6]     & $-0.755\pm0.080 $  & $-0.710\pm0.084 $  &  $-0.503\pm0.120 $  & $-0.489\pm0.110 $  & $-0.717\pm0.092 $   & $-0.846\pm0.059 $\\
& [1.1, 6]   & $-0.440\pm0.093 $  & $-0.339\pm0.107 $  &  $-0.120\pm0.104 $  & $-0.104\pm0.115 $  & $-0.354\pm0.106 $   & $-0.458\pm0.094 $\\
& [15, 19]    & $-0.594\pm0.035 $  & $-0.590\pm0.050 $  &  $-0.542\pm0.051 $  & $-0.539\pm0.059 $  & $-0.591\pm0.041 $   & $-0.575\pm0.038 $\\ \hline
\multirow{5}{*}{$\langle \rm P'_5 \rangle(B^0 \pm\to K^\star\, \mu^+\,\mu^-)$}
& [1.1, 2.5] & $ 0.140\pm0.077 $  & $ 0.245\pm0.077 $  &  $ 0.413\pm0.071 $  & $ 0.424\pm0.069 $  & $ 0.247\pm0.085 $   & $ 0.403\pm0.072 $\\
& [2.5, 4]   & $-0.501\pm0.105 $  & $-0.390\pm0.122 $  &  $-0.135\pm0.126 $  & $-0.117\pm0.132 $  & $-0.388\pm0.109 $   & $-0.152\pm0.124 $\\
& [4, 6]     & $-0.759\pm0.063 $  & $-0.713\pm0.089 $  &  $-0.505\pm0.105 $  & $-0.491\pm0.120 $  & $-0.712\pm0.091 $   & $-0.525\pm0.126 $\\
& [1.1, 6]   & $-0.447\pm0.094 $  & $-0.346\pm0.102 $  &  $-0.124\pm0.119 $  & $-0.108\pm0.124 $  & $-0.344\pm0.102 $   & $-0.139\pm0.110 $\\
& [15, 19]    & $-0.595\pm0.040 $  & $-0.591\pm0.045 $  &  $-0.543\pm0.054 $  & $-0.539\pm0.053 $  & $-0.591\pm0.043 $   & $-0.550\pm0.050 $\\ 
 \hline
\end{tabular}}
\caption{The angular observable $P'_5$ in $B^0 \to K^\star\, e^+\,e^-$ and $B^0 \to K^\star\, \mu^+\,\mu^-$ in SM and in the presence of various NP scenarios. Each entry denotes the central value and the corresponding $1\sigma$ error.}  
 \label{tab_ap10}
 \end{table}

\begin{table}[ht]
 \centering
 \setlength{\tabcolsep}{6pt} 
 \renewcommand{\arraystretch}{1.5} 
 \resizebox{\columnwidth}{!}{
 \begin{tabular}{|c|c|c|c|c|c|c|c|}
 \hline
Observable & $q^2$ bin (GeV$^2$) & SM & S3 & S9 & S10 & S18 & S42 \\
\hline \hline
\multirow{5}{*}{$\langle \rm P_3 \rangle(B^0 \to K^\star\, e^+\,e^-)$}
& [1.1, 2.5] & $ 0.004\pm0.023 $  & $ 0.004\pm0.028 $  &  $ 0.002\pm0.024 $  & $ 0.002\pm 0.024$  & $ 0.004\pm0.027 $   & $ 0.005\pm0.031 $\\
& [2.5, 4]   & $ 0.004\pm0.009 $  & $ 0.004\pm0.013 $  &  $ 0.003\pm0.016 $  & $ 0.003\pm 0.017$  & $ 0.004\pm0.016 $   & $ 0.009\pm0.022 $\\
& [4, 6]     & $ 0.003\pm0.017 $  & $ 0.003\pm0.014 $  &  $ 0.002\pm0.011 $  & $ 0.002\pm 0.011$  & $ 0.003\pm0.021 $   & $ 0.005\pm0.036 $\\
& [1.1, 6]   & $ 0.003\pm0.008 $  & $ 0.003\pm0.011 $  &  $ 0.002\pm0.013 $  & $ 0.002\pm 0.014$  & $ 0.003\pm0.011 $   & $ 0.006\pm0.016 $\\
& [15, 19]    & $-0.000\pm0.015 $  & $-0.001\pm0.017 $  &  $-0.001\pm0.015 $  & $-0.001\pm0.016 $  & $-0.001\pm0.015 $   & $-0.001\pm0.026 $\\ \hline
\multirow{5}{*}{$\langle \rm P_3 \rangle(B^0 \to K^\star\, \mu^+\,\mu^-)$}
& [1.1, 2.5] & $ 0.004\pm0.024 $  & $ 0.004\pm0.026 $  &  $ 0.002\pm0.023 $  & $ 0.002\pm0.019 $  & $ 0.004\pm0.025 $   & $ 0.002\pm0.025 $\\
& [2.5, 4]   & $ 0.004\pm0.010 $  & $ 0.004\pm0.015 $  &  $ 0.003\pm0.018 $  & $ 0.003\pm0.019 $  & $ 0.004\pm0.015 $   & $ 0.003\pm0.019 $\\
& [4, 6]     & $ 0.003\pm0.015 $  & $ 0.003\pm0.017 $  &  $ 0.002\pm0.009 $  & $ 0.002\pm0.011 $  & $ 0.003\pm0.018 $   & $ 0.002\pm0.011 $\\
& [1.1, 6]   & $ 0.003\pm0.009 $  & $ 0.003\pm0.012 $  &  $ 0.002\pm0.015 $  & $ 0.002\pm0.013 $  & $ 0.003\pm0.010 $   & $ 0.003\pm0.013 $\\
& [15, 19]    & $-0.000\pm0.014 $  & $-0.001\pm0.018 $  &  $-0.001\pm0.015 $  & $-0.001\pm0.014 $  & $-0.001\pm0.018 $   & $-0.001\pm0.017 $\\ 
 \hline
\end{tabular}}
\caption{The angular observable $P_3$ in $B^0 \to K^\star\, e^+\,e^-$ and $B^0 \to K^\star\, \mu^+\,\mu^-$ in SM and in the presence of various NP scenarios. Each entry denotes the central value and the corresponding $1\sigma$ error.}  
 \label{tab_ap8}
 \end{table}

\begin{table}[ht]
 \centering
 \setlength{\tabcolsep}{6pt} 
 \renewcommand{\arraystretch}{1.5} 
 \resizebox{\columnwidth}{!}{
 \begin{tabular}{|c|c|c|c|c|c|c|c|}
 \hline
Observable & $q^2$ bin (GeV$^2$) & SM & S3 & S9 & S10 & S18 & S42 \\
\hline \hline
\multirow{5}{*}{$\langle \rm P'_6 \rangle(B^0 \to K^\star\, e^+\,e^-)$}
& [1.1, 2.5] & $-0.068\pm0.072 $  & $-0.069\pm0.079 $  &  $-0.069\pm0.078 $  & $-0.069\pm0.070 $  & $-0.069\pm0.069 $   & $-0.064\pm0.077 $\\
& [2.5, 4]   & $-0.052\pm0.098 $  & $-0.058\pm0.107 $  &  $-0.059\pm0.127 $  & $-0.059\pm0.116 $  & $-0.057\pm0.114 $   & $-0.062\pm0.129 $\\
& [4, 6]     & $-0.030\pm0.119 $  & $-0.035\pm0.142 $  &  $-0.038\pm0.141 $  & $-0.039\pm0.156 $  & $-0.034\pm0.138 $   & $-0.035\pm0.130 $\\
& [1.1, 6]   & $-0.046\pm0.104 $  & $-0.051\pm0.101 $  &  $-0.053\pm0.108 $  & $-0.054\pm0.112 $  & $-0.050\pm0.097 $   & $-0.050\pm0.107 $\\
& [15, 19]    & $-0.002\pm0.075 $  & $-0.003\pm0.068 $  &  $-0.003\pm0.079 $  & $-0.003\pm0.088 $  & $-0.002\pm0.073 $   & $-0.002\pm0.059 $\\ \hline
\multirow{5}{*}{$\langle \rm P'_6 \rangle(B^0 \to K^\star\, \mu^+\,\mu^-)$}
& [1.1, 2.5] & $-0.069\pm0.075 $  & $-0.069\pm0.077 $  &  $-0.070\pm0.082 $  & $-0.070\pm0.077 $  & $-0.069\pm0.077 $   & $-0..70\pm0.174 $\\
& [2.5, 4]   & $-0.052\pm0.103 $  & $-0.058\pm0.111 $  &  $-0.059\pm0.126 $  & $-0.060\pm0.122 $  & $-0.058\pm0.112 $   & $-0.060\pm0.121 $\\
& [4, 6]     & $-0.030\pm0.115 $  & $-0.035\pm0.129 $  &  $-0.038\pm0.134 $  & $-0.039\pm0.112 $  & $-0.034\pm0.140 $   & $-0.039\pm0.118 $\\
& [1.1, 6]   & $-0.046\pm0.113 $  & $-0.051\pm0.111 $  &  $-0.054\pm0.106 $  & $-0.054\pm0.105 $  & $-0.051\pm0.109 $   & $-0.054\pm0.116 $\\
& [15, 19]    & $-0.002\pm0.064 $  & $-0.003\pm0.079 $  &  $-0.003\pm0.100 $  & $-0.003\pm0.081 $  & $-0.003\pm0.072 $   & $-0.003\pm0.084 $\\ 
 \hline
\end{tabular}}
\caption{The angular observable $P'_6$ in $B^0 \to K^\star\, e^+\,e^-$ and $B^0 \to K^\star\, \mu^+\,\mu^-$ in SM and in the presence of various NP scenarios. Each entry denotes the central value and the corresponding $1\sigma$ error.}  
 \label{tab_ap11}
 \end{table}

\begin{table}[ht]
 \centering
 \setlength{\tabcolsep}{6pt} 
 \renewcommand{\arraystretch}{1.5} 
 \resizebox{\columnwidth}{!}{
 \begin{tabular}{|c|c|c|c|c|c|c|c|}
 \hline
Observable & $q^2$ bin (GeV$^2$) & SM & S3 & S9 & S10 & S18 & S42 \\
\hline \hline
\multirow{5}{*}{$\langle \rm P'_8 \rangle(B^0 \to K^\star\, e^+\,e^-)$}
& [1.1, 2.5] & $-0.018\pm0.037 $  & $-0.016\pm0.041 $  &  $-0.009\pm0.029 $  & $-0.009\pm0.032 $  & $-0.016\pm0.035 $   & $-0.024\pm0.046 $\\
& [2.5, 4]   & $-0.017\pm0.037 $  & $-0.018\pm0.044 $  &  $-0.012\pm0.035 $  & $-0.012\pm0.035 $  & $-0.018\pm0.040 $   & $-0.031\pm0.068 $\\
& [4, 6]     & $-0.012\pm0.035 $  & $-0.013\pm0.041 $  &  $-0.009\pm0.034 $  & $-0.010\pm0.031 $  & $-0.013\pm0.036 $   & $-0.020\pm0.050 $\\
& [1.1, 6]   & $-0.015\pm0.033 $  & $-0.015\pm0.036 $  &  $-0.010\pm0.033 $  & $-0.010\pm0.034 $  & $-0.015\pm0.037 $   & $-0.024\pm0.051 $\\
& [15, 19]    & $ 0.001\pm0.018 $  & $ 0.001\pm0.025 $  &  $ 0.001\pm0.022 $  & $ 0.001\pm0.019 $  & $ 0.001\pm0.023 $   & $ 0.001\pm0.031 $\\ \hline
\multirow{5}{*}{$\langle \rm P'_8 \rangle(B^0 \to K^\star\, \mu^+\,\mu^-)$}
& [1.1, 2.5] & $-0.018\pm0.033 $  & $-0.016\pm0.039 $  &  $-0.009\pm 0.029$  & $-0.009\pm0.028 $  & $-0.016\pm0.032 $   & $-0.010\pm0.032 $\\
& [2.5, 4]   & $-0.017\pm0.037 $  & $-0.018\pm0.041 $  &  $-0.012\pm0.032 $  & $-0.012\pm0.037 $  & $-0.018\pm0.044 $   & $-0.013\pm0.034 $\\
& [4, 6]     & $-0.012\pm0.036 $  & $-0.013\pm0.040 $  &  $-0.010\pm0.031 $  & $-0.010\pm0.031 $  & $-0.013\pm0.039 $   & $-0.010\pm0.036 $\\
& [1.1, 6]   & $-0.015\pm0.028 $  & $-0.015\pm0.036 $  &  $-0.010\pm0.033 $  & $-0.010\pm0.029 $  & $-0.015\pm0.037 $   & $-0.011\pm0.032 $\\
& [15, 19]    & $ 0.001\pm0.023 $  & $ 0.001\pm0.022 $  &  $ 0.001\pm0.022 $  & $ 0.001\pm0.021 $  & $ 0.001\pm0.024 $   & $ 0.001\pm0.022 $\\ 
 \hline
\end{tabular}}
\caption{The angular observable $P'_8$ in $B^0 \to K^\star\, e^+\,e^-$ and $B^0 \to K^\star\, \mu^+\,\mu^-$ in SM and in the presence of various NP scenarios. Each entry denotes the central value and the corresponding $1\sigma$ error.}  
 \label{tab_p8p}
 \end{table}

\begin{table}[ht]
 \centering
 \setlength{\tabcolsep}{6pt} 
 \renewcommand{\arraystretch}{1.5} 
 \begin{tabular}{|c|c|c|c|c|c|c|}
 \hline
Observable & $q^2$ bin $\rm(GeV^2)$  & $\Delta A_{FB}$ & $\Delta F_L$ & $\Delta P_1$ & $\Delta P_2$ & $\Delta P_3$ \\
\hline \hline
\multirow{4}{*}{SM}
& [1.1, 2.5] & $0.003\pm0.042$ & $-0.028\pm0.066$ & $-0.000\pm0.073$ & $-0.006\pm0.018$ & $0.000\pm0.033$ \\
& [2.5, 4]   & $0.000\pm0.044$ & $-0.014\pm0.051$ & $-0.000\pm0.051$ & $-0.000\pm0.145$ & $0.000\pm0.014$\\
& [4, 6]     & $-0.001\pm0.056$& $-0.007\pm0.068$ & $-0.000\pm0.066$ & $0.001\pm0.104$ & $0.000\pm0.023$\\
& [1.1, 6]   & $0.001\pm0.049$ & $-0.015\pm0.056$ & $-0.001\pm0.047$ & $0.003\pm0.124$ & $0.000\pm0.014$\\
 \hline
 \multirow{4}{*}{S42}
& [1.1, 2.5] & $-0.065\pm0.042$ & $-0.111\pm0.081$ & $-0.043\pm0.082$ & $-0.045\pm0.023$ & $-0.003\pm0.039$ \\
& [2.5, 4]   & $-0.089\pm0.048$ & $-0.118\pm0.051$ & $ 0.023\pm0.073$ & $-0.198\pm0.153$ & $-0.006\pm0.028$\\
& [4, 6]     & $-0.093\pm0.065$ & $-0.086\pm0.067$ & $ 0.052\pm0.078$ & $-0.303\pm0.130$ & $-0.003\pm0.027$\\
& [1.1, 6]   & $-0.084\pm0.051$ & $-0.102\pm0.067$ & $ 0.014\pm0.048$ & $-0.201\pm0.116$ & $-0.003\pm0.022$\\
 \hline
\end{tabular}
 \caption{The binned values for $\Delta A_{FB}$, $\Delta F_L$, $\Delta P_1$, $\Delta P_2$ and $\Delta P_3$ observables in SM and NP scenarios S42}  
 \label{tab_delta1}
 \end{table}

 \begin{table}[ht]
 \centering
 \setlength{\tabcolsep}{6pt} 
 \renewcommand{\arraystretch}{1.5} 
 \begin{tabular}{|c|c|c|c|c|c|}
 \hline
Observable & $q^2$ bin (GeV$^2$) & $\Delta P'_4$ & $\Delta P'_5$ & $\Delta P'_6$ & $\Delta P'_8$ \\
\hline \hline
\multirow{4}{*}{SM}
& [1.1, 2.5] & $-0.001\pm0.068 $  & $ 0.001\pm0.112 $  &  $-0.001\pm0.112 $  & $-0.000\pm0.050 $\\
& [2.5, 4]   & $-0.000\pm0.063 $  & $-0.004\pm0.138 $  &  $-0.000\pm0.157 $  & $ 0.000\pm0.052 $\\
& [4, 6]     & $-0.000\pm0.039 $  & $-0.003\pm0.102 $  &  $-0.000\pm0.160 $  & $ 0.000\pm0.053 $\\
& [1.1, 6]   & $-0.002\pm0.055 $  & $-0.007\pm0.120 $  &  $-0.000\pm0.147 $  & $ 0.000\pm0.041 $\\
 \hline
 \multirow{4}{*}{S42}
& [1.1, 2.5] & $-0.166\pm0.052$ & $ 0.259\pm0.099$ & $-0.007\pm0.109$ & $ 0.014\pm0.061$\\
& [2.5, 4]   & $-0.026\pm0.089$ & $ 0.416\pm0.167$ &  $0.002\pm0.165$ & $ 0.018\pm0.073$\\
& [4, 6]     & $ 0.022\pm0.057$ & $ 0.321\pm0.135$ & $-0.004\pm0.200$ & $ 0.009\pm0.066$\\
& [1.1, 6]   & $-0.053\pm0.067$ & $ 0.319\pm0.161$ & $-0.004\pm0.148$ & $ 0.013\pm0.054$\\
 \hline
\end{tabular}
 \caption{The binned values for $\Delta P'_4$, $\Delta P'_5$, $\Delta P'_6$ and $\Delta P'_8$ observables in SM and NP scenarios S42}  
 \label{tab_delta2}
 \end{table}

\FloatBarrier
\section{Prediction of $B_s \to \phi \ell^+ \ell^-$ and $\Lambda_b \to \Lambda\, \ell^+\,\ell^-$ ($\ell \in e,\mu$) observables in SM and in the presence of various NP scenarios at different $q^2$ bins}
\label{ap4}

\begin{table}[h]
 \centering
 \setlength{\tabcolsep}{6pt} 
 \renewcommand{\arraystretch}{1.5} 
 \resizebox{\columnwidth}{!}{
 \begin{tabular}{|c|c|c|c|c|c|c|c|}
 \hline
Observable & $q^2$ bin (GeV$^2$) & SM & S3 & S9 & S10 & S18 & S42 \\
\hline \hline
\multirow{5}{*}{$\langle \rm BR \rangle (B_s \to \phi e^+ e^-)\times 10^{8}$}
& [1.1, 2.5] & $5.49\pm0.62 $  & $4.51\pm0.53 $  &  $4.28\pm0.52 $  & $4.23\pm0.45 $  & $4.63\pm0.51 $   & $3.76\pm0.51 $\\
& [2.5, 4]   & $5.18\pm0.59 $  & $4.08\pm0.50 $  &  $3.80\pm0.43 $  & $3.74\pm0.42 $  & $4.22\pm0.54 $   & $3.30\pm0.49 $\\
& [4, 6]     & $5.54\pm0.67 $  & $4.31\pm0.61 $  &  $3.94\pm0.47 $  & $3.87\pm0.51 $  & $4.46\pm0.60 $   & $3.48\pm0.58 $\\
& [1.1, 6]   & $5.42\pm0.65 $  & $4.30\pm0.50 $  &  $3.99\pm0.51 $  & $3.93\pm0.49 $  & $4.44\pm0.54 $   & $3.51\pm0.58 $\\
& [15,19]    & $5.59\pm 0.57 $  & $4.37\pm 0.46 $  &  $3.86\pm 0.33 $  & $3.78\pm 0.36 $  & $4.53\pm 0.43 $   & $3.75\pm 0.42 $\\ \cline{1-1}
\multirow{5}{*}{$\langle \rm BR \rangle (B_s \to \phi\, \mu^+\,\mu^-)\times 10^{8}$}
& [1.1, 2.5] & $5.47\pm0.61 $  & $4.49\pm0.52 $  &  $4.27\pm0.44 $  & $4.22\pm0.44 $  & $4.48\pm0.51 $   & $4.21\pm0.45 $\\
& [2.5, 4]   & $5.17\pm0.59 $  & $4.07\pm0.52 $  &  $3.78\pm0.43 $  & $3.72\pm0.46 $  & $4.06\pm0.50 $   & $3.72\pm0.49 $\\
& [4, 6]     & $5.53\pm0.76 $  & $4.30\pm0.67 $  &  $3.92\pm0.48 $  & $3.85\pm0.53 $  & $4.28\pm0.60 $   & $3.86\pm0.50 $\\
& [1.1, 6]   & $5.40\pm0.63 $  & $4.28\pm0.53 $  &  $3.98\pm0.42 $  & $3.92\pm0.38 $  & $4.27\pm0.50 $   & $3.91\pm0.44 $\\
& [15,19]    & $5.58\pm0.57 $  & $4.37\pm0.48 $  &  $3.85\pm0.42 $  & $3.78\pm0.40 $  & $4.35\pm0.43 $   & $3.80\pm0.36 $\\
 \hline
\end{tabular}}
\caption{The branching ratios of $B_s \to \phi\, e^+\,e^-$ and $B_s \to \phi\, \mu^+\,\mu^-$ in SM and in the presence of various NP scenarios at different $q^2$ bins. Each entry denotes the central value and the corresponding $1\sigma$ error.}  
 \label{tab_ap13}
 \end{table}

\begin{table}[ht]
 \centering
 \setlength{\tabcolsep}{6pt} 
 \renewcommand{\arraystretch}{1.5} 
 \resizebox{\columnwidth}{!}{
 \begin{tabular}{|c|c|c|c|c|c|c|c|}
 \hline
Observable & $q^2$ bin (GeV$^2$) & SM & S3 & S9 & S10 & S18 & S42 \\
\hline \hline
\multirow{5}{*}{$\langle \rm R_\phi \rangle $}
& [1.1, 2.5] & $0.997\pm0.000 $  & $0.997\pm0.000 $  &  $0.997\pm0.000 $  & $0.997\pm0.000 $  & $0.968\pm0.002 $   & $1.119\pm0.029 $\\
& [2.5, 4]   & $0.997\pm0.000 $  & $0.997\pm0.000 $  &  $0.997\pm0.000 $  & $0.996\pm0.000 $  & $0.961\pm0.002 $   & $1.127\pm0.049 $\\
& [4, 6]     & $0.997\pm0.000 $  & $0.997\pm0.000 $  &  $0.997\pm0.000 $  & $0.997\pm0.000 $  & $0.959\pm0.002 $   & $1.106\pm0.056 $\\
& [1.1, 6]   & $0.997\pm0.000 $  & $0.997\pm0.000 $  &  $0.997\pm0.000 $  & $0.997\pm0.000 $  & $0.963\pm0.002 $   & $1.116\pm0.037 $\\
& [15,19]    & $0.998\pm0.000 $  & $0.998\pm0.000 $  &  $0.998\pm0.000 $  & $0.998\pm0.000 $  & $0.961\pm0.001 $   & $1.015\pm0.019 $ \\
 \hline
\end{tabular}}
\caption{The ratio of branching ratio $R_\phi$ in SM and in the presence of various NP scenarios at different $q^2$ bins. Each entry denotes the central value and the corresponding $1\sigma$ error.}  
 \label{tab_ap14}
 \end{table}

\begin{table}[ht]
 \centering
 \setlength{\tabcolsep}{6pt} 
 \renewcommand{\arraystretch}{1.5} 
 \resizebox{\columnwidth}{!}{
 \begin{tabular}{|c|c|c|c|c|c|c|c|}
 \hline
Observable & $q^2$ bin (GeV$^2$) & SM & S3 & S9 & S10 & S18 & S42 \\
\hline \hline
\multirow{2}{*}{$\langle \rm BR \rangle(\Lambda_b \to \Lambda \, e^+\,e^-)\times 10^{8}$}
& [1.1, 6]   & $1.03\pm0.63 $  & $0.83\pm0.50 $   &  $0.79\pm0.41 $  & $0.78\pm0.43 $  & $0.85\pm0.45 $   & $0.66\pm0.39 $\\
& [15, 20]   & $7.09\pm0.83 $  & $5.55\pm0.59 $   &  $4.90\pm0.53 $  & $4.80\pm0.53 $  & $5.75\pm0.61 $   & $4.76\pm0.62 $\\ \cline{1-1}
\multirow{2}{*}{$\langle \rm BR \rangle(\Lambda_b \to \Lambda \, \mu^+\,\mu^-)\times 10^{8}$}
& [1.1, 6]   & $1.03\pm0.59 $  & $0.83\pm0.44 $   &  $0.79\pm0.33 $  & $0.78\pm0.35 $  & $0.82\pm0.43 $   & $0.77\pm0.39 $\\
& [15, 20]   & $7.09\pm0.82 $  & $5.55\pm0.59 $   &  $4.90\pm0.57 $  & $4.80\pm0.50 $  & $5.53\pm0.61 $   & $4.84\pm0.59 $\\
 \hline
\end{tabular}}
\caption{The branching ratios of $\Lambda_b \to \Lambda\, e^+\,e^-$ and $\Lambda_b \to \Lambda\, \mu^+\,\mu^-$ in SM and in the presence of various NP scenarios at different $q^2$ bins. Each entry denotes the central value and the corresponding $1\sigma$ error.}  
 \label{tab_ap15}
 \end{table}

\FloatBarrier

\bibliographystyle{ieeetr}
\bibliography{ref}

\end{document}